\documentclass[a4paper,fleqn,usenatbib]{mnras}

\usepackage{newtxtext,newtxmath}

\usepackage[T1]{fontenc}
\usepackage{ae,aecompl}

\usepackage{graphicx}	
\usepackage{amsmath}	
\usepackage{amssymb}	
\usepackage{multirow}

\usepackage[utf8]{inputenc}
\usepackage[export]{adjustbox}
\usepackage{wrapfig}
\usepackage{dutchcal}
\usepackage{braket}

\graphicspath{ {images/} }

\newcommand{\lya}{\mbox{$\rmn{Ly}\alpha$}}

\newcommand{\llya}{\mbox{$\rm L_{{\rm Ly}\alpha}$}}
\newcommand{\galform}{\texttt{GALFORM}}
\newcommand{\pmill}{\texttt{P-Millennium}}
\newcommand{\flareon}{\texttt{FLaREON}}
\newcommand{\fesc}{\mbox{$f_{\rmn esc}^{\rm Ly \alpha}$}}

\newcommand{\slae}{SLAE}
\newcommand{\nlae}{PLAE}
\newcommand{\flae}{FLAE}
\newcommand{\ZZ}{ZZ11}
\newcommand{\BC}{BC17}
\newcommand{\Munits}{ {\rm M_{\odot} }h^{-1} }

\title [Ly$\it \alpha$ emitters in a cosmological volume II]{\lya\ emitters in a cosmological volume II: the impact of the intergalactic medium}

\author[S. Gurung-L\'opez. et al.]{
Siddhartha Gurung-L\'opez,$^{1}$\thanks{E-mail: sidgurung@cefca.es}
\'Alvaro A. Orsi,$^{1}$
Silvia Bonoli,$^{1,2}$ 
Nelson Padilla,$^{3}$
\newauthor
Cedric G. Lacey,$^{4}$
and
Carlton M. Baugh$^{4}$.
\\
$^{1}$Centro de Estudios de F\'isica del Cosmos de Arag\'on, Plaza San Juan 1, piso 2, Teruel, 44001, Spain. \\
$^{2}$ DIPC, Manuel Lardizabal Ibilbidea, 4, 20018 San Sebastian, Spain. \\
$^{3}$ Centro de Astro-Ingenier\'ia, Pontificia Universidad Cat\'olica de Chile, Santiago, Chile \\
$^{4}$ Institute for Computational Cosmology, Durham University.
}

\date{Accepted XXX. Received YYY; in original form ZZZ}

\pubyear{2018}

\begin{document}
\label{firstpage}
\pagerange{\pageref{firstpage}--\pageref{lastpage}}
\maketitle

\begin{abstract}
In the near future galaxy surveys will target Lyman alpha emitting galaxies (LAEs) to unveil the nature of  dark energy. It has been suggested that the observability of LAEs is coupled to the large scale properties of the intergalactic medium. Such coupling could introduce distortions into the observed clustering of LAEs, adding a new potential difficulty to the interpretation of upcoming surveys. We present a model of LAEs that incorporates \lya\ radiative transfer  processes in the interstellar and intergalactic medium. The model is implemented in the \galform\ semi-analytic model of galaxy of formation and evolution. We find that the radiative transfer inside galaxies produces selection effects over galaxy properties. In particular, observed LAEs tend to have low metallicities and intermediate star formation rates. At low redshift we find no evidence of a correlation between the spatial distribution of LAEs and the intergalactic medium properties. However, at high redshift the LAEs are linked to the line of sight velocity and density gradient of the intergalactic medium. The strength of the coupling depends on the outflow properties of the galaxies and redshift. This effect modifies the clustering of LAEs on large scales, adding non linear features. In particular, our model predicts modifications to the shape and position of the baryon acoustic oscillation peak. This work highlights the importance of including radiative transfer physics in the cosmological analysis of LAEs.
\end{abstract}

\begin{keywords}
Radiative transfer -- Intergalactic medium -- ISM -- High-redshift -- Emission lines
\end{keywords}



    
\begin{figure*} 
\includegraphics[width=6.9in]{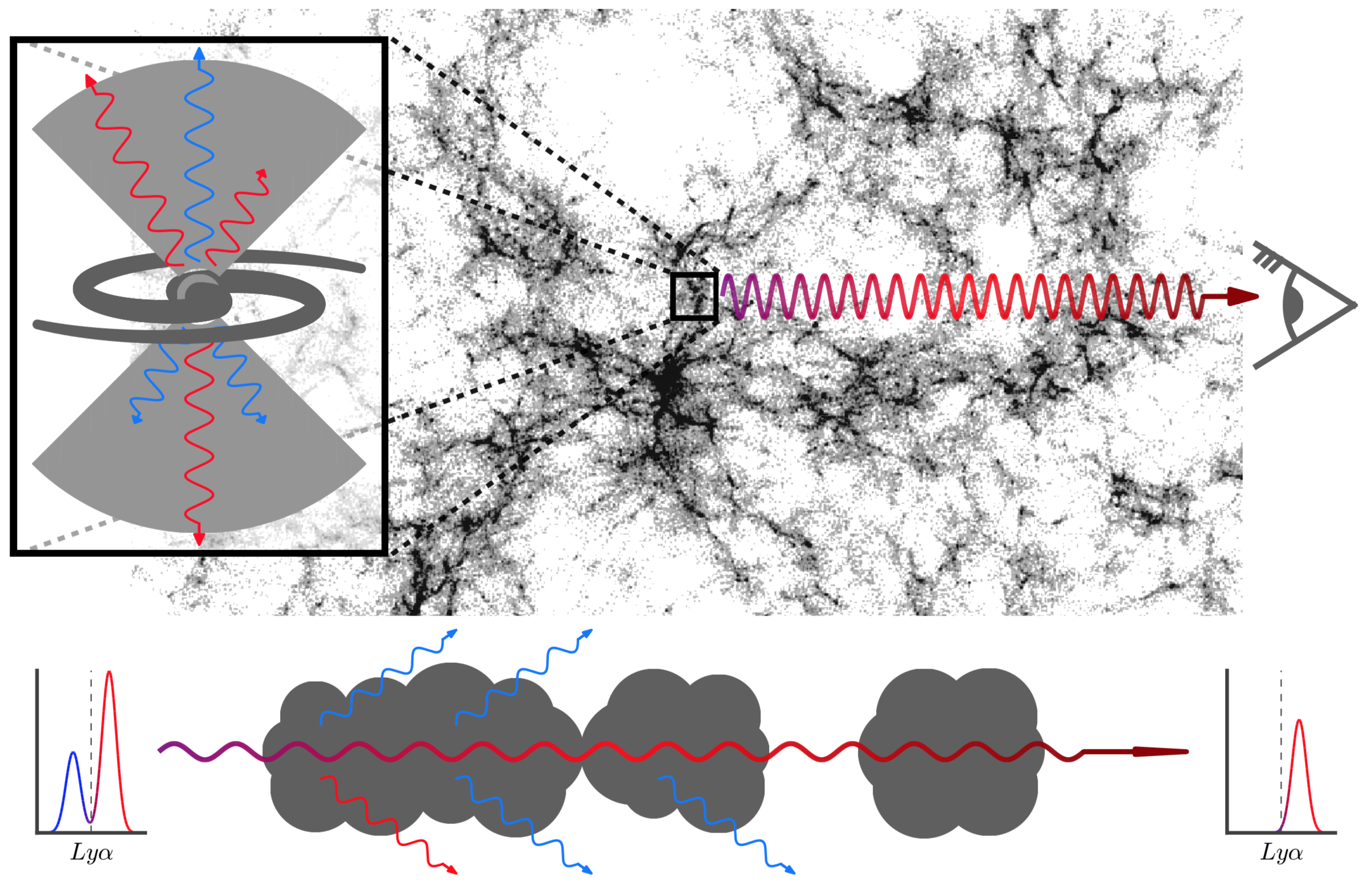} 
\caption{ Illustration of the journey of \lya\ photons  since they are emitted until they reach the observer. Since \lya\ photons are generated in star forming regions they go through RT processes until escaping through galaxy outflows. Then they enter into the IGM, where photons interacting with \ion{H}{I} are scattered out of the line of sight and never reach the observer.    }\label{fig:cartoon}
\end{figure*}


\section{Introduction}

    Galaxies exhibiting strong Lyman-$\alpha$ emission, the so-called LAEs, are one of the most important tracers of the high redshift Universe in modern astrophysics. After their first detection about 20 years ago \citep[e.g.][]{steidel96, hu98,rhoads00, malhotra02}  studies have focused on finding LAEs at low \citep{Orlitova_2018,Henry_2018} and high redshift \citep{ouchi08,oyarzun17,Matthee_2017_boot,Caruana_2018}  including at the epoch of reionization \citep{Sobral_2015_CR7, Ouchi2018a, Shibuya_2018}.

    In the coming years, cosmological surveys such as HETDEX \citep{Hill2008} and J-PAS \citep{J-PAS} aim to unveil the mystery of  the dark energy. These surveys will scan the sky chasing LAEs to trace the underlying dark matter density fluctuations and make clustering measurements. Additionally, these surveys will contribute notably to our knowledge of galaxy formation and evolution. Thus, it is becoming timely to understand properly the selection function of LAEs and how it might affect the  apparent spatial distribution and galaxy properties.

    The large cross-section of neutral hydrogen atoms around \lya\ wavelengths makes these photons suffer multiple scattering events that modify their frequency and direction. These frequency changes produce the characteristic \lya\ line profiles widely studied \citep{verhamme06,Gronke_2016}. The typical distance covered by \lya\ photons inside atomic hydrogen is increased drastically due to multiple scattering events, making these photons very sensitive to dust absorption. 
    
    In order to understand the complex escape of \lya\ photons consider the illustration shown in Fig.\ref{fig:cartoon}. The journey of \lya\ photons begins when a star formation episode takes place inside a galaxy. During these events, hot massive stars (mostly {\it O-type} and {\it B-type}) emit high energy photons capable of ionizing neutral hydrogen. A fraction of these photons dissociates the \ion{H}{I} in the surroundings of the star forming region. Another part is absorbed by dust grains, while the rest escapes from the galaxy and ionizes \ion{H}{I} in the intergalactic medium.
    
    Then, the free electrons within the ionized region surrounding the young stars recombine with  \ion{H}{II} ions in excited energy levels, causing a cascade until electrons reach the ground energy level. \lya\ photons are emitted when an electron decays from the first excited  level to the ground level, an event that occurs with probability $\sim2/3$ \citep{Spitzer_1978}) per ionizing photon. 

    \lya\ photons have to get through the intricate interstellar medium (ISM) before escaping the galaxy (Fig. \ref{fig:cartoon} top left panel)\citep{neufeld91}. The ISM morphology includes dusty gas rich regions such as bars \citep{Spinoso_2017}, arms \citep{Kormendy_2013}, \ion{H}{II} bubbles, outflows \citep{Cazzoli_2016} and other structures that complicate the radiative transfer of \lya\ photons. Resonant scattering inside the ISM enhances dust absorption, thus only a fraction of the emitted \lya\ photons  manages to escape from the galaxy. Additionally, in this process the \lya\ line profile is modified due to consecutive \ion{H}{I} interactions \citep{harrington73}. The final \lya\ escape fraction and line profiles depend strongly on the ISM topology and kinematics \citep{verhamme06,Gurung_2018b}. 
    
    After emerging from galaxies (Fig.\ref{fig:cartoon} zoom panel), \lya\ photons enter the IGM and interact with the \ion{H}{I} atoms within, producing further \lya\ scattering events (Fig.\ref{fig:cartoon} main panel). While inside galaxies the flux reduction is due to dust absorption, in the IGM the \lya\ photons are scattered out of the line of sight, as illustrated in the bottom panel of Fig.\ref{fig:cartoon}. The Hubble flow redshifts the emitted photons, causing further IGM absorptions at  wavelengths bluewards of \lya\, while redder photons travel freely.
    
    The RT processes, occurring in both the ISM and IGM, complicate the selection function of galaxy surveys using \lya\ line detection as a tracer. Galaxies selected with this technique tend to have low metallicity and high specific star formation rates \citep{Sobral_2018_dust}. Observational evidence also suggests that this galaxy population lies in low density environments \citep{Shimakawa_2017}. Further observations are needed to cast light on this matter.
    
    The RT in the ISM has been modeled and explored using Monte Carlo Radiative Transfer codes \citep{zheng02, ahn03, verhamme06, orsi12, Gronke_2016, Gurung_2018b}. These tools generate photons in \ion{H}{I} structures and track the subsequent interactions, changes in direction, frequency and possible absorptions. Monte Carlo Radiative Transfer codes have been implemented in cosmological simulations in the $\Lambda$CDM scenario  to understand the effect of the ISM on the selection function of LAEs \citep{orsi12, garel12,gurung18a}.
    
    Radiative transfer inside the IGM has been implemented in several ways. For example, \cite{dijkstra07}  made use of  analytic expressions to determine the velocity field, density field and ionization state of the IGM around galaxies as a function of some galaxy properties like the host halo mass or the circular velocity. Meanwhile, \cite{laursen11} studied the IGM transmission around the \lya\ wavelength in a hydrodynamic simulation tracking \lya\ rays along different lines of sight. A different approach was taken by \cite{zheng10}, where they run a Monte Carlo radiative transfer code (similar to the algorithms used to model the ISM) to study the  observability of LAEs as a function of the IGM large scale properties.  In particular, \cite{zheng10} found that the resonant nature of \lya\ might introduce new clustering features in  galaxy samples selected by \lya\ detection. However, using a higher resolution simulation,  \cite{Behrens2017} claimed to find only a marginal coupling between LAEs and the IGM. 
    
    If the IGM impacts significantly on the observed spatial distribution of LAEs, this could introduce dramatic biases \citep[e.g.][]{Wyithe_2011} into the cosmological interpretation of clustering data from surveys like HETDEX \citep{hill08}, and future space mission concepts, such as the ATLAS Probe \citep{Atlas_2018} and the Cosmic Dawn Intensity Mapper \citep{Cooray_2016}. Hence, understanding the role of the IGM in shaping the observed properties of high redshift LAEs is of crucial importance.

    This is the second in a series of papers that tackle the selection effects on LAE caused by the RT of \lya\ resonant scattering nature. In our first paper \citep{gurung18a} we focused in the RT inside the ISM and how it determines the properties of galaxies observed as LAEs. Here, we expand our model and include the RT in the nearby IGM. We study the coupling between \lya\ observability and different IGM large scale properties and how this modifies the clustering of LAEs.
    
    This work is structured as follows : In \S \ref{sec:model} we present our model and its calibration (\S \ref{sec:Calibration}). Then, in \S \ref{sec:gal_prop} we briefly study the selection function of galactic properties while in \S \ref{sec:IGM_prop} we focus on the selection effects of the IGM and its impact on LAE clustering (\S \ref{sec:clustering}). Finally, we compare our work with the literature (\S\ref{sec:discussion}) and present our conclusions (\S\ref{sec:conclusion}).

\section{A comprehensive model for LAEs.}\label{sec:model}

The model presented here is a follow up of the work presented in \cite{gurung18a}. We combine a wide range of physical scenarios in order to assemble a realistic LAE model. Our model is built upon four main pillars:\\

\begin{enumerate}
    \item The \pmill\ N-Body simulation \citep{Baugh_2019}, a state-of-the-art dark matter $N$-body simulation with box size $\rm (542.16 cMpc/h)^3$ with $5040^3$ dark matter particles of mass ${\rm M_p = 1.061 \times 10^8 M_{\odot} } h^{-1} $ and Planck cosmology : $\rm H_{0}=67.77 \;km\;s^{-1}Mpc^{-1}$, $\Omega_{\Lambda}=0.693$, $\Omega_{M} = 0.307$ , $\sigma_{8}=0.8288$ \citep{Planck_2016}. The \pmill\ models the hierarchical growth of structures in the $\rm \Lambda CDM$ scenario in our model. 

    \item \galform, a semi-analytic model of galaxy formation and evolution. The \galform\ version used in this work is detailed in \cite{lacey16} and \cite{Baugh_2019} and it is based on \citet{cole00}. 

    In short, initially \galform\ populates the dark matter halos extracted from a high redshift output of the \pmill\ with gas. Then, the gas is evolved tracking the merger histories of halos and several physical mechanisms are included, such as i) shock-heating and radiative cooling of gas inside halos; ii) formation of galactic disk with quiescent star formation ; iii) the triggering of starburst episodes due to disk instabilities and mergers in bulges; iv)  AGN, supernovae and photoionization feedback to regulate the star formation rate; v) the chemical evolution of gas and stars. Additionally, the \galform\ version used in this work implements different stellar initial mass functions (IMFs) for quiescent star formation and starburst  episodes (for further details in \cite{lacey16}). To ensure a proper resolution, in our model we consider only galaxies with stellar masses higher than $10^{7}\Munits$, which roughly corresponds with dark halo masses around $10^{10}\Munits$ ($\sim 100$ dark matter particles).

    \item \flareon \footnote{ \texttt{ https://github.com/sidgurun/FLaREON}} \citep{Gurung_2018b}, an open source python code to predict  Lyman $\alpha$ escape fractions and line profiles in minimal computational time. \flareon\ is based on the radiative transfer Monte Carlo code \texttt{LyaRT}\footnote{ \texttt{ https://github.com/aaorsi/LyaRT} } \citep{orsi12} that fully tracks the  trajectory of \lya\ photons in outflows with different gas geometries, hydrogen column densities ($\rm N_{H}$), macroscopic expansion velocities ($\rm V_{exp}$) and dust optical depths ($\rm \tau_a$). Briefly, \flareon\ combines precomputed grids of \fesc\ and \lya\ line profiles in the $\rm N_H$-$\rm V_{exp}$-$\rm \tau_a$ space for several outflow geometries and different algorithms, such as multi-dimensional interpolation and machine learning, achieving high accuracy at a very low computational cost. We make use of \flareon\ to include the RT inside galaxies in our model (see \S \ref{sec:Galaxy_transmission.}).

    \item Radiative transfer of \lya\ photons in the intergalactic medium. We estimate the IGM transmission for every galaxy depending on the local environment properties, such as the density, velocity and ionization state of the IGM (see \S \ref{sec:IGM_transmission.}). 
\end{enumerate}

In reality there is an important interface between the ISM and the IGM: the  circumgalactic medium (CGM). In principle, the CGM is also populated by neutral hydrogen, implying that \lya\ photons are prone to continue experiencing the resonant scatter. Some observational studies have shown that LAEs present extended \lya\ halos around them \cite[e.g.][]{Leclercq_2017}. Also, theoretical works have studied the impact of the CGM in the RT of \lya\ photons \cite[e.g.][]{zheng11,Wyithe_2011,Behrens2017}. In this work, our simulation lacks the enough spatial resolution to include properly a CGM contribution to our analysis.  However, we expect that the changes introduced by the CGM in our model would be small. In fact, the CGM could modify slightly the line frequency distribution of \lya\ photons entering into the IGM, which in principle could increase the IGM-LAE coupling found in this work (see \S\ref{ssec:IGM_LAE_coupling}) if the CGM is infalling into the galaxy or it could decrease it if the CGM was being ejected from the galaxy. 

In the following subsections we describe in detail the design of our LAE models.

\subsection{Modeling the radiative transfer of ${\bf Ly \alpha}$ photons inside galaxies.}\label{sec:Galaxy_transmission.}

The physics of \lya\ photons escaping galaxies through galactic outflows are implemented with \flareon. We focus on two outflow geometries: i) expanding homogeneous Thin Shell \citep[e.g.][]{verhamme06,orsi12, gurung18a}; ii) expanding Galactic Wind \citep{orsi12, gurung18a} with a density gradient. Both geometries exhibit an empty cavity in the center of the geometry, where monochromatic \lya\ photons are generated. We assume a constant temperature $ T=10^4 K$. These outflow geometries are detailed in \cite{Gurung_2018b}.  

\galform\ galaxies are divided into two components, disk and bulge, with distinct galactic properties, such as metallicity, cold gas mass or star formation rate. Therefore, each component is assigned a unique \fesc\ and \lya\ line profile. This assumes that photons generated in a certain galaxy component only interact with that galaxy component. 

We calculate outflow properties as in \cite{gurung18a}. In particular, motivated by observational studies \citep[e.g.][]{Cazzoli_2016}, the outflow expansion velocity is computed as  

\begin{align} 
\label{eq:recipe-V_exp}
\rm
V_{exp,c}  & =  \kappa_{V,c} {{\rm SFR}_{c}} \frac{ r_{c} }{ M_{*} }, 
\end{align}
where the subscript $c$ denotes the galaxy component (disk or bulge), $\rm SFR_{c}$ is the star formation rate in ${\rm M_{\odot} Gyr ^{-1} } h^{-1}$ units,  $\rm r_{c}$ is the half stellar mass radius in ${\rm pMpc \;} h^{-1}$, $\rm M_{*}$ is the total stellar mas of the galaxy  in ${\rm M_{\odot} } h^{-1}$ units. Additionally, $\rm \kappa_{V,c}$ are free dimensionless parameters regulating the efficiency of gas ejection. 

\begin{figure}
    \includegraphics[width=3.2in]{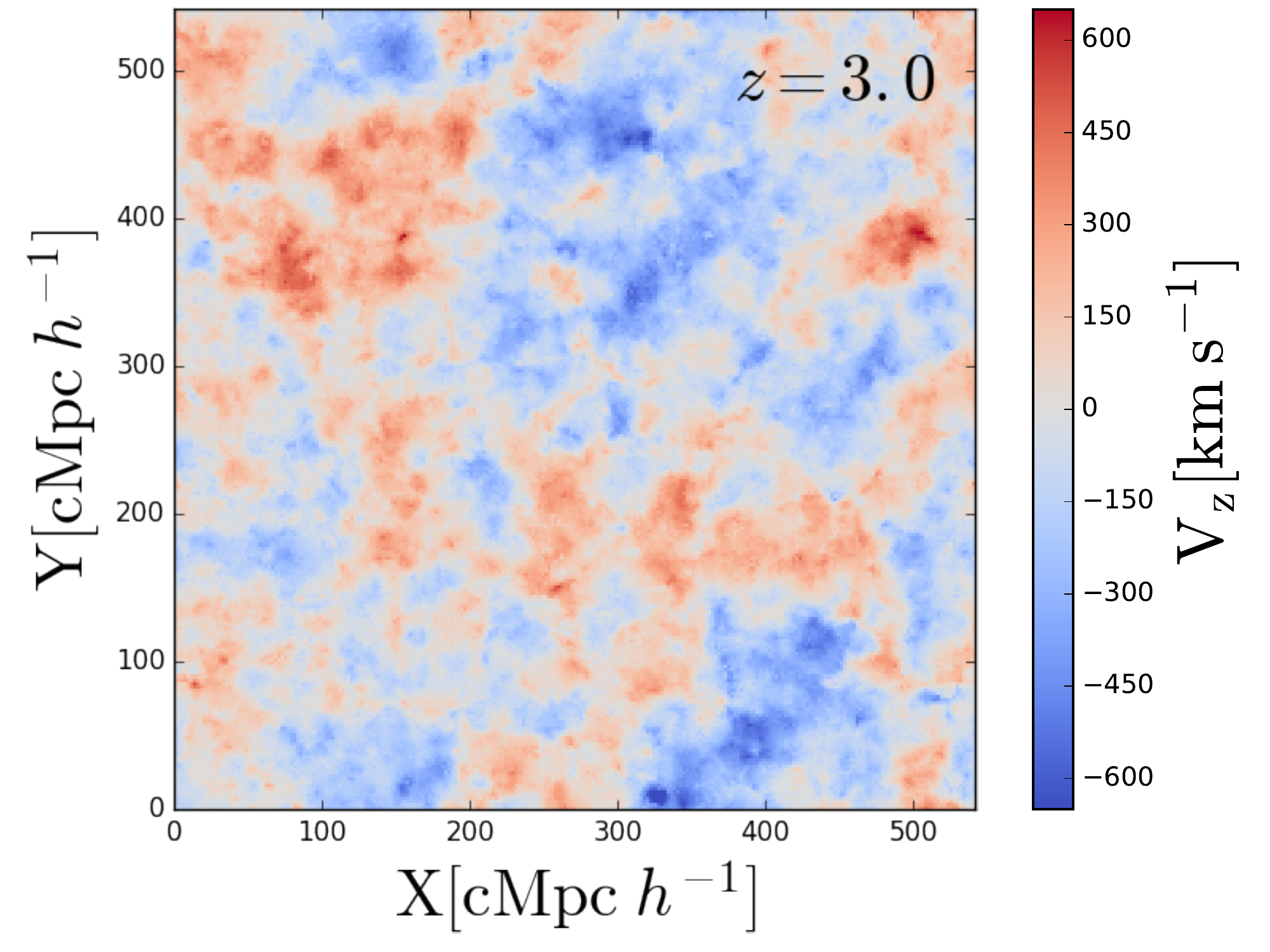} 
    \caption{  Velocity along the line of sight ($Z$ axis) spatial distribution at redshift $z=3$ in a slice of  $\sim 2 {\rm cMpc}h^{-1}$ width. Velocities pointing to the observer are shown in blue, while regions distancing from the observer are drawn in red. }\label{fig:v_field}
\end{figure}

In \flareon, the Thin Shell and the Galactic Wind geometries present different density profiles, thus different column densities. The outflow neutral hydrogen column density of each component is computed as

\begin{equation}\label{eq:recipe-column_density}
{\rm N_{H,c}} = \left\{
\begin{array}{c l}
\kappa_{N , c }  \frac{\rm M_{\rm cold, c}}{ \rm r^2_c} & {\rm Thin \; Shell} \\
\kappa_{N , c }  \frac{\rm   M_{{\rm cold}, c}}{\rm r_c V_{{\rm exp},c} } & {\rm Galactic \; Wind} 
\end{array}
\right. ,
\end{equation}
where $\rm M_{cold ,c}$ is the cold gas mass in ${\rm M_{\odot} } h^{-1}$ and $\rm \kappa_{N , c }$ are dimensionless free parameters. These are calibrated later in \S\ref{sec:Calibration}. Note that in the Galactic Wind we have used $M_{\rm cold} $ as  proxy for the ejected cold gas mass. Therefore, $\kappa_{N , c } $  absorbs the relation between these two galaxy properties and has dimension of $\rm [Time]^{-1}$ . 

The free parameters presented above outline the outflow properties, regulating the \lya\ luminosity distribution. In general, the parameters related to the quiescent star formation (galactic disks) give form to the faint end of the \lya\ luminosity function. Meanwhile, the bright end is shaped by the free parameters regulating the starburst episodes (galactic bulges) \citep{lacey16}.

Finally, the dust absorption optical depth is simply computed as

\begin{align}
\label{eq:recipe-ta}
\rm 
\tau_{a,c} & =  ( 1 - A_{{\rm Ly}\alpha} ) \frac{E_{\odot}}{Z_{\odot}}{N_{{\rm H},c}} Z_c,
\end{align}
where $\rm E_{\odot} = 1.77 \times 10^{-21 } {\rm cm^{-2}}$ is the ratio $\rm \tau_a/N_H$ for solar metallicity, $ A_{\rm Ly\alpha} = 0.39 $ is the albedo at the Ly$\alpha$ wavelength, the solar metallicity is  $\rm Z_{\odot} = 0.02$  \citep{granato00} and  $\rm Z_c$ is the metallicity of the cold gas in  $\rm Z_{\odot}$ units.

\begin{figure}
    \centering
    \includegraphics[width=3.05in]{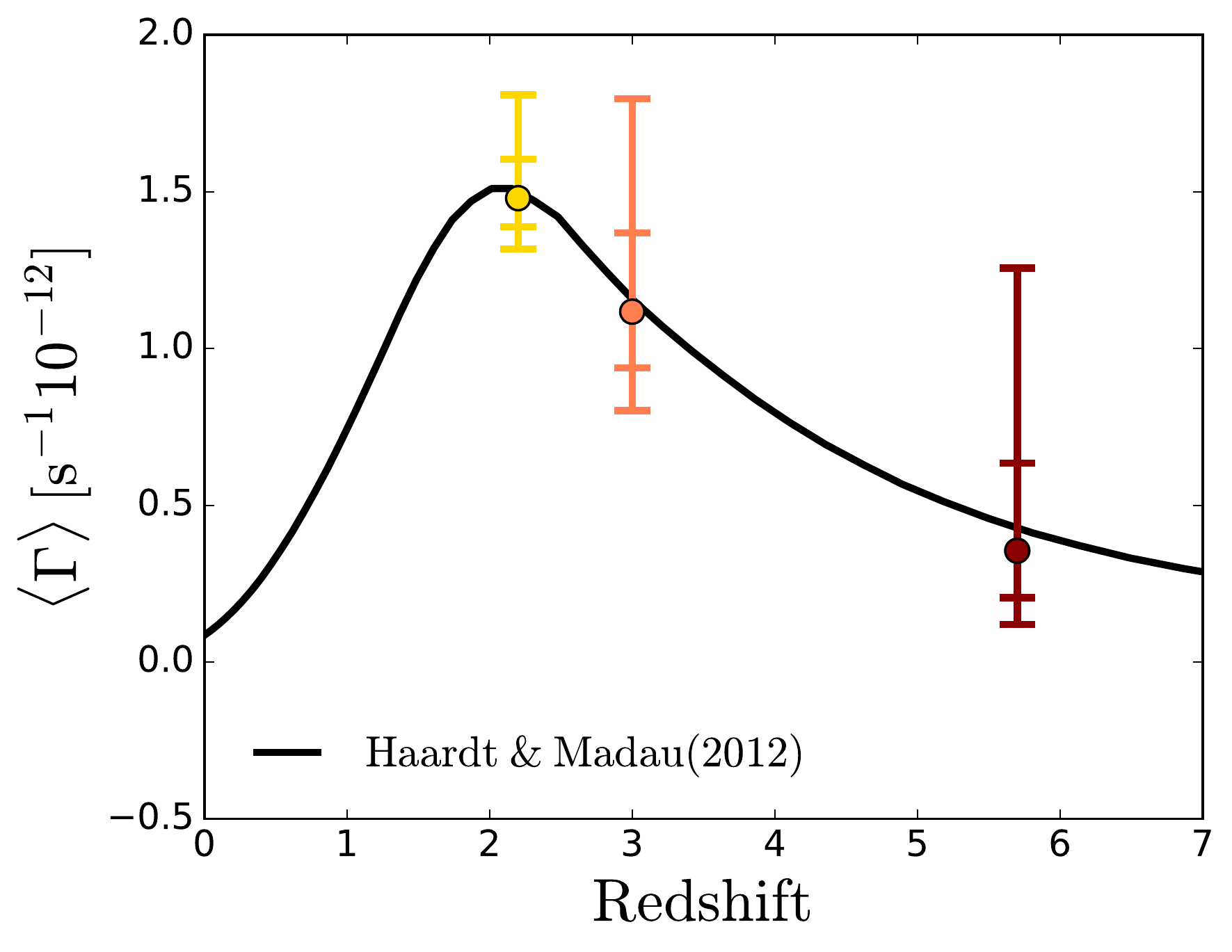} 
    \caption{  Median photoionization rate though the cosmic history. In solid black we show the photoionization predictions \citep{Haardt_2012}. In dots we show the median photoionization at different redshifts. The bars indicate the 2.5, 16 , 84, 97.5 percentiles of the distribution from bottom to top respectively. }\label{fig:ion_haardt_field}
\end{figure}


\begin{figure*}
    \includegraphics[width=7.0in]{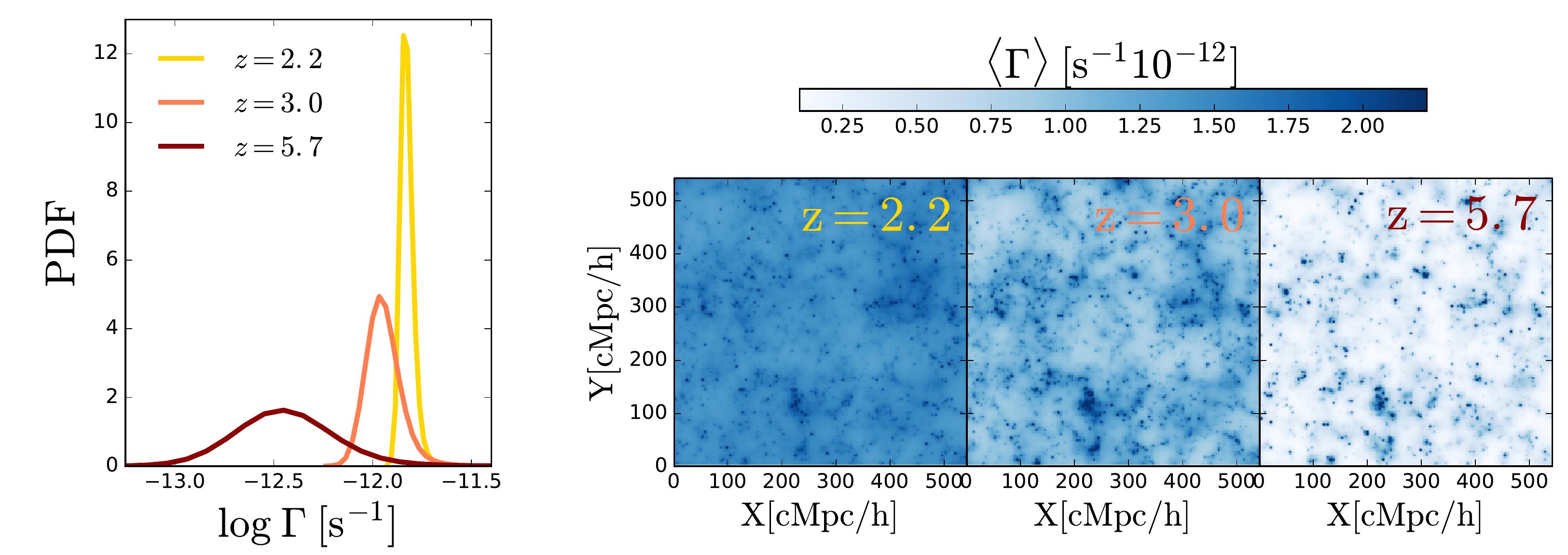}
    \caption{ \textbf{Left:} Probability distribution function of the cosmic photoionization rate ($\Gamma$) at redshift 2.2 (yellow), 3.0 (orange) and 5.7 (brown). \textbf{Right:} Spatial distribution of the photoionization rate  in different snapshots of the simulation (as labeled) in slices of $\sim 2 {\rm cMpc}h^{-1}$ width.  }\label{fig:ionization}
\end{figure*}


\subsection{Modeling the radiative transfer of \lya\ photons inside the IGM.}\label{sec:IGM_transmission.}

While inside galaxies the losses of \lya\ flux are due to dust absorption, in the IGM photons are scattered out of the line of sight by the neutral hydrogen. The total opacity of the IGM is given by \citep{dijkstra07}

\begin{equation}\label{eq:IGM_opacity_anlytic}
\tau_{IGM}(\lambda) = \sigma_{0} \displaystyle\int_{R_{vir}}^{\infty} n_{\rm H}(s) \;\;\chi_{\rm H_I}(s)\;\;\phi(\lambda,V_{\rm shift},T_{\rm gas}) \; ds , 
\end{equation} 
where $\lambda$ is the wavelength, $s$ is the proper distance to the galaxy where the photon is emitted, $n_{\rm H}$ is the IGM hydrogen number density, $\chi_{\rm H_I}$ is the fraction of neutral hydrogen in the IGM and $\phi(\lambda,V_{\rm shift}, T_{\rm gas})$ is the Voigt profile Doppler-shifted by the velocity between the emitting galaxy and the IGM at the  temperature $T_{\rm gas}$. 

We compute the IGM transmission of each galaxy, which depends on the local environment. To do so we compute $n_{\rm H}(\overrightarrow{x})$ , $\chi_{\rm H_I}(\overrightarrow{x})$ and the hydrogen velocity field, $V_{\rm H}(\overrightarrow{x})$ from our simulation. Due to disk storage limitations, the dark matter particles of \pmill\ at the snapshots used in this work were not saved. Hence, these quantities are computed from the halo catalogs as shown in the next subsections.

\subsubsection{ Hydrogen number density field. } 

We assume that density of hydrogen, $\rho_{\rm H}(\overrightarrow{x})$, is coupled to the dark matter density, $\rho_{\rm DM}(\overrightarrow{x})$. Thus, 

\begin{equation}\label{eq:hydrogen_density_field}
n_{\rm H}(\overrightarrow{x}) = x_{\rm H}{ {\Omega_{\rm b}} \over {\Omega_{\rm DM}}} \rho_{\rm DM}(\overrightarrow{x})/m_{\rm H} , 
\end{equation} 
where $x_{\rm H}=0.74$ is the hydrogen fraction of baryonic matter in the Universe, $\Omega_{\rm b}$ and $\Omega_{\rm DM}$ are the densities of baryons and dark matter respectively and $m_{\rm H}$ is the hydrogen mass.\\

In order to compute $\rho_{\rm DM}(\overrightarrow{x})$, it is useful to define the overdensity field of a given quantity,  $\delta_{a}(\overrightarrow{x})$, as

\begin{equation}\label{eq:overdensity_definition}
\delta_{a}(\overrightarrow{x}) = { { \rho_{a}(\overrightarrow{x}) - \langle \rho_{a} \rangle} \over {\langle \rho_{a} \rangle} } , 
\end{equation}
where $\rho_{a}(\overrightarrow{x})$ is the density field of some quantity and $\langle \rho_{a} \rangle$ is its average. Additionally, the definition of bias for a given dark matter halo mass, $b(M)$,  is 

\begin{equation}\label{eq:bias_definition}
\delta_{\rm halos}(\overrightarrow{x}) = b(M) \delta_{\rm DM}(\overrightarrow{x}) , 
\end{equation}
where $\delta_{\rm halos}(\overrightarrow{x})$ is the overdensity field of the dark matter halos and  $\delta_{\rm DM}(\overrightarrow{x})$ is the dark matter overdensity field.\\

By combining eq. \ref{eq:overdensity_definition} and eq. \ref{eq:bias_definition}, $\rho_{\rm DM}(\overrightarrow{x})$  can be expressed as 

\begin{equation}\label{eq:rho_dark_mater_expression}
\rho_{\rm DM}(\overrightarrow{x}) =  \left(
\frac{\delta_{\rm halos}(\overrightarrow{x})}
{b_{\rm eff}(\overrightarrow{x})} 
+ 1 \right)  \;
\langle \rho_{\rm DM} \rangle \; , 
\end{equation}
where we have defined the effective bias $b_{\rm eff}(\overrightarrow{x})$ in each cell as

\begin{equation}\label{eq:effective_bias_definition}
b_{\rm eff}(\overrightarrow{x}) = 
\frac{\displaystyle\int b(M) { {dN} \over {dM} }(\overrightarrow{x})\,dM}
{\displaystyle\int { {dN} \over {dM} }(\overrightarrow{x}) \; dM} ,
\end{equation}
where ${{dN} \over {dM}}(\overrightarrow{x})$ is the halo mass distribution in each cell.

\subsubsection{ Velocity field. }

We assume that the motion of the dark matter and gas are the same, i.e., $V_{\rm H}(\overrightarrow{x}) = V_{\rm DM}(\overrightarrow{x})$. In practice, we divide our simulation box into smaller volumes and compute $V_{\rm DM}(\overrightarrow{x})$ from the halo catalog  simply as the median of the velocity distribution of halos within the subvolume. 

In Fig.\ref{fig:v_field} we show the matter velocity along the line of sight (chosen arbitrarily as the $Z$ coordinate of the simulation box)  at redshift 3.0 in a slice of width 2${\rm cMpc}h^{-1}$ of our simulation box. The volume is divided in big chunks with coherent positive motion along the line of sight (red) and negative motion (blue).  The typical scales of these areas are  hundreds of comoving megaparsecs. Meanwhile, the transition between these regions is relatively small. This causes great contrasts of velocity along the line of sight on scales below  $\sim 10{\rm cMpc}h^{-1}$. We have checked that this behaviour is also found at the other redshifts studied in this work.


\begin{figure*}
    \includegraphics[width=7.0in]{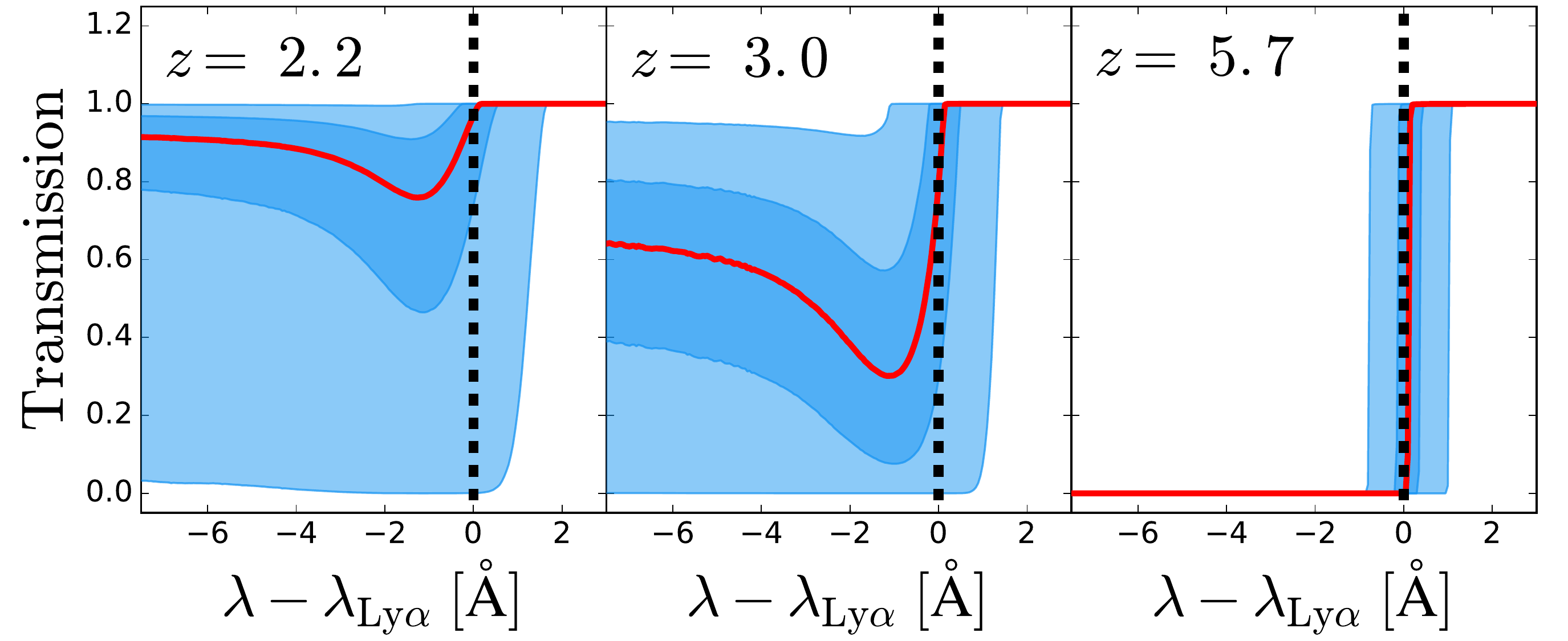} 
    \caption{  Average IGM transmission around \lya\ for different redshift bins (2.2, 3.0 and 5.7 from left to right). In red we show the median IGM transmission, while in dark blue and light blue we show the $1\sigma$ and $2\sigma$ ranges of the distribution.  }\label{fig:transmission}
\end{figure*}


\subsubsection{ Fraction of neutral hydrogen field. }

Star forming galaxies and active galactic nuclei (AGN) are the main sources of ionizing photons (photons wavelength shorter than 912\AA{}) in the Universe \citep{kimm_2014}. Thus the photoionization field, $\rm \Gamma (\overrightarrow{x})$, is coupled to the number density field of these objects. In principle, regions with a high rate of ionizing photons, i.e. close to these sources, will be more ionized, thus lowering the IGM opacity to Ly$\alpha$ photons. 

In order to compute $\Gamma(\overrightarrow{x})$ we assume that the ionizing radiation is produced only by galaxies. Whether or not AGNs and QSO contribute significantly to $\Gamma(\overrightarrow{x})$ at high redshift is still an open debate. In fact, some works in the literature \citep[e.g.][]{Parsa_2018} have shown that  AGNs or QSOs, are not sufficiently abundant to significantly contribute to $\Gamma(\overrightarrow{x})$, meanwhile, other works in the literature suggest the opposite \citep[e.g.][]{Giallongo_2019} . However, we do not expect this to significantly affect  our result since, as described later, $\Gamma(\overrightarrow{x})$ is calibrated to reproduce the observed mean $\Gamma$ at the different epochs studied here.

We compute $\Gamma(\overrightarrow{x})$ as the superposition of every ionizing field generated by each galaxy, i.e,

\begin{equation}
\label{eq:gamma_field}
\Gamma(\overrightarrow{x}) =  \displaystyle\sum_{i} { {\sigma_{0}\dot Q_{{\rm H},i} f^{\rm ion}_{\rm esc} } \over {4\pi | \overrightarrow{x} -\overrightarrow{x}_{i} |^2} } { {\beta} \over {\beta-3} } G( | \overrightarrow{x} -\overrightarrow{x}_{i} | ) , 
\end{equation} 
where the sum is over all the galaxies in the box, $\overrightarrow{x}_{i}$ is the location of each galaxy, $\overrightarrow{x}$ is the position where ionizing field in evaluated. Additionally, we assume a global escape fraction of ionizing photons $f^{\rm ion}_{\rm esc}=0.1$ \citep{kimm_2014} and $\dot Q_{{\rm H},i}$ is the total luminosity of ionizing photons given by \galform\ for each galaxy. We assume that the SED of the galaxy takes the form $J(\nu)=\nu^{\beta}$ in bluer parts of the hydrogen ionization frequency threshold, $\nu_{\rm H}$. We  also assume that the photoionization cross section as $\sigma_{\rm H}(\nu) = \sigma_{0}(\nu/\nu_{\rm H})^{-3}$ with $\sigma_0 = 6.3\times10^{-18}cm^{-2}$ . Finally, the function $G( | \overrightarrow{x} -\overrightarrow{x}_{i} | )$ takes into account the fact that photons emitted by a single galaxy do not reach every point in space and is given by

\begin{equation}
     \label{eq:G_equation}
     G( | \overrightarrow{x} -\overrightarrow{x}_{i} | ) = \left\{
	       \begin{array}{ll}
		 0      & \mathrm{if\ } | \overrightarrow{x} -\overrightarrow{x}_{i} | > R_{ion,i} \\
		 1      & \mathrm{if\ } | \overrightarrow{x} -\overrightarrow{x}_{i} | < R_{ion,i} 
	       \end{array}
	     \right. , 
   \end{equation}
where $R_{ion,i}$ is the radius of the sphere centered in the location of the galaxy $i$. We use a similar expression to the Str\"{o}mgren radius to compute $R_{ion,i}:$

\begin{equation}
\label{eq:R_ion}
R_{ion,i} = K_{a} ( \dot Q_{H,i} / 10^{55} s^{-1} )^{1/3} ,
\end{equation}
where $K_{a}$ is a free parameter. Increasing (decreasing)  $K_{a}$ leads to greater (lower) $R_{ion,i}$. Hence, at each point, more (less) galaxies  contribute to the the ionizing radiation field, which augments (lowers) the $\braket{\Gamma}$ of the simulation box. \\

We determine the $K_{a}$ values at each redshift by fitting the $\braket{\Gamma}$ of our simulations to the $\braket{\Gamma}$ given by \cite{Haardt_2012}. The best fitting values are listed in Table \ref{tab:gamma_fit}. We find that $K_{a}$ decreases with redshift. Since the number density of ionizing sources decreases from $z=2.2$ to 5.7, the volume ionized by a single galaxy must be increased with redshift in order to fit observations.

Fig.\ref{fig:ion_haardt_field} displays a comparison between the observed $\braket{\Gamma}$  and the $\Gamma(\overrightarrow{x})$ distribution (percentiles 2.5, 16, 50, 84 and 97.5) of our model. The right panel  shows the spatial variations of $\Gamma(\overrightarrow{x})$ over cosmic time in the same slice of width $\sim2 {\rm cMpc}h^{-1}$. By construction, $\Gamma(\overrightarrow{x})$ reproduces  \cite{Haardt_2012} quite well. 

Additionally, in Fig.\ref{fig:ionization} it is shown the PDF (left) and spatial distribution of the photoionization rate of the IGM. At $z=2.2$ $\braket{\Gamma}$ peaks and decreases towards higher redshifts. The dispersion of $\Gamma(\overrightarrow{x})$ evolves with redshift too. While at high redshift the photoionization field exhibits a  complex structure with high contrasts (broader PDF), as the Universe evolves, it becomes smoother (tighter PDF).

\begin{table}
\caption{Values of the best fitting $K_{a}$ used to compute the ionization field in our model.}\label{tab:gamma_fit}
\centering
\begin{tabular}{cc}
redshift & $ \log K_{a} [{\rm cMpc} \; h^{-1}]$ \\ \hline 
2.2      & 2.79         \\
3.0      & 3.65         \\
5.7      & 4.48        
\end{tabular}
\end{table}

Finally, we compute the neutral hydrogen fraction field as in \cite{dijkstra09} :

\begin{equation}\label{eq:fraction_field}
X_{\rm H_{I}}(\overrightarrow{x}) = 1 + a(\overrightarrow{x})/2 - \big[ a + (a(\overrightarrow{x})/2)^2 \big]^{1/2} ,
\end{equation} 
 with $a(\overrightarrow{x}) = \Gamma(\overrightarrow{x}) / n_{\rm H}(\overrightarrow{x})\alpha_{\rm rec}(\overrightarrow{x})$ where the case-A\footnote{Case-A recombination assumes that the medium is optically thin to ionizing photons.} recombination coefficient is given by  $\alpha_{\rm rec}(\overrightarrow{x}) = 4.2 \times 10^{-13}(T_{\rm gas}(\overrightarrow{x})/10^{4}K)^{0.7}cm^{3}s^{-1}$. We assume that the IGM temperature  behave as

\begin{equation}\label{eq:temperature_field}
\displaystyle\frac{T_{\rm gas}(\overrightarrow{x})}{T_0} = 
\left( \displaystyle\frac{ \rho_{\rm DM}(\overrightarrow{x})}{\langle \rho_{\rm DM} \rangle} \right)^{\gamma-1 }
\end{equation}
where $T_0=2\times10^{4}K$ and $\gamma-1=0.45$ \citep{Rudie_2012}. We have tested that our results do not depend on our choice of $T_0$ by repeating our analysis for several values of $T_0$ spawning between $10^2 K$ and $10^5  K$. Additionally, we have implemented different models for $T_{\rm gas}(\overrightarrow{x})$ and we find that our results depend very weakly (or not at all) on how $T_{\rm gas}(\overrightarrow{x})$ is computed. In fact, we have also made the whole analysis presented in this work assuming  $T_{\rm gas}(\overrightarrow{x})=T_0$, finding very similar results, even for the different values of $T_0$ discussed above.   
    
\subsubsection{ IGM transmission. }\label{sssection:IGM_model_transmission}
  
We compute $n_{H}(\overrightarrow{x})$, $V_{H}(\overrightarrow{x})$ and $\chi_{H_I}(\overrightarrow{x})$ on a grid of $250^3$ cells with side ${\rm \sim 2 cMpc}\; h^{-1} $. Then, by linear interpolation, we reevaluate these fields on a grid with higher spatial resolution along an arbitrary direction chosen as the line of sight. In particular, the new grid is composed of  250$\times$250$\times$3000 cells, where the size of the cells along the line of sight is $\rm L_{cells} \sim 0.2 cMpc/h $. We choose this grid size to ensure good signal to noise in the computed field and the IGM transmission curves with high enough resolution in frequency space.

\begin{figure}
    \includegraphics[width=3.2in]{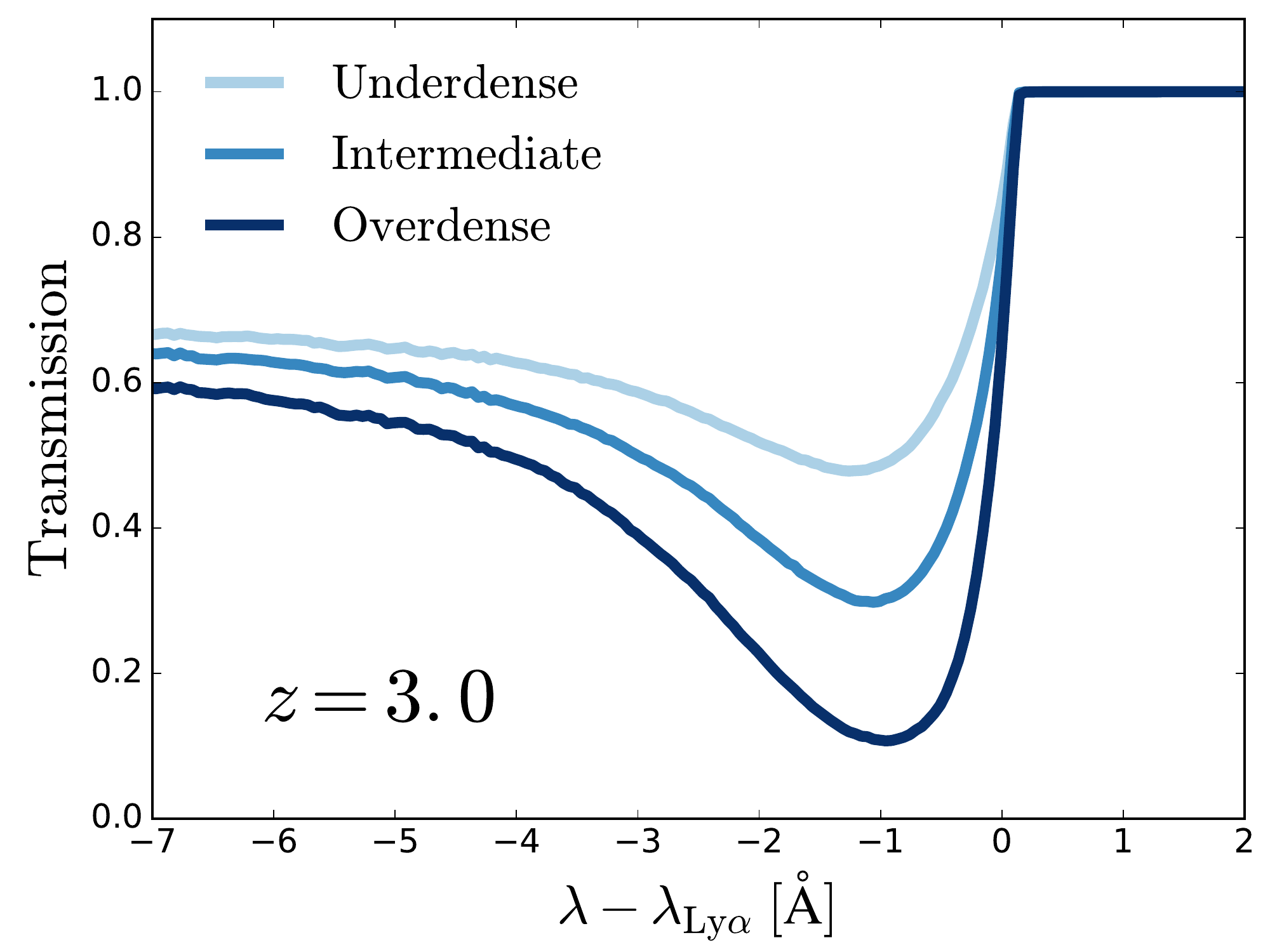}
    \caption{ Median IGM transmission around the \lya\ wavelength for LAE samples hosted regions with different density. LAE living at the 1/3 of the lowest densities are shown in light blue, while in blue from the 1/3 to the 2/3 and dark blue for the LAE lying in the 1/3 densest environments. }\label{fig:density_trans}
\end{figure}

Following eq. \ref{eq:IGM_opacity_anlytic}, the IGM optical depth in each cell in the simulation box rest frame is given by

\begin{equation}\label{eq:optic_depth_IGM_cells}
\begin{split}
\tau_{\rm IGM}(\lambda) = \sigma_{0} \; L_{\rm cells} \displaystyle\sum_{l=l_{\rm gal}}  & 
\left[ n_{\rm H_{I}}(i,j,l) \;  \times \;X_{\rm H_I}(i,j,l)\; \times \right.  \\
& \left. \times \; \phi(\lambda,V_{\rm shift}(i,j,l),T_{\rm gas}(i,j,l)) \right],
\end{split}
\end{equation}
where the cell indices \{$i$, $j$\} and \{$l$\} are perpendicular and parallel, respectively, to the line of sight. Moreover, $l$ starts at the cell where the galaxy lies, $l_{\rm gal}$, and iterates towards the observer along the line of sight direction. Additionally, $V_{\rm shift}$ is the relative velocity along the line of sight between the cell containing the galaxy ($i$,$j$,$l=l_{\rm gal}$) and the IGM in the iterated cell ($i,j,l$), given by

\begin{equation}\label{eq:V_shift}
V_{\rm shift}(i,j,l) = V_{\rm H}(i,j,l) + H_{\rm Hubble}(z) \times (l_{\rm gal}-l) \times L_{\rm cells} , 
\end{equation}
where $H_{\rm Hubble}(z)$ is the Hubble constant at redshift $z$.

Finally, the IGM transmission at each position is computed as $T(\lambda) = e^ {-\tau_{\rm IGM}(\lambda)}$. In Fig. \ref{fig:transmission} we show the median and dispersion of the IGM transmission at different redshifts for a sample of the 150000 \galform\ galaxies with the highest specific star formation rate (sSFR) as representative of an emission line galaxy population. In general, we find a good qualitative agreement between the shape of our mean transmission curves and the  mean transmission curves obtained by \cite{laursen11} computed from hydrodynamical simulation with RT physics implemented. The IGM absorbs photons bluer than $\lesssim1216$\AA{}. Moreover, as galaxies lie in overdense regions, the IGM opacity is higher close to the galaxy, causing the drop in the transmission close to \lya\ wavelength. Then the IGM transmission flattens to the IGM cosmic transmission.

Additionally, the optical depth of the IGM evolves with redshift, producing higher transmissions at lower redshifts. This becomes dramatic at $z=5.7$, where the IGM transmission goes below  1\% at bluer frequencies than \lya. 

The large dispersion of the IGM transmission reflects the complex variety of environments surrounding galaxies. In order to test this idea, we rank our LAE samples in IGM density $\rho_{\rm H}$ and split them into 3 subsamples: underdense (below the percentile 33rd of density), intermediate (between the 33 and 66 percentiles) and overdense (above the 66 percentile). In Fig. \ref{fig:density_trans} we show the median IGM transmission at $z=3.0$ for these subsamples. In general the behaviour of the three population is the same. At redder wavelengths than \lya\, the transmission is 1.  Meanwhile,  at bluer wavelengths far from \lya\ ($\rm \lambda \sim 1210$\AA) the transmission converges to the mean IGM transmission.  However, at blue wavelengths around \lya\ ($\rm \lambda \sim 1214$\AA) the density has a great impact on the IGM transmission. We find that LAEs hosted in denser environments exhibit lower IGM transmission than their counterparts in low density regions. The transmission is $\sim 0.4$, $\sim 0.2$ and $\sim 0.1$ for the underdense, intermediate and overdense environments. For completeness, we also did this analysis at the other snapshots of our simulations, finding the same trend. The typical transmission at $z=2.2$ around \lya\ is $\sim0.9$, $\sim0.85$ and $\sim0.8$ for the underdense, intermediate and overdense environments. Meanwhile, at $z=5.7$ the transmission remains below  1\% even in the underdense regions.


\begin{figure}
    \includegraphics[width=3.2in]{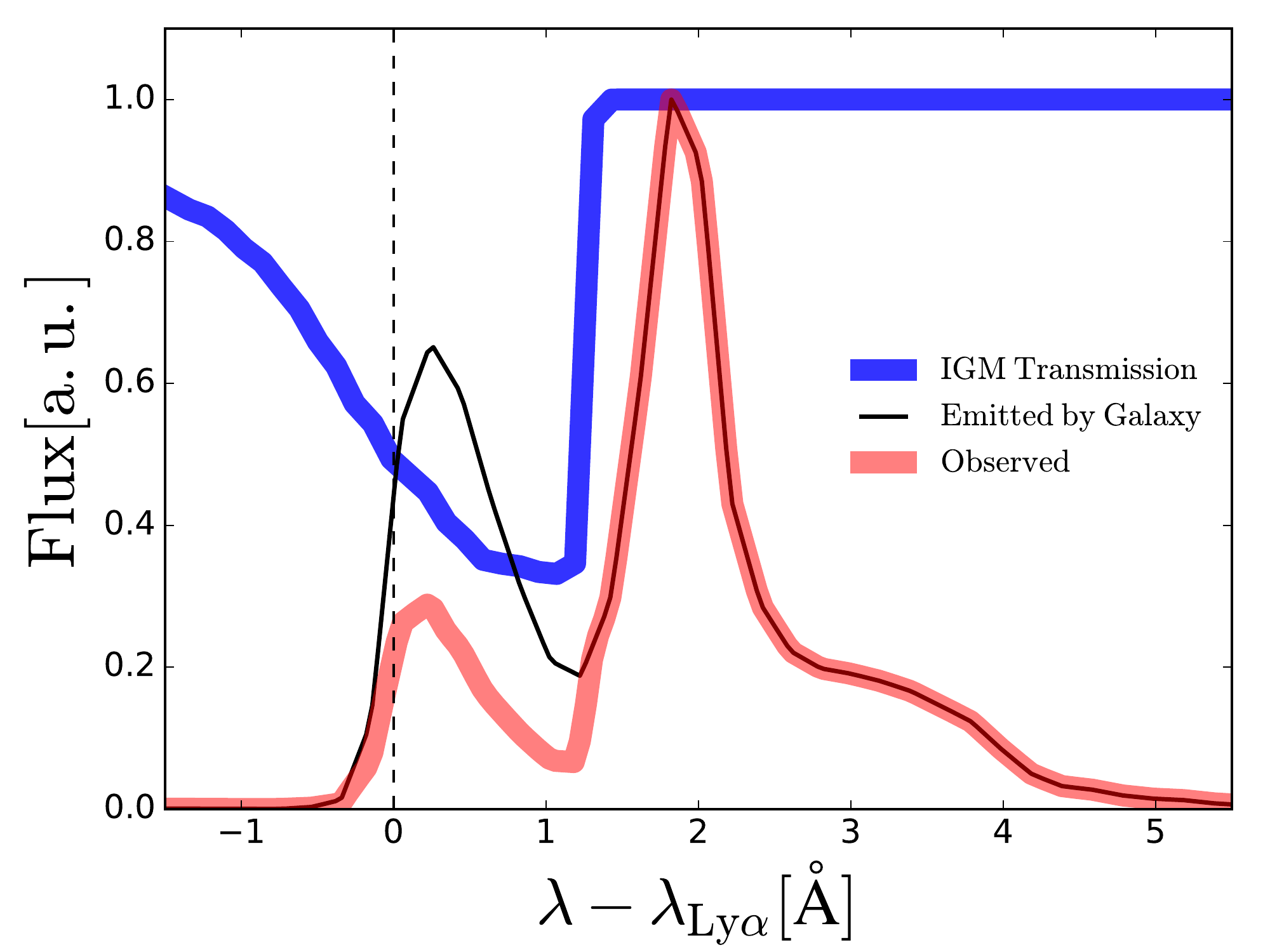}
    \caption{ Example of the interaction between the IGM and a galaxy approaching the IGM along the line of sight. We show in blue the IGM transmission at the galaxy position, in solid black the \lya\ line profile emerging from the galaxy after the interaction with the ISM and in thick red the observed \lya\ line profile after the IGM absorption. }\label{fig:transmission_line_profile}
\end{figure}



\begin{figure*}
    \includegraphics[width=7.0in]{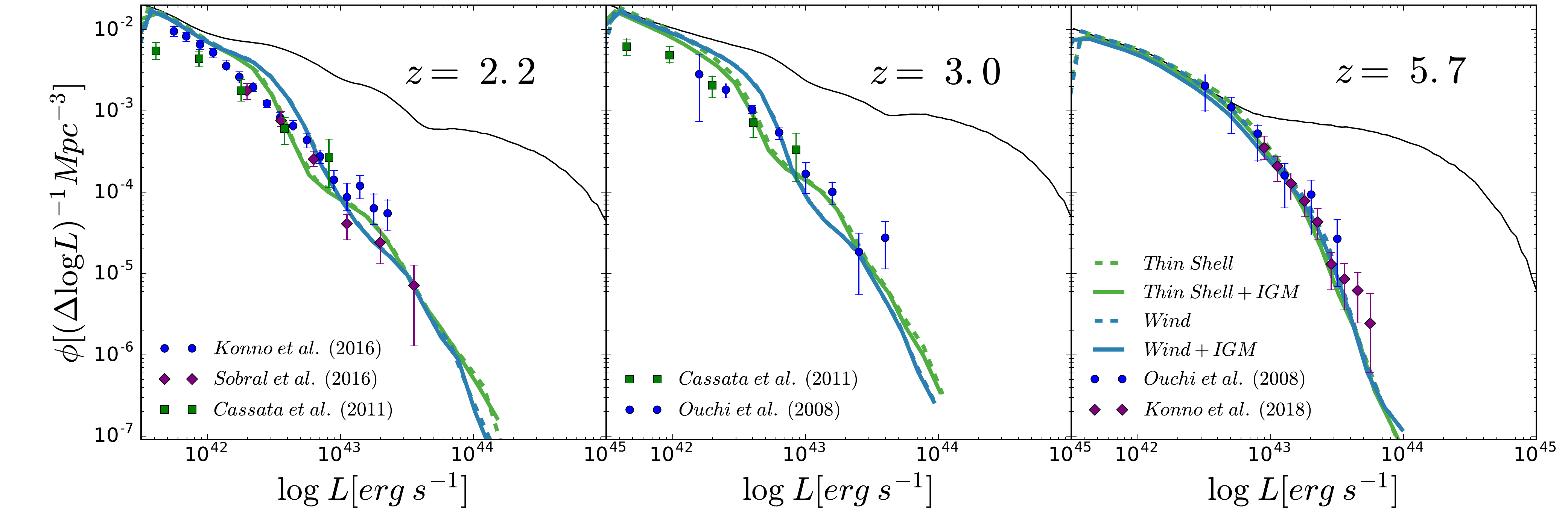}
    \caption{ Comparison between our model's LAE luminosity function (LF) and observations at different redshifts. The intrinsic \lya\ LF is shown in thin black, the Thin Shell and Galactic Wind geometries  are shown in green and blue dashed lines respectively. The colored dashed lines correspond to models with RT in the ISM but without RT in the IGM, while the colored solid lines include the RT in the ISM ad IGM. Different observational data sets are shown in symbols according to the legends.   }\label{fig:LFs}
\end{figure*}


\subsection{The observed \lya\ luminosity.}\label{ssec:Lya_luminosity}

The observed \lya\ luminosity is a convolution of galactic and IGM properties. We compute the observed \lya\ luminosity as follows.

Physical processes taking place inside galaxies are implemented as in \citep{gurung18a}. For each galactic component: i) the intrinsic \lya\ luminosity, ${\rm L_{Ly\alpha}^{0,c}}$ (where the superindex "$c$" denotes the disk or bulge component), is predicted by \galform\ from the instantaneous star formation rate; ii) the outflow properties are computed with equations \ref{eq:recipe-V_exp}, \ref{eq:recipe-column_density} and \ref{eq:recipe-ta}; iii) the outflow \lya\ escape fraction $f_{\rm esc} ^{c}$ and line profile $\rm \Phi ^{c} (\lambda)$ are predicted with \flareon; iv) the \lya\ luminosity escaping the ISM is computed as 

\begin{equation}
\label{eq:galaxy_lum}
{\rm L_{Ly\alpha} ^{c}}  =  {\rm L_{Ly\alpha} ^{0,c}}  f_{\rm esc} ^{\rm c} .
\end{equation} 

The IGM transmission is calculated in the simulation rest frame, as discussed in the previous section. However,  the galaxy rest frame mismatches, in general, the simulation rest frame due to galaxy peculiar velocities. Hence, the IGM transmission for a given galaxy is Doppler shifted by the peculiar velocity of the galaxy along the line of sight.  Then, for each component, the fraction of photons that travel unscattered through the IGM, $f_{esc}^{\rm IGM,c}$, is computed as 

\begin{equation}
\label{eq:IGM_escape_fraction}
f_{\rm esc} ^{\rm IGM , c}  =  
{
{\int  {\rm \Phi ^{c} (\lambda)} \times {\rm T ^{gal}(\lambda)} d{\rm \lambda}} 
\over 
{\int  {\rm \Phi ^{c} (\lambda)} d{\rm \lambda} } 
}
,
\end{equation} 
where ${\rm T ^{gal}(\lambda)}$ is the IGM transmission at the galaxy position and rest frame and $\rm \Phi ^ c $ is the \lya\ line profile of each galaxy component. 

In Fig. \ref{fig:transmission_line_profile} we illustrate how $f_{\rm esc} ^{\rm IGM , c}$ is computed. In this particular case the IGM and the galaxy are approaching along the line of sight. As shown previously, in the rest frame, the IGM absorbs photons bluer than \lya . However, due to the relative motion between the galaxy and the IGM along the line of sight, the IGM transmission (blue curve) is redshifted in the galaxy rest frame. Hence, the \lya\ photons emerging from the ISM (black curve) are partially absorbed. As a result the observed \lya\ line profile (red curve) is modified. $f_{\rm esc} ^{\rm IGM , c}$ is calculated as the ratio between the integrals of the observed and the emerging line profiles.  

Then, for each galaxy, the observed \lya\ luminosity is computed as 

\begin{equation}
\label{eq:observed_luminosty}
{\rm L_{Ly\alpha} = } \; 
{\rm L_{Ly\alpha} ^{Disk }}   \; \times  \; 
f_{\rm esc} ^{\rm IGM , Disk}   \; +      \;
{\rm L_{Ly\alpha} ^{Bulge} }     \; \times \; 
f_{\rm esc} ^{\rm IGM,Bulge}  .
\end{equation} 


Moreover, we estimate the rest frame \lya\ equivalent width, $\rm EW$, as

\begin{equation}
\label{eq:EW}
\rm
EW = L_{Ly\alpha} / L_{continuum } ,
\end{equation} 
where $\rm  L_{continuum }$ is the continuum luminosity per unit of wavelength around \lya\ computed by \galform. This quantity is based on the evolution of the composite stellar population of each modeled galaxy. From now on, unless stated otherwise, we define LAEs in our model as galaxies with $\rm EW>20$\AA{}.

\begin{table*}
\centering
\caption{ Free parameters as defined in equations \ref{eq:recipe-V_exp} and \ref{eq:recipe-column_density} after the calibration with the observed luminosity function for different geometries and redshifts. }
\label{tab:parameters}
\begin{tabular}{l|c|cccc}
redshift & Geometry   & $\log \kappa_{V,disk}$                      & $\log \kappa_{V,bulge}$                     & $\log \kappa_{N,disk}$                        & $\log \kappa_{N,bulge}$  \\ \hline
$z=2.2$  & Thin Shell & \multicolumn{1}{c|}{ 4.43 } & \multicolumn{1}{c|}{ 4.27 } & \multicolumn{1}{c|}{ -12.33 } &   -12.11   \\
         & Galactic Wind       & \multicolumn{1}{c|}{ 4.14 } & \multicolumn{1}{c|}{ 4.80 } & \multicolumn{1}{c|}{ -8.01 } &   -5.58   \\ \hline
$z=3.0$  & Thin Shell & \multicolumn{1}{c|}{ 3.78 } & \multicolumn{1}{c|}{ 4.23 } & \multicolumn{1}{c|}{ -12.73 } &   -12.06   \\
         & Galactic Wind       & \multicolumn{1}{c|}{ 4.04 } & \multicolumn{1}{c|}{ 4.47 } & \multicolumn{1}{c|}{ -8.10 } &   -5.69   \\ \hline
$z=5.7$  & Thin Shell & \multicolumn{1}{c|}{ 4.55 } & \multicolumn{1}{c|}{ 3.48 } & \multicolumn{1}{c|}{ -14.03 } &   -12.24   \\
         & Galactic Wind       & \multicolumn{1}{c|}{ 4.35 } & \multicolumn{1}{c|}{ 3.28 } & \multicolumn{1}{c|}{ -9.62 } &   -6.85   \\
\end{tabular}
\end{table*}


\section{ Lyman-$\alpha$ emitters }\label{sec:Calibration}

\subsection{ Model calibration.}

In these section we briefly discuss how the free parameters described in Eq.\ref{eq:recipe-V_exp} and  \ref{eq:recipe-column_density} are calibrated. For further details we refer the reader to \cite{gurung18a}.

In short, for each redshift and outflow geometry, we perform an MCMC analysis using the open source \texttt{Python} library \texttt{emcee} \citep{emcee}  to estimate the values of  $\rm \kappa_{N,c}$ and $\kappa_{V,c}$ that best reproduce the observed LAE luminosity function (LF) at different redshifts. In particular, at $z=2.2$ we fit the observed LF from  \cite{Cassata_2011}, \cite{Konno2016} and \cite{Sobral2017}, while at $z=3.0$ we fit the LFs from \cite{ouchi08} and \cite{Cassata_2011}. Finally, at $z=5.7$ we model the LFs of \cite{ouchi08} and \cite{Konno_2018}. In order to fit the different observed LFs at the same same redshift, we combine them at each redshift by fitting a 5th-order polynomial (in $\log$\llya -$\log$LF space) taking into account the uncertainties of each data set. The functional form of a 5th-order polynomial was chosen due to the fact that some recent works suggest that observed LFs of LAEs are more complex than the typical Schechter function \citep{Konno2016,Sobral_2018a}. The free parameter values are listed in Table \ref{tab:parameters}. Additionally, the resulting $\rm N_{H}$, $\rm V_{\rm exp}$ and $\tau_a$ distributions are discussed in Appendix \ref{Ap:A}.

In Fig. \ref{fig:LFs} we compare the observed LF at several redshifts with our models. For illustration we show the \galform\ intrinsic LAE LF (black thin line), which over predicts the number of LAEs over the full luminosity range at $z=2.2$ and $z=3.0$. Hence, the total \lya\ escape fraction $f _{\rm esc} ^ {\rm Ly\alpha}$ (galaxy+IGM) must be $<1$. However at redshift $z=5.7$, while bright intrinsic LAE surpass observations, the intrinsic number counts of faint  LAEs resembles observations, implying $f _{\rm esc} ^ {\rm Ly\alpha} <1 $ and $f _{\rm esc} ^ {\rm Ly\alpha} \sim 1$ respectively. 

In general, after calibration, our models (colored solid lines) match the observed LAE LF at all redshifts by construction. The good agreement is remarkable at $z=5.7$. The Thin Shell geometry matches slightly better observations than the Galactic Wind. Additionally, we show the LAE LF of our calibrated model excluding the IGM absorption (colored dashed lines). Although the LAE number counts of the model without IGM exceeds the abundance of LAE in the complete models at every \lya\ luminosity, their LAE LF are very similar. This points to the fact that the RT inside galaxies is the main driver shaping the observed LAE LF.

\subsection{ LAE samples }

Throughout this work we analyze and compare the properties of different LAE samples to highlight the RT selection effects. Here we describe how the samples are built.

\subsubsection{ Full Ly$\alpha$ emitters samples. }

Full \lya\ emitters (\flae). These samples represent the observed LAE population.  They include all the radiative transfer processes explained above (ISM + IGM). The \flae\ samples are derived from the full \galform\ population.  The outflow properties of the \galform\ galaxies are computed with the calibrated free parameters (listed in Table \ref{tab:parameters}). Then we assign to each galaxy a \lya\ luminosity (as described in \S \ref{ssec:Lya_luminosity}). Note that these samples, by construction, reproduce the observed LF. We rank these populations by  \lya\ luminosity and perform a number density cut of $4 \times 10^{-3} ( {\rm cMpc} \; h^{-1})^{-3}$. The chosen number density cut is arbitrary. We obtain similar results for higher and lower number density cuts. 

\subsubsection{ Partial Ly$\alpha$ emitters samples. }

Partial \lya\ emitters (\nlae). These samples include  galactic RT physics but lack the IGM absorption. They are also subsamples of the full \galform\ galaxy population. \nlae s can be seen as the LAE population that would be observed if the IGM was completely transparent. In particular, $\rm N_{H}$ and $\rm V_{exp}$ are computed with the same calibration as the models with full RT (\flae\ samples). However, in contrast to \flae\ populations where the \lya\ luminosity is computed through Eq. \ref{eq:observed_luminosty}, in the \nlae\ samples, the observed \llya\ is computed simply as
    
\begin{equation}
\label{eq:observed_luminosty_no_igm}
{\rm L_{Ly\alpha} = } \; 
{\rm L_{Ly\alpha} ^{Disk }}   +
{\rm L_{Ly\alpha} ^{Bulge} }    
,
\end{equation} 
where the ISM RT is included in $\rm L_{Ly\alpha} ^{Disk }$ and $\rm L_{Ly\alpha} ^{Bulge}$. 

Finally, we rank galaxies by \llya\ and make the same number density cut as in the \flae\ samples. Note that, by construction, the intrinsic galaxy population and properties are identical for \flae\ and \nlae\ samples. However, the \flae\ samples include the IGM selection effects. Therefore, the comparison between these samples sheds light on the impact of the IGM.


\begin{figure*}
        \centering
        \includegraphics[width=1.\linewidth]{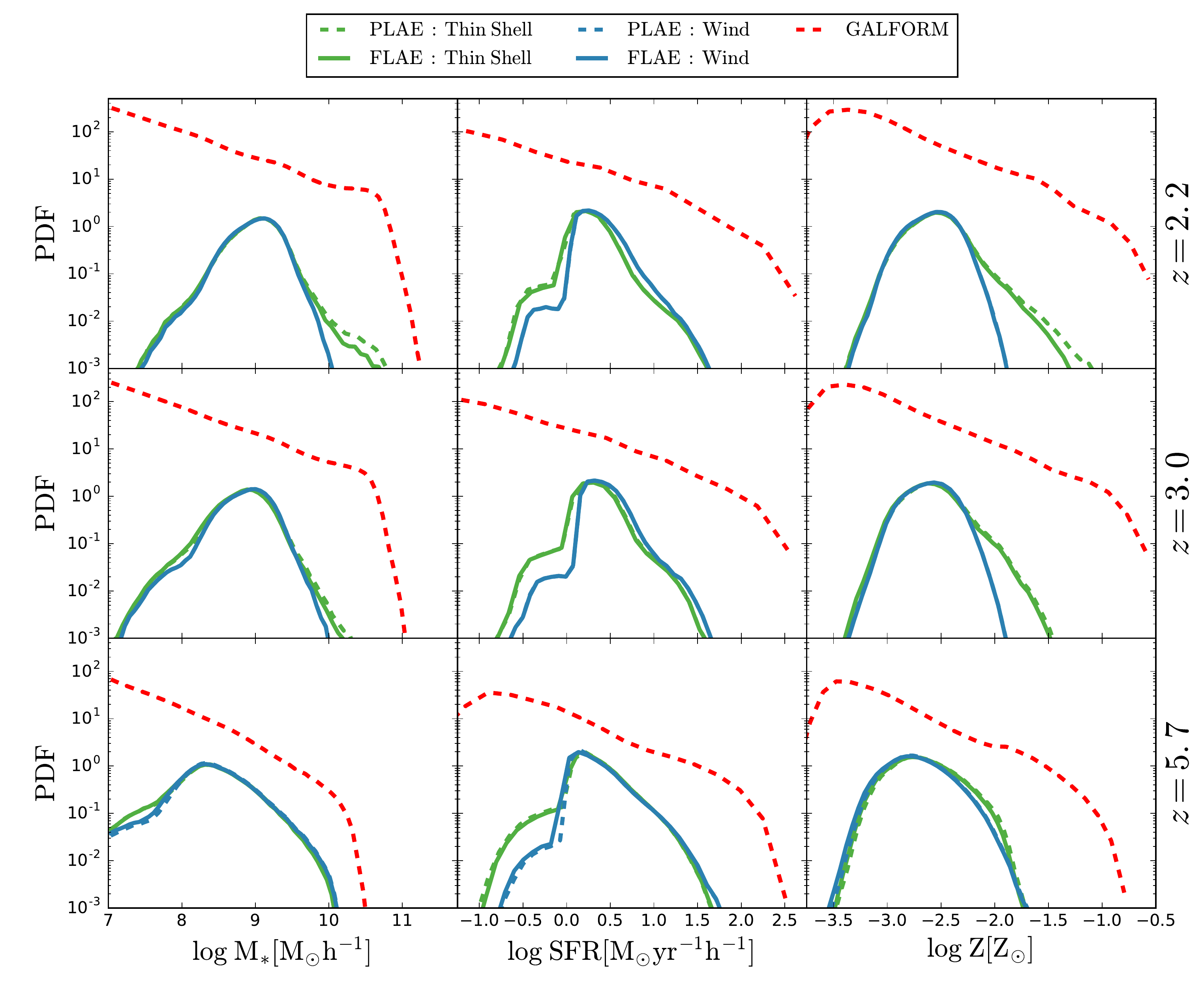}

        \caption{ Comparison between the galaxy properties distribution (stellar mass, star formation rate and metallicity from left to right) of our different galaxy samples at several redshifts (2.2, 3.0, 5.7 from top to bottom). In dashed red we show the galaxy property distributions of the full GALFORM catalog. In dashed and solid green we display the galaxy properties distribution of the Thin Shell excluding and including the RT in the IGM. Meanwhile,  blue shows the same as green but implementing the Galactic Wind geometry.   } \label{fig:galaxy_prop}
\label{fig:LFbreakdown}
\end{figure*}
 

\subsubsection{ Shuffled Ly$\alpha$ emitters samples. }

Shuffled \lya\ emitters (\slae) samples are built to exhibit the same clustering as the \flae\ samples if no LAE-IGM coupling is found. Throughout this work we analyze the clustering and the galactic properties of LAEs and how the IGM affects them. Since the \flae\ and \nlae\ are computed from the same galaxy population they are useful to understand how the IGM shapes the galactic properties. As seen later on, the \flae\ and \nlae\ exhibit different mass functions. This causes that each population displays a different clustering, and in particular, different bias. Additionally, if the large scale IGM  properties are coupled to the \lya\ observability, further clustering distortions are expected \citep{zheng11}. Therefore, in order to study the LAE-IGM coupling it would be desirable to use samples with the same bias.

\slae\ samples are derived from the \pmill\ halo catalog mimicking the \flae\ halo mass functions. In detail:

\begin{enumerate}
    \item We separate our full RT LAE population into centrals and satellites. We find that at all redshifts and gas geometries, central galaxies constitute  $\sim 98\%$ of the full RT LAE samples. 
    
    \item Comparing with the dark matter halo catalogs of the simulation we compute the fraction of halos occupied by LAEs as a function of mass (HOD) individually for the central and satellite population.
    
    \item From the halo catalog we select central halos reproducing the HOD of the central LAE sample using a uniform random distribution. 
    
    \item We determine the number of LAE satellites  hosted in each dark matter halo with a random Poisson distribution with mean equal to the satellite's HOD evaluated at the mass of the halo. Following J\'imenez et al. (in prep.), we  assign to the satellite the same location as the central halo and restrict our clustering analysis to the 2-halo-term scales. 
    
    \item We combine the new satellite and central population in a single \slae\ sample.
\end{enumerate}

By construction, \slae\  are free of the IGM selection effects and exhibit the same halo mass distribution as their full RT progenitors. Therefore, if the IGM is not shaping the LAE spatial distribution, the clustering in the \slae s and full RT LAE samples should be the same\footnote{ Note that, by construction, the \slae\ samples do not isolate the IGM-LAE coupling, but it also includes the effects of assembly bias \citep{Contreras_2019}. This might cause differences between the clustering of the \flae\ and \slae\ populations. To check the assembly bias impact on the clustering of our LAE populations we built \slae\ samples from the \nlae\ samples. No significant difference was found between the clustering of these populations, i.e., there was no assembly bias evidence. Therefore, we assume that the difference between the \flae\ and \slae\ (computed from the \flae)  samples are due to the IGM-LAE coupling. } on scales greater than the 1-halo term. 



 
\section{ LAE galaxy properties. }\label{sec:gal_prop}

In this section we briefly study the selection function of LAEs. First, we analyze which galaxies would be observed as LAEs if the IGM was completely transparent to photons around \lya. For this goal, we contrast the full \galform\ galaxy population with the \nlae\ samples, which  includes RT only in the ISM. In the second case, we characterize the IGM impact by directly comparing the \nlae\ and the full RT LAE sample. The variations among these samples are caused by the IGM, since the only difference between the \nlae\ and \flae\ samples is that \flae\ also include RT in the IGM.

In Fig.\ref{fig:galaxy_prop} we compare the galaxy property distributions between our \lya\ flux selected samples and the full galaxy population predicted by \galform\ at different redshifts.  We define a \lya -weighted average gas metallicity for a galaxy as

\begin{equation}
\label{eq:observed_luminosty_no_igm}
{\rm Z = } \; 
{{
\rm Z ^{Disk} L_{Ly\alpha} ^{Disk } + 
Z ^{Bulge} L_{Ly\alpha} ^{Bulge}
}
\over{
\rm L_{Ly\alpha} ^{Disk } + L_{Ly\alpha} ^{Bulge}
}}
.
\end{equation} 

We find very little difference between the \flae\ and \nlae\ populations. This indicates that the IGM does not induce significant selection effects on the galaxy properties of LAEs. Moreover, independently of the outflow geometry and cosmic time,  galaxies with strong \lya\ emission present moderate stellar mass and SFR and low metallicity. We find that models implementing different outflow geometries behave in a similar fashion. Still there are tiny differences between them. 


The typical stellar mass of galaxies observed as LAEs also evolve over cosmic history. In general, we find the same trend in the LAEs samples as in the full galaxy population: LAEs at lower redshift exhibit higher stellar content than their homologous at higher redshift. In detail, the $\rm M_{*}$ distributions peak around moderate masses; $10^{8.5}\Munits$, $10^{9}\Munits$ and $10^{9.2}\Munits$ at redshift 5.7, 3.0 and 2.2 respectively. The shapes of the distributions are very similar between $z=2.2$ and $z=3.0$. The dynamical range of $\rm M_{*}$ at these redshifts is very similar to the observed LAEs \citep{oyarzun17}.  Additionally, very small differences can be found between the two outflow geometries implemented in this work. At $z=2.2$ and 3.0 the Thin Shell predicts a greater abundance of massive galaxies in comparison with the Galactic Wind. This trend disappears at $z=5.7$, when both stellar mass distribution are almost identical.


\begin{figure*}
        \centering
        \includegraphics[width=1.\linewidth]{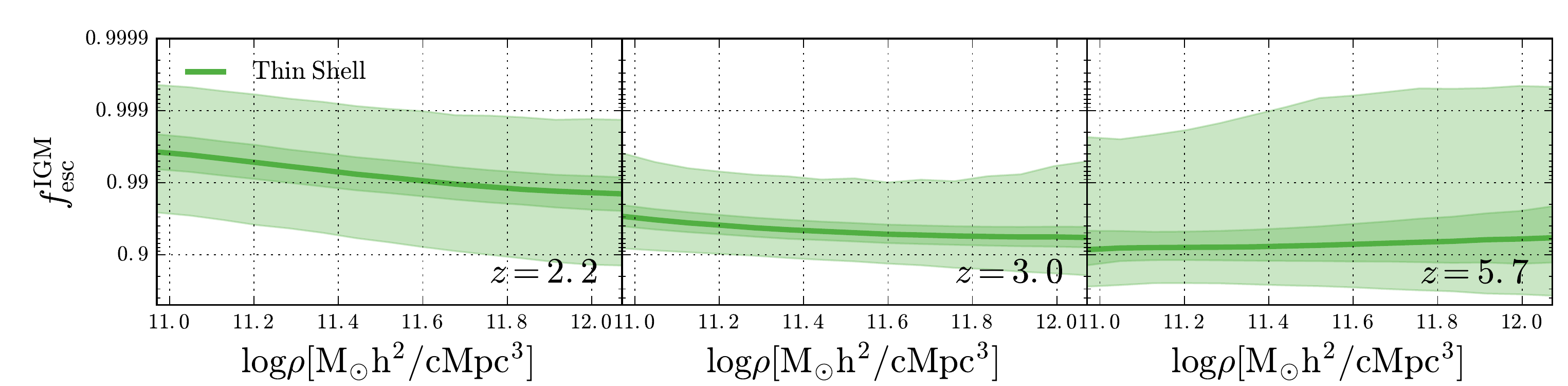}
        
        \includegraphics[width=1.\linewidth]{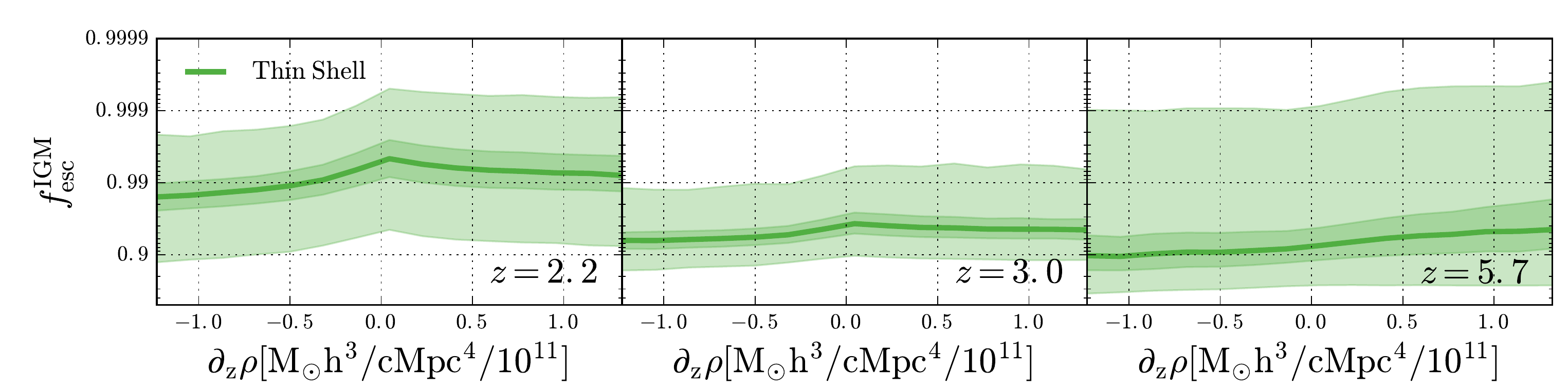}
        
        \includegraphics[width=1.\linewidth]{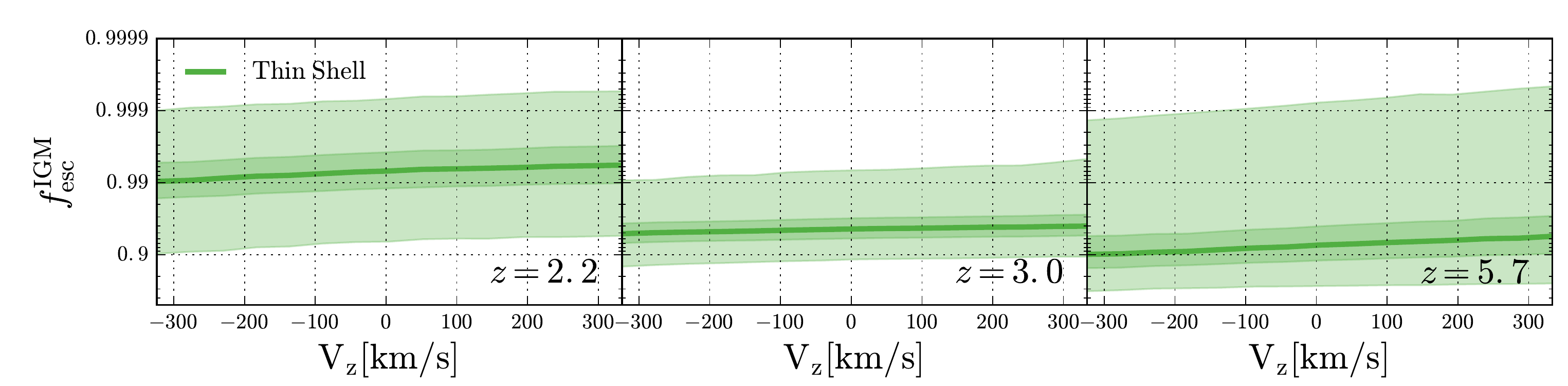}
        
        \includegraphics[width=1.\linewidth]{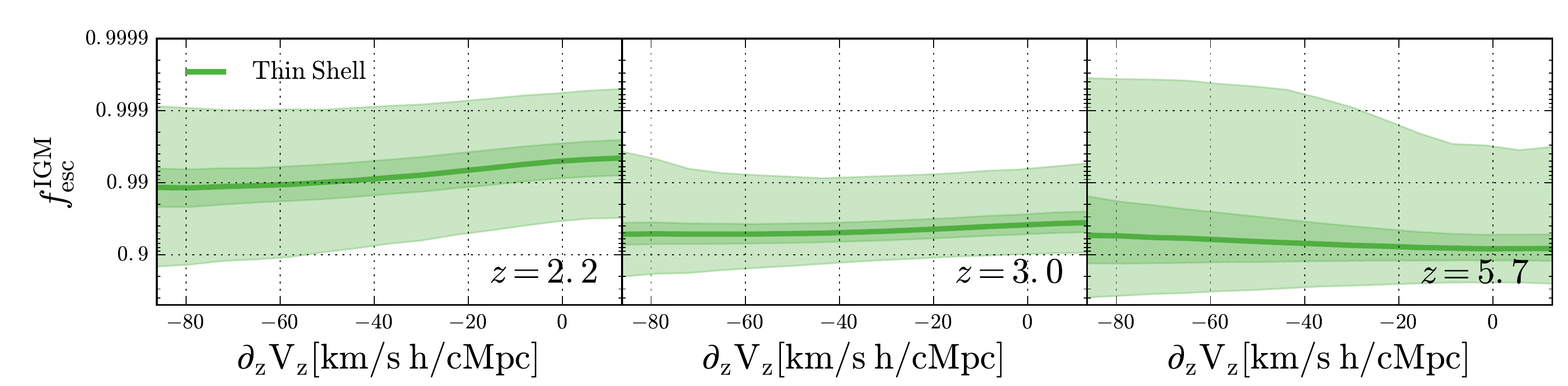}

        \caption{ IGM escape fraction for the Thin Shell as a function of the density, density gradient, velocity along the line of sight and velocity along the line of sight gradient (from top to bottom) for redshift 2.2, 3.0 and 5.7 (from left to right). The dark solid line shows the median, while the $1\sigma$ and $2\sigma$ values of the distributions are shown in dark and light shaded regions respectively.} \label{fig:IGM_transmission_Thin}
\end{figure*}


\begin{figure*}
        \centering
        \includegraphics[width=1.\linewidth]{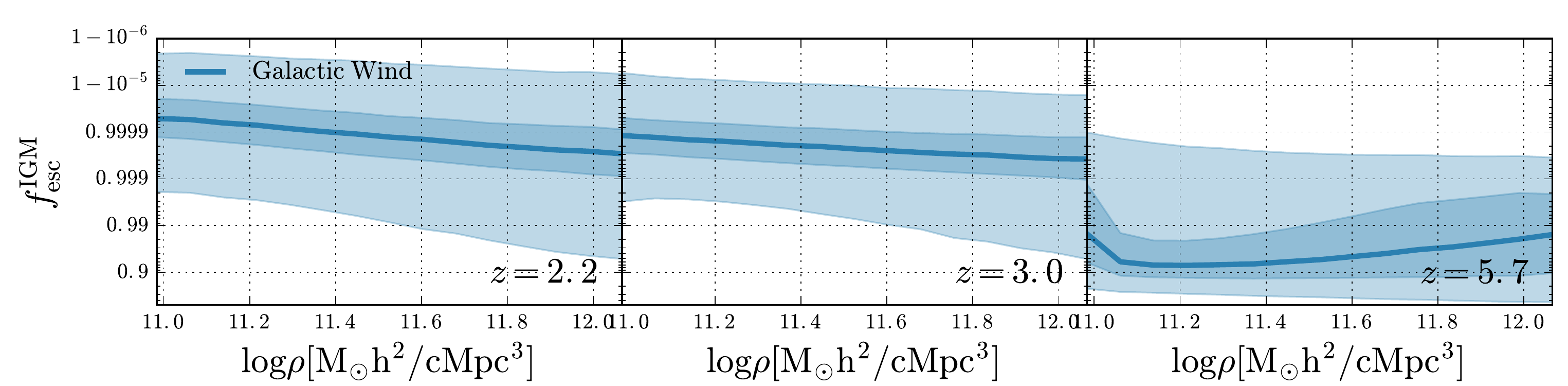}
        
        \includegraphics[width=1.\linewidth]{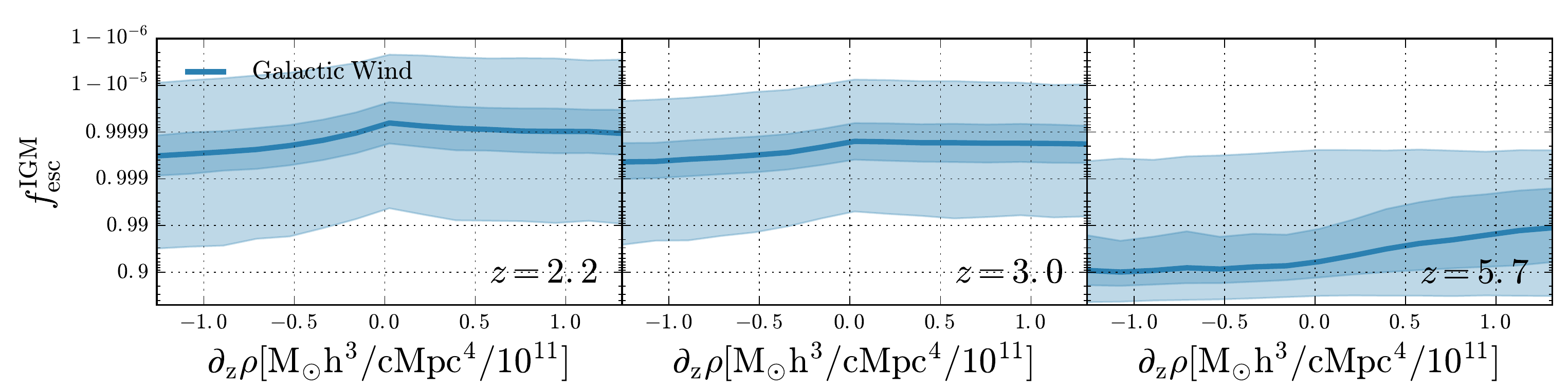}
        
        \includegraphics[width=1.\linewidth]{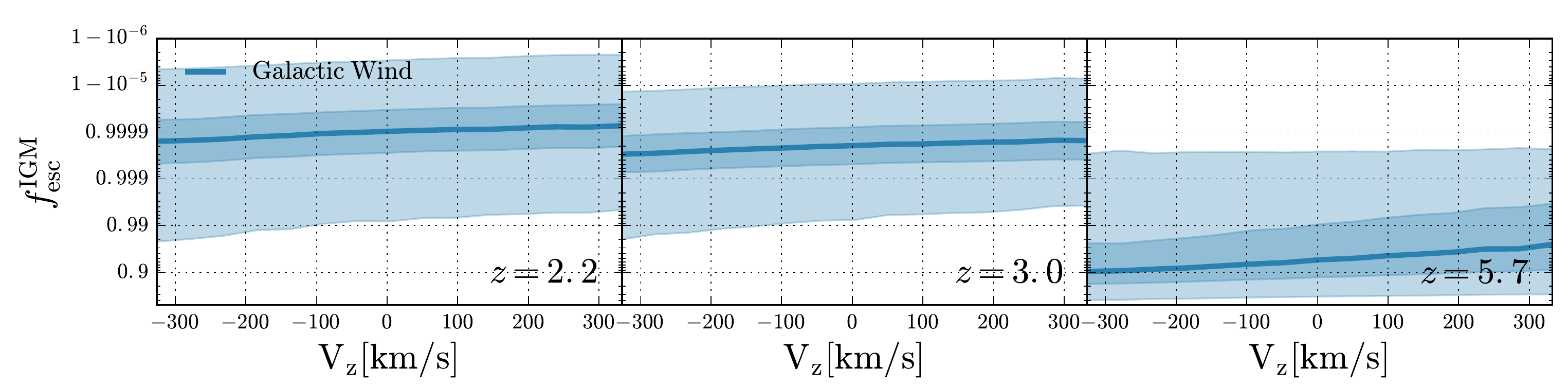}
        
        \includegraphics[width=1.\linewidth]{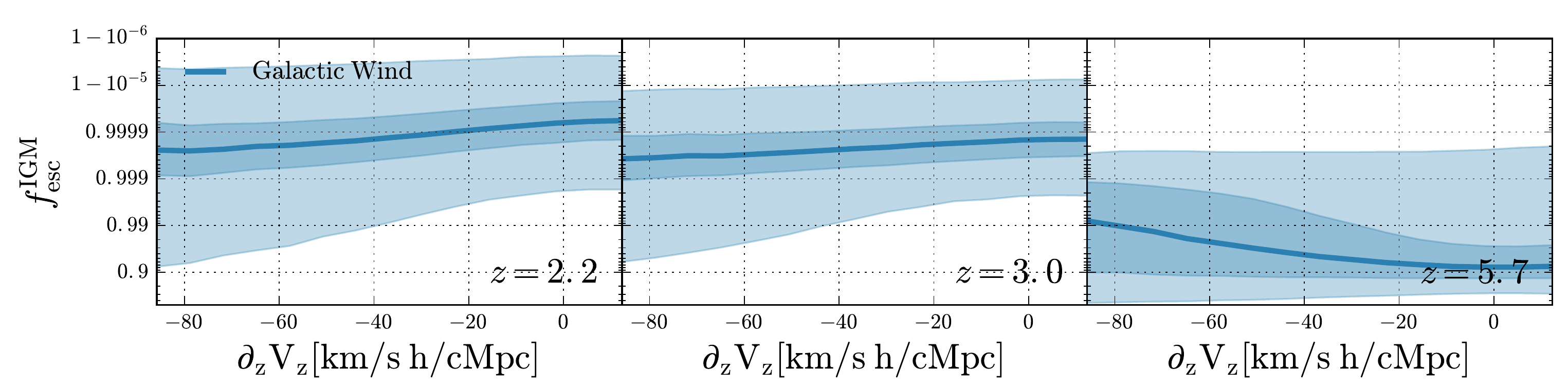}

        \caption{ Same as Fig. \ref{fig:IGM_transmission_Thin} but for the Galactic Wind geometry.} \label{fig:IGM_transmission_Wind}
\end{figure*}


We note that at $z=5.7$ the galaxy stellar mass distributions for our LAEs samples truncates abruptly at $\rm M_{*}=10^{7}\Munits$. This is caused by the cut  in \galform\ at this stellar mass imposed to ensure a good resolution of galaxies. This suggests that a fraction of the LAE population at  $z=5.7$ would inhabit galaxies with $\rm M_{*}<10^{7}\Munits$. However, as our galaxy population lacks these low mass galaxies, other more massive galaxies are selected as LAEs. Computing the precise number of LAEs that should be hosted by galaxies with $\rm M_{*}<10^{7}\Munits$ is challenging. As a simple calculation, we rescale the stellar mass distribution found  in our model at $z=3.0$  towards smaller $\rm M_{*}$ such as the peak of the distribution matches the one at $z=5.7$. Then, the fraction of galaxies below the resolution limit is $\sim 0.04$. This hints that only a small part of the LAE population a would lie in galaxies not resolved in our model. 

Meanwhile, the SFR of the full galaxy population also evolves with redshift. \galform\ predicts a progressive increase in the SFR from redshift 5.7 to 2.2 (middle column of Fig.\ref{fig:galaxy_prop}). However, LAE populations exhibit a moderate SFR distribution that remains almost frozen through cosmic time. For all redshifts the SFR peaks at $10^{0.25}\Munits yr^{-1}$, well below the galaxies with the highest SFR ($\sim 10^{2.5}\Munits yr^{-1}$). We find no significant evolution from $z=2.2$ to 3.0, while at $z=5.7$ the distribution becomes a little bit broader.   

The outflow geometry has a significant impact on the SFR distribution of LAEs. In particular, at all redshifts, the LAE samples characterized by the Thin Shell exhibit lower SFR than those using the Galactic Wind. This is caused by several reasons. First, the recipes to link galaxy proprieties to outflow properties are different between both geometries (see Eq.\ref{eq:recipe-V_exp} and Eq.\ref{eq:recipe-column_density}). Second, the escape fraction of \lya\ photons depends strongly  on the geometry \citep{gurung18a, Gurung_2018b}.

\galform\ also predicts an evolution through cosmic time of the metal abundance in galaxies (right column of Fig.\ref{fig:galaxy_prop}). In fact, galaxies at lower redshifts exhibit higher metallicity. This is a consequence of the consecutive events of SFR that pollute the initially pristine interstellar medium. The typical maximum metallicities are $\rm Z \sim 10^{-0.75}Z_{\odot}$, $\rm \sim 10^{-0.6}Z_{\odot}$ and $\rm \sim 10^{-0.5}Z_{\odot}$ at redshift 5.7, 3. and 2.2 respectively. 

The galaxies observed as LAEs exhibit a low metallicity independently of redshift or cosmic time. Additionally, LAEs also show the $Z$ evolution present in the full galaxy population. Their metallicity distributions peak around $\rm \sim 10^{-2.5}Z_{\odot}$ at $z=2.2$ and 3.0, while at $z=5.7$ it peaks at $\rm \sim 10^{-2.75}Z_{\odot}$. There also differences between the Thin Shell and Galactic Wind models at low redshift. At $z=2.2$ and 3.0 the Thin Shell predicts a higher number of LAEs with  $\rm Z\sim 10^{-1.25}Z_{\odot}$. Finally, very little differences are found between the LAE samples including RT in the IGM and excluding it.

\section{ LAE IGM properties. }\label{sec:IGM_prop}

In this section we study how large scale properties of the IGM affect our full RT LAE samples (\flae). Later, in \S \ref{sec:discussion}, we compare our results with previous works. 

We analyze the \lya\ IGM escape fraction computed as the ratio of the observed \lya\ emission and the total \lya\ flux escaping galaxies, i.e,

\begin{equation}
\label{eq:EW}
f_{\rm esc}^{\rm IGM} =
{
{ 
{\rm L_{Ly\alpha}^{\rm Disk}} \times
f_{\rm esc}^{\rm IGM, Disk}
+
{\rm L_{Ly\alpha}^{\rm Bulge}} \times
f_{\rm esc}^{\rm IGM, Bulge}
}
\over
{
{\rm L_{Ly\alpha}^{\rm Disk}}
+
{\rm L_{Ly\alpha}^{\rm Bulge}}
}} ,
\end{equation} 
as a function of the density $\rho$ (Eq. \ref{eq:rho_dark_mater_expression}), the IGM line of sight (LoS) velocity $\rm V_{z}$, the IGM gradient along the LoS of the velocity $\partial _{z}V_{z}$\footnote{ Throughout this paper we use the Einstein notation for partial derivatives, i.e., $\partial_{z}\coloneqq {{\partial}\over{  \partial z }}$.} and density $\partial_{z}\rho$. The gradients are computed from their respectively fields by a 3-point derivative method.

    
\begin{figure*}
        \centering
        \includegraphics[width=3.2in]{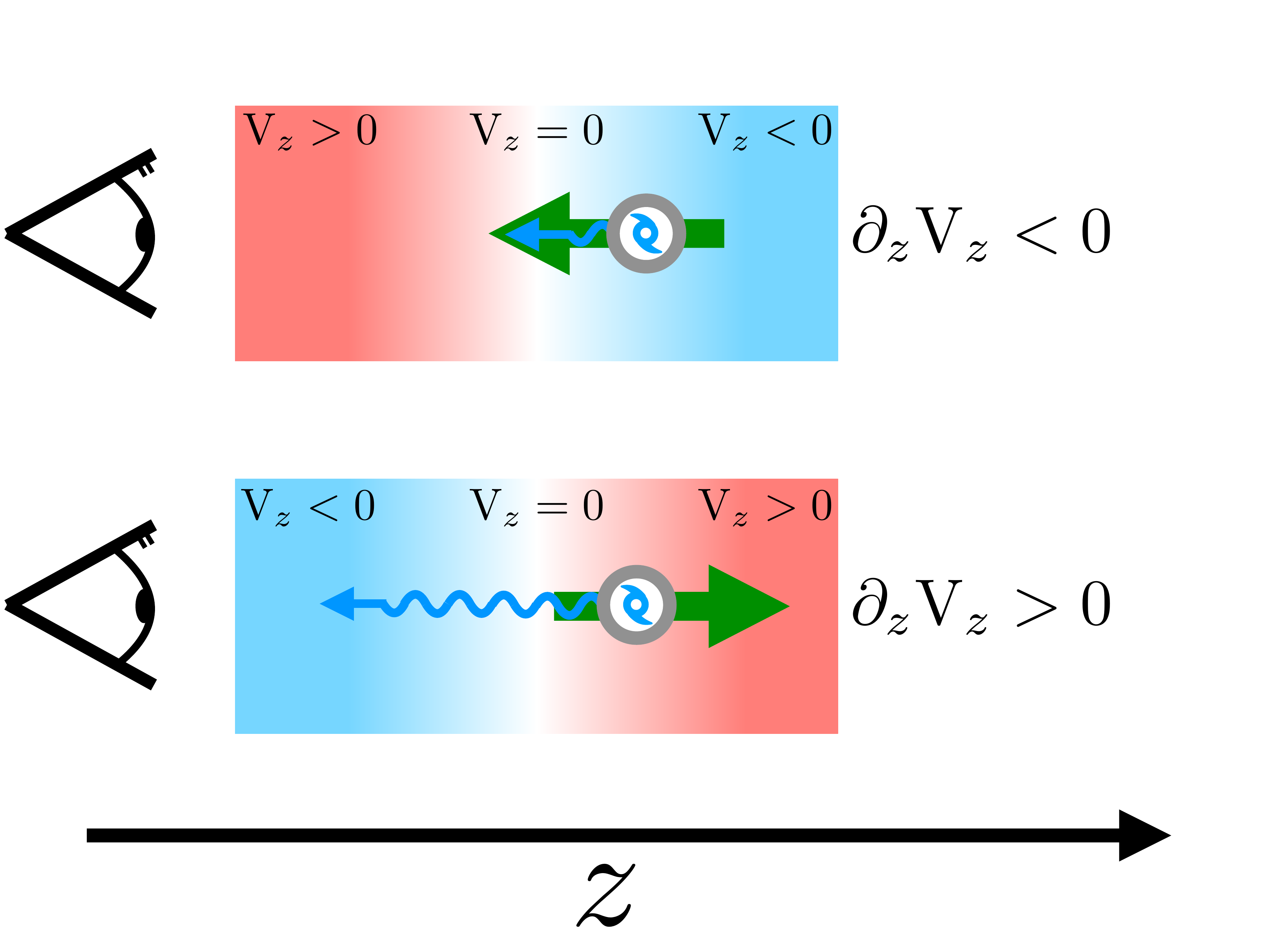}
        \includegraphics[width=3.2in]{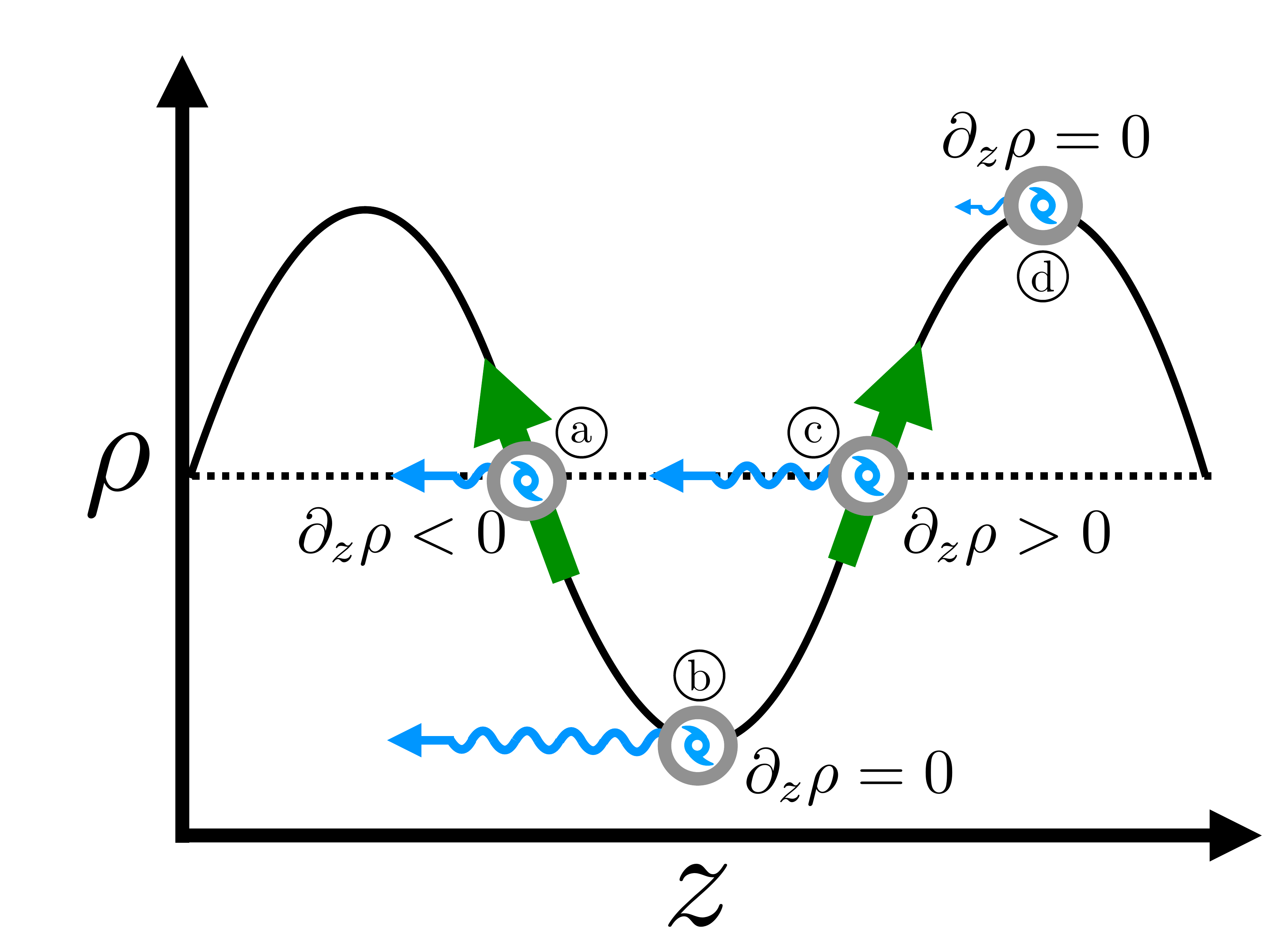}
        \caption{ Schematic representation of the effect of density gradients. In this figure the observer is always at the left edge. Galaxies are represented as grey circles with a spiral inside. The green arrows indicate the gradient. \lya\ photons emitted toward the observer are shown with sinusoidal curves. The longer the photon arrows are, the higher the received flux and IGM transmission.  \textbf{Left}: Illustration of the $\rm \partial_z V_z$ influence on the \lya\ observability. The background color maps represent IGM regions with a velocity gradient. The velocity along the line is color coded from negative (blue) to positive (red). \textbf{Right}: Cartoon of how the $\rho$ and $\partial_z \rho$ modify the \lya\ transmission. The dark line represents a density fluctuation along the line of sight.  } \label{fig:sketch_fields}
\label{fig:LFbreakdown}
\end{figure*}

    \subsection{ The Transmission of the IGM }
    
In Fig.\ref{fig:IGM_transmission_Thin} and Fig.\ref{fig:IGM_transmission_Wind} we show the $f_{\rm esc}^{\rm IGM}$ behaviour of  the Thin Shell and Galactic Wind  models against the IGM properties at different redshifts. Overall, we find that the Thin Shell and Galactic Wind exhibit the same global trends.  Since most of the \lya\ photons are redshifted by the ISM, the IGM absorbs only a small fraction of the \lya\ flux that escaped from galaxies. For both geometries, $f_{\rm esc}^{\rm IGM}$ is close to unity. The median $f_{\rm esc}^{\rm IGM}$ increases with the  age of the Universe, as the IGM becomes more transparent to \lya\ photons. We find that at $z=2.2$ and 3.0 the IGM absorbs more photons in the Thin Shell model than in the Galactic Wind, while at $z=5.7$ the absorption is comparable. In particular, for the Thin Shell, $f_{\rm esc}^{\rm IGM} \gtrsim 0.99$ , $\sim0.96$ , $\sim0.92$ at redshift $2.2$, $3.0$ and $5.7$ respectively. Meanwhile, the median $f_{\rm esc}^{\rm IGM}$ for the Galactic Wind rounds $\sim 0.9999$ , $\sim 0.999$ and $\sim 0.92$ at redshift 2.2, 3.0 and 5.7 respectively. 

The differences between the Thin Shell and Galactic Wind  are mainly due to their different \lya\ line profiles. Our model is calibrated by fitting the luminosity function of our model with observations. There are two actors converting the intrinsic \lya\ LF into the observed one: the RT in the ISM and in the IGM. As seen in Fig.\ref{fig:LFs} the IGM effect in the LF is small at any redshift, which puts most (but not all) of the weight of the fitting in the ISM component. Roughly speaking, the shape of the observed LF determines the $\rm V_{exp}$ and $\rm N_H$  distributions (see Appendix) in our model through equations  \ref{eq:recipe-V_exp} and \ref{eq:recipe-column_density}. Moreover, these distributions determine the properties of the \lya\ line profiles and, therefore, the IGM absorption.  Galaxies emitting more flux bluewards of \lya\ would be more obscured than if they were emitting only redwards of \lya . In practice, we find that the  $\rm V_{exp}$ and $\rm N_H$ distributions make the Thin Shell to be more coupled to IGM at $z=2.2$ and $3.0$ while the opposite happens at $z=5.7$, where the Galactic Wind is more affected. 


The median $f_{\rm esc}^{\rm IGM}$ varies over the dynamical range of the large scale IGM properties studied in this work. We note that the $f_{\rm esc}^{\rm IGM}$ variation is smaller than the dispersion around the median. However, the scatter is not caused by uncertainties, but by the great diversity of combinations of $\rho$, $\rm \partial_z \rho$, $\rm V_z$ and $\rm \partial_z V_z$. This points in the same direction as Fig. \ref{fig:transmission}, where the great variety of the IGM transmission curves is shown. These trends are statistically significant, i.e., they are not caused by noise or sample variance in our data set. In general, we find the same trends in the Thin Shell and Galactic Wind models. The strength of the correlations evolves with redshift. As the IGM becomes more transparent the trends become weaker. In fact, the lower the redshift, the weaker are the dependencies on $f_{\rm esc}^{\rm IGM}$. In particular, for the Thin Shell, the typical changes in $f_{\rm esc}^{\rm IGM}$ are $\lesssim 1 \%$, $\sim 2\%$ and $\sim 5\%$ at $z=2.2$, 3.0 and 5.7 respectively. Meanwhile, in the Galactic Wind modes, the   $f_{\rm esc}^{\rm IGM}$ variations are $\lesssim 0.01 \%$, $\lesssim 0.1 \%$ and $\sim 10\%$.

We find an anti-correlation between the IGM transmission and the local density. This  is more apparent at $z=2.2$ and dilutes at higher redshift. In particular, at $z=5.7$ the median $f_{\rm esc}^{\rm IGM}$ is quite flat and the dispersion becomes greater at higher densities.

In general, $f_{\rm esc}^{\rm IGM}$ correlates with $\partial _z \rho $ at all redshifts. In addition to this trend, at redshift $2.2$ and $3.0$, $f_{\rm esc}^{\rm IGM}$ peaks at $\partial _z \rho = 0$, where the IGM transmission is slightly higher. However this peak is not present at $z=5.7$.

We find a clear correlation between the IGM transmission and the IGM velocity along the line of sight at all redshifts. Galaxies in IGM regions moving towards the observer ($\rm V_{z}<0$) suffer, statistically, greater absorption than galaxies moving away from the observer ($\rm V_{z}>0$). The amplitude of the correlation augments towards higher redshifts, when the IGM becomes more optically thick. In particular, the IGM transmission variation in the $\rm V_{z}$ dynamical range rounds $<1\%$, $\sim1\%$ and $\sim5\%$ in the Thin Shell at redshift 2.2, 3.0 and 5.7 respectively.  

Finally, at $z=2.2$, $f_{\rm esc}^{\rm IGM}$ correlates with $\rm \partial_z V_{z}$. However, at $z=3.0$ no correlation is found.  At $z=5.7$ we find a small anti-correlation that produces variations of $\sim 2\%$ in the IGM transmission in the $\rm \partial_z V_{z}$ dynamical range.


    \subsection{The IGM-LAE coupling. }\label{ssec:IGM_LAE_coupling}

In \cite{zheng11} (from now on \ZZ) the authors present a model that features the LAEs coupling with the large scale properties of the IGM. The LAE models presented in this work exhibit similar trends to the relations found by \ZZ . In this section we discuss the origin of the  IGM-LAE coupling, that matches \ZZ\ interpretation for some IGM properties ($\rho$, $\partial_z \rho$, $\rm V_{z}$) while differs in others ($\rm \partial_z V_z$).


\begin{figure*} 
\includegraphics[width=6.8in]{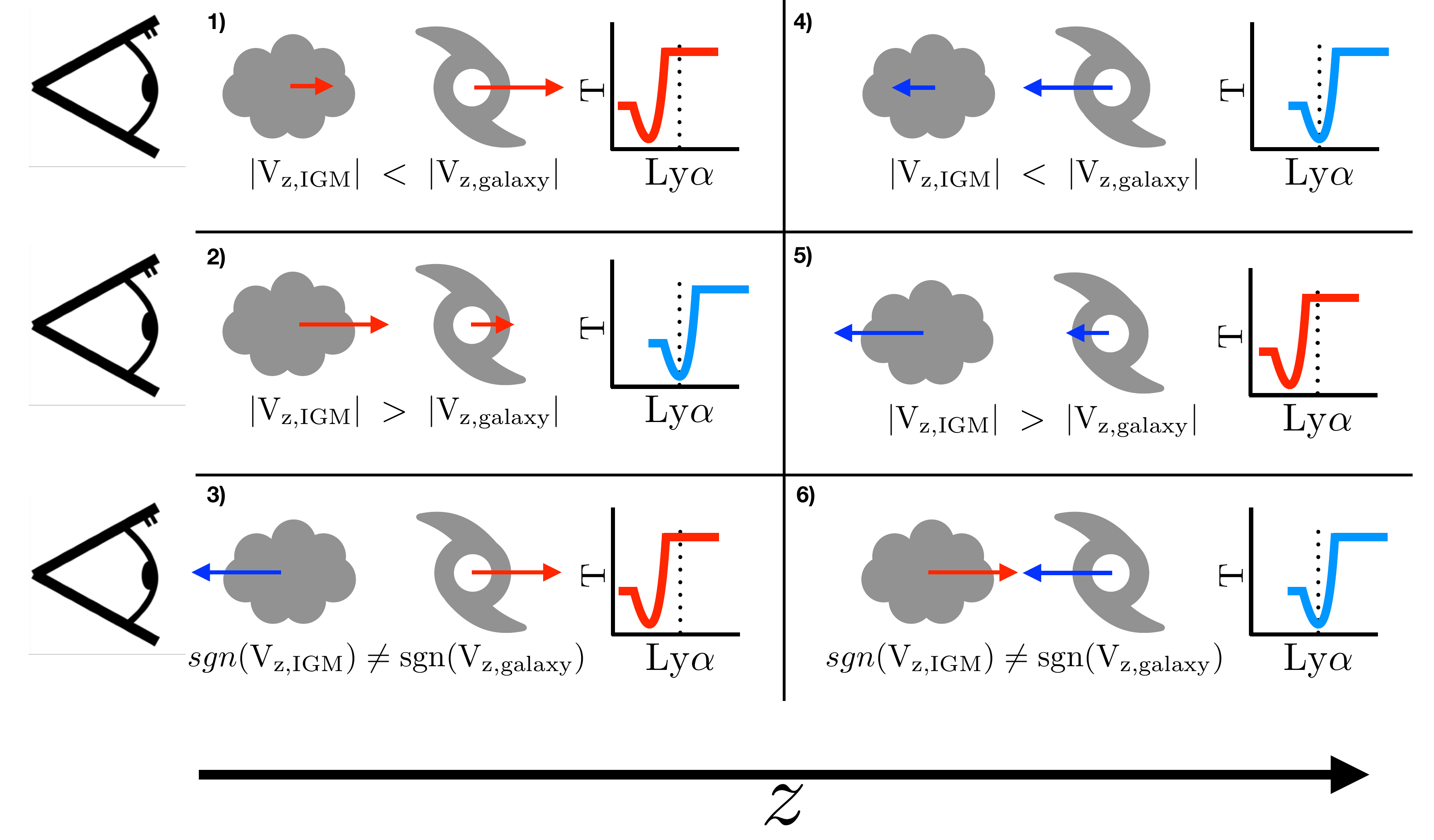} 
\caption{ Illustration of how the peculiar velocities of galaxies and the IGM change the \lya\ observability. The observer is at the left edge. The figure is divided in six panels, one for each of the different combinations of IGM and galaxy motion (see \S \ref{sssec:motion}). The clouds represent the IGM close to the galaxy along the LoS direction. The arrows are the velocity vectors of each component. The small plots are the IGM transmission in the galaxy rest frame versus wavelength. The dotted line within indicates the \lya\ wavelength.  }\label{fig:velocities_cartoon}
\end{figure*}


    \subsubsection{IGM Density.}

At low redshift, $f_{\rm esc}^{\rm IGM}$ anticorrelates with the density. This is caused by the column density along the line of sight between the galaxy and the distance where the IGM becomes transparent due to the Hubble flow. In regions of high density the amount of \ion{H}{I} that \lya\ photons have to go through is greater than in environments with lower density. This causes a greater number of scattering events and absorptions, in the high density regions. Actually, this can be seen in  Fig.\ref{fig:density_trans}, where underdense regions exhibit a higher transmission than denser regions. 


However, at $z=5.7$ the trend reverses, and $f_{\rm esc}^{\rm IGM}$ is higher the greater the IGM density is. It is still true that the \ion{H}{I} column density is systematically higher in over dense regions.  However, independently of the environment the IGM transmission at wavelengths bluer than \lya\ is below 1\%. This causes that the IGM selection effect on the density less important. At the same time, the impact of the IGM increases with redshift, causing strong selection effects on other IGM properties. Therefore, the correlation found between $f_{\rm esc}^{\rm IGM}$ and density is caused by the other IGM properties.

        \subsubsection{IGM density gradient.}

Our models predict a correlation between $f_{\rm esc}^{\rm IGM}$ and $\partial _ z \rho$. This trend  is caused by the difference in \ion{H}{I} column density between the galaxy and the distance where the IGM becomes transparent. This scenario is illustrated in the right panel of Fig.\ref{fig:sketch_fields}. On one hand, for a source of \lya\ photons at a fixed $\rho$, if the density gradient is positive (galaxy 'c'), then the IGM density decreases as the photons travel through it, allowing more photons to escape. On the other hand, in the case that  $\partial _ z \rho<0$ (galaxy 'a'), the IGM density increases as photons go through the IGM, causing a higher number of lost photons.

Additionally, there are two main different regimes with $\partial _ z \rho = 0$: i) the peak of overdensities (galaxy 'd' of Fig.\ref{fig:sketch_fields}) with low IGM transmission due to the high column density; and ii) the bottom of underdensities (galaxy 'b') with high $f_{\rm esc}^{\rm IGM}$.  The relative occupancy of these regions changes due to the RT in the IGM and ISM. LAE are preferentially observed in underdense environments. This is caused by two main reasons. First, the low IGM transmission in overdense regions. Second, the RT in the ISM prevents very massive galaxies (which would lie in overdense regions) to be observed as LAEs (see Fig.\ref{fig:galaxy_prop}).  Therefore, most of the galaxies observed in environments with $\partial _ z \rho = 0$ are in underdense regions. Meanwhile, the IGM transmission is higher the lower the density is. This causes the transmission peak found at $\partial _ z \rho = 0$ for $z=2.2$, $3.0$.

\begin{figure*} 
\includegraphics[width=7.0in]{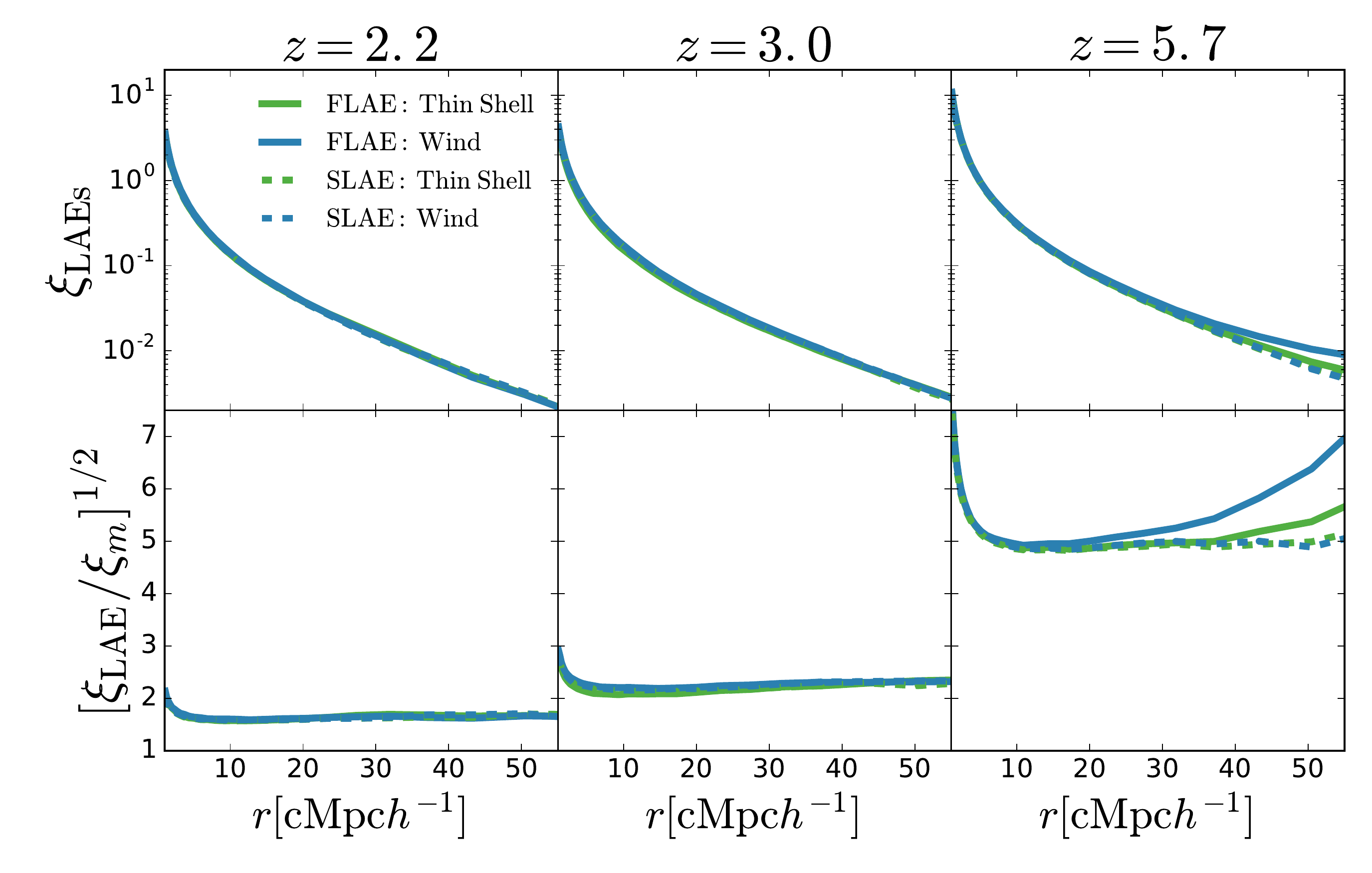} 
\caption{ \textbf{Top:} 2-point Correlation function in real space for different samples at different redshits (2.2, 3.0 and 5.7 from left to right). In dashed blue and green we show the SLAE Galactic Wind and Thin Shell respectively, while on the colors and solid line show the full RT models for the Galactic Wind and Thin Shell. \textbf{Bottom:} Bias of the different models as a function of distance. }\label{fig:xcor}
\end{figure*}

\subsubsection{IGM kinematics }\label{sssec:motion}

When the relative velocity between the galaxy and the IGM is zero only photons bluewards of \lya\ are absorbed, as seen in Fig. \ref{fig:transmission}. However, if the IGM and the galaxy are moving towards each other, the \lya\ line is blueshifted in the IGM rest frame, causing greater absorption. In the opposite scenario, where the galaxy and the IGM are moving away from each other, the \lya\ photons are redshifted and escape the nearby IGM more easily. 

The relative motion between the IGM and galaxies causes  selection effects. On average, galaxies moving  away from the observer are more likely to be observed as LAE. In Fig. \ref{fig:velocities_cartoon}  we list the different combinations of velocities between IGM and galaxies along the LoS.  

\begin{itemize}
    \item{Case 1) Both the galaxy and the IGM are moving away form the observer. However, the galaxy velocity is greater. The IGM only absorbs bluer photon than \lya. } 
    \item{Case 2) Both the galaxy and the IGM are moving away form the observer. Additionally, the IGM moves faster than the galaxy. In this case, redder wavelengths than \lya\ are absorbed.  }
    \item{Case 3) The galaxy goes away while the IGM approaches the observer. \lya\ photons are redshifted when they reach the IGM. Only bluer frequencies than \lya\ suffer absorption. }
    \item{Case 4) Both the IGM and the galaxy approaches the observer. In this case the galaxy moves faster and the IGM absorbs redder wavelengths than \lya.}
    \item{ Case 5) Both the galaxy and IGM come towards the observer, but the IGM travels faster. The absorption only happens in bluer frequencies than \lya.}
    \item{ Case 6) The IGM moves away from the observer while the galaxy approaches it. The \lya\ photons are blueshifted when they reach the IGM. Wavelengths redder than \lya\ get absorbed. }
\end{itemize}

Additionally, we have computed the relative abundance of these six scenarios in the galaxy population of \galform. We find that the cases 1) and 4) are equally probable, as well as 2) with 5) and 3) with 6). In particular, cases 1) and 4) constitute the 25\% of the galaxy population each. Moreover, 2) and 4) represent 20\% each, while 3) and 6) only 5\% each. 

On one hand, among the three scenarios where galaxies are moving away from the observer (cases 1, 2 and 3), in two of them (cases 1 and 3) the relative velocity between the IGM and galaxy is positive. In these cases the \lya\ line profile is received redshifted in the IGM frame, causing low absorption. On the other hand, when the galaxy is approaching the observer (cases 4, 5 and 6) in two scenarios (4 and 6) the IGM sees \lya\ blueshifted and absorbs \lya\ photons.  In addition, 60\% of galaxies moving away from the observer are redshifted in the frame of the IGM. Meanwhile the 60\% of galaxies approaching the observer are seen blueshifted by the IGM. This asymmetry causes that galaxies with $\rm V_{z}>0$ (getting away from the observer) are more likely to be observed as LAEs.


\subsubsection{IGM velocity gradient.}

A further level of complexity is given by the cosmological velocity structure of the Universe (see Fig.\ref{fig:v_field}). The IGM is divided into regions of coherent motion of hundreds of $\rm cMpc$ that can be collapsing or moving away from each other. Additionally, the areas with $V_{\rm LoS}\sim 0$ between these regions are small (${\rm \sim 5 cMpc} h^{-1}$), causing a great contrast of velocity on scales critical\footnote{ Typically, the Hubble flow redshifts the \lya\ line 1\AA{} per ${\rm \sim 3  cMpc}$ at $z=3.0$.} to the IGM absorption.  

There are two opposite effects controlling the IGM transmission dependence on $\partial_z {\rm V}_z$. On the one hand, the transition between coherent motion regions facilitates the escape of \lya\ photons emitted in  clouds with $\partial_z V_z > 0$, as illustrated in the left panel of  Fig.\ref{fig:sketch_fields}. For example, if a given galaxy lies in the border of the IGM cloud with  $\rm V_{z}> 0$ (bottom panel), \lya\ photons are observed strongly redshifted in static ($\rm V_{z} \sim 0$) regions, and even more in clouds with $\rm V_{z}< 0$. In this case the \lya\  escapes more easily. Meanwhile, in the opposite scenario, where the galaxy lies in a region with $\partial_z V_z < 0$, the neutral hydrogen increases the velocity towards the galaxy as the photon travels. This would result in a greater absorption. On the other hand, in regions with $\partial_z V_z < 0$ the Hubble flow effect is enhanced and photons escape more easily.  

The Hubble parameter determines which of these effects prevail. If $H(z)$ is low, the typical distance that \lya\ photons have to travel before the IGM becomes transparent is greater. In this case, this distance is compatible with the transition region between $\rm V_{z}<0$ and $\rm V_{z}>0$. Therefore, the \lya\ transmission correlates with  $\partial_z {\rm V}_z < 0$. We find that in our model this happens at z=2.2 and 3.0. Meanwhile, if $H(z)$ is high, then the distance at which the IGM becomes transparent is smaller than the transition region. In this scenario regions with $\partial_z {\rm V}_z > 0$ exhibit greater transmission. We find that this last scenario dominates $z=5.7$ in our models.

\begin{figure*}
        \centering
        \includegraphics[width=3.2in]{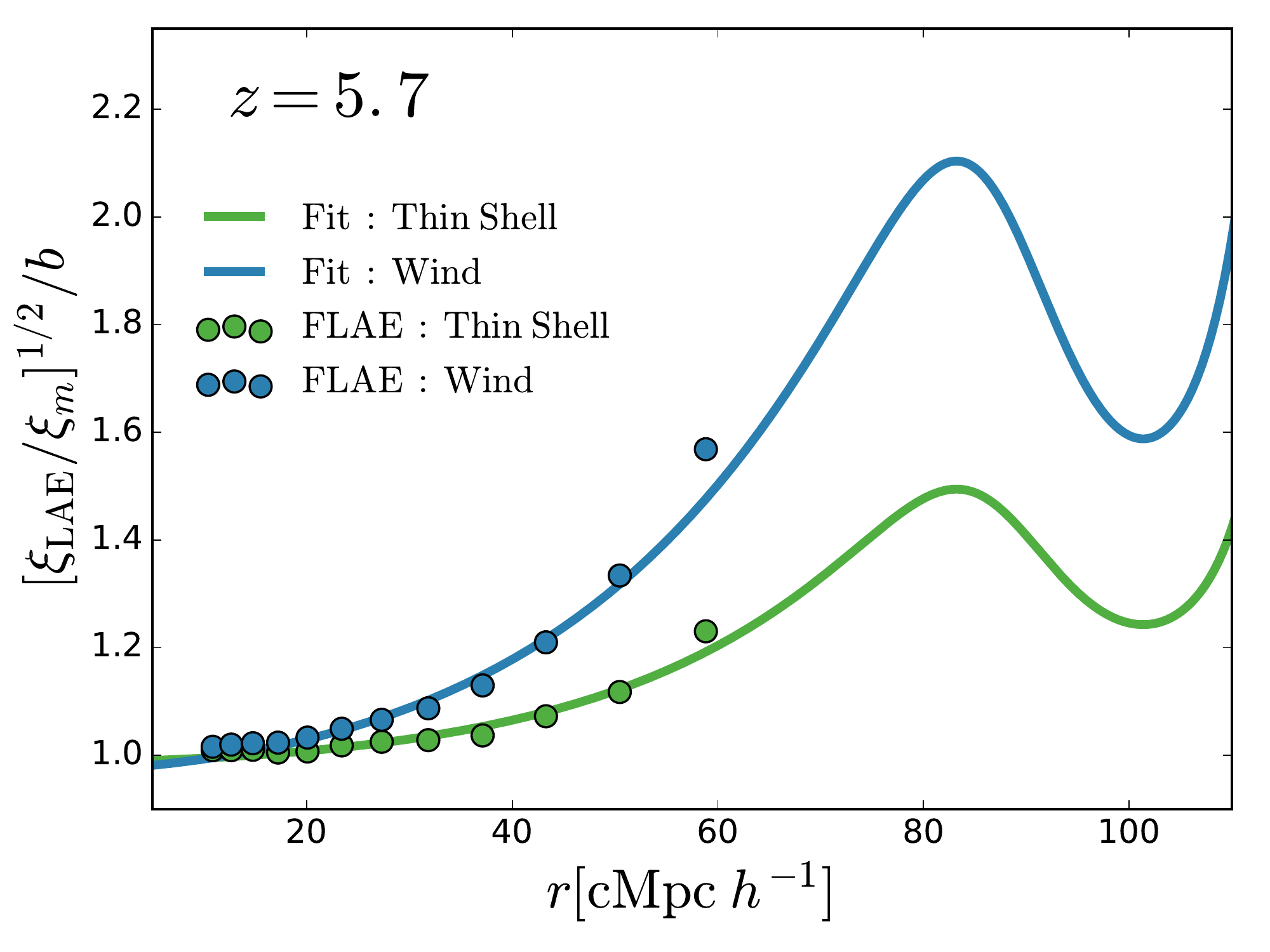}
        \includegraphics[width=3.2in]{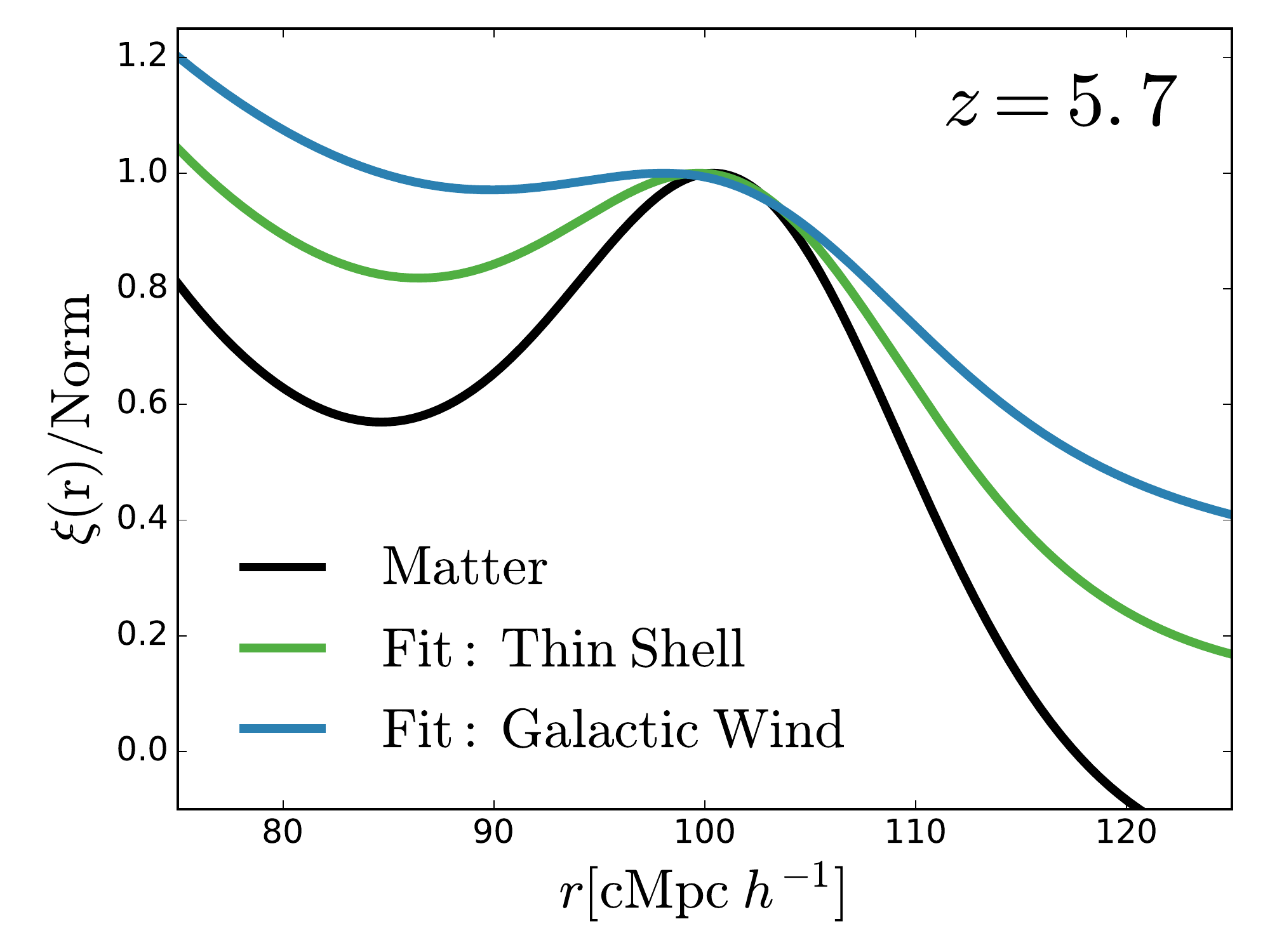}
        \caption{ \textbf{Left:} Ratio between the bias as a function of distance and the median bias measured from 10 to ${30 \rm cMpc }h^{-1}$ at redshift 5.7. In dots we show our simulation clustering while in solid curves the output of the MCMC analysis fitting the \ZZ\ clustering description is illustrated. In blue and green we show the Galactic Wind and Thin Shell respectively. \textbf{Right:} Comparison of the 2PCF of our models and the matter 2PCF around the BAO peak. The clustering amplitudes are renormalized so that the maximun of the 2PCF between 90 and ${110 \rm cMpc }h^{-1}$ match unity.} \label{fig:ZZ_fits}
\label{fig:LFbreakdown}
\end{figure*}

\section{ The clustering of LAEs }\label{sec:clustering}

In this section we analyze how the coupling of \lya\ detectability  and the IGM large scale properties modify the clustering of LAEs. To do so we compare the full RT LAE sample (\flae) and the \slae. The \slae\ populations inherit the mass function from the \flae\ population. In contrast with the \flae\ galaxies, by construction, the \slae\ positions do not depend on the IGM transmission.

\subsection{ 2-point 3D Correlation function. }

In this section we analyze the clustering of our samples in real space. Note that similar result are found in redshift space. 

In Fig. \ref{fig:xcor} we show the real-space 3D 2-point correlation function $\xi(r)$ and the bias computed as 

\begin{equation}
\label{eq:bias_from_xcor}
b(r)={ \left[  { { \xi(r)} \over {\xi_{m}(r)}} \right] ^ {1/2} } , 
\end{equation} 
where  $\xi_{m}(r)$ is the matter correlation function. 

By comparing LAE samples that include the full RT processes (\flae) and their shuffled samples (\slae) we can understand the IGM impact. We note that, by construction, our \slae\ samples do not isolate the IGM effects but also includes assembly bias processes. However, we have checked that if we create shuffled samples from the \nlae\ populations (only RT in the ISM) the clustering measurements are identical, i.e., no evidence of assembly bias is found. Therefore, we attribute the differences in the clustering between \flae\ and \slae\ samples to only the IGM impact. 

Overall, we find that the \slae\ populations (both, Thin Shell and Galactic Wind geometry) behave in the same way in every redshift bin. Below $\sim 5 {\rm cMpc}h^{-1}$ the bias of \slae\ is not constant and decays with distance. From  $\sim 5 {\rm cMpc}h^{-1}$ on, their bias becomes scale-independent at a very similar value for both geometries. The bias increases with redshift. In detail, the bias of the \slae\ on scales larger than  $10 {\rm cMpc}h^{-1}$ is $\sim 1.8$, $\sim 2.4$ and $\sim 5.2$ at redshift 2.2, 3.0 and 5.7 respectively.

Meanwhile, the \flae\ samples exhibit different behaviours at different epochs. On one hand, at redshifts 2.2 and 3.0, the \flae\ and \slae\ clustering exhibit the same trends and they are almost indistinguishable. Therefore, we find  that the IGM does not shape the LAE clustering at these redshifts. 

On the other hand, at larges distances, we find that the IGM increases the clustering of \flae s at $z=5.7$. At scales smaller than $\sim 5 {\rm cMpc}h^{-1}$ the \flae\ and \slae\ clustering are identical. However, the \flae\ samples including RT in the IGM exhibit a scale dependent clustering excess on scales ${\rm >20 cMpc }h^{-1}$. The boost is present in both outflow geometries. However, its amplitude changes with the geometry. We find that the Galactic Wind geometry exhibit a more powerfull boost than the Thin Shell. We attribute this to the greater coupling with the IGM in the Galactic Wind than in the Thin shell. In particular, the clustering at $ 50 {\rm cMpc}h^{-1}$ is boosted a factor of $\sim 1.1$ and $\sim 1.3$ in the Thin Shell and Galactic Wind geometry respectively. We study the origin of the clustering boost in next section.

\subsection{ The LAE clustering at large scales. }\label{ssec:largE_scales}

In this section we interpret our clustering measurements with the physical model presented in \ZZ . They described the overdensity field  of galaxy population correlated with the large scale properties of the IGM as

\begin{align} 
\label{eq:over_density_laes}
\delta_{g} = b \delta_{m} \times 
\left[ 1 + \tilde{\alpha}_1 \delta_m +
\tilde{\alpha}_2 {{1}\over{aH}} \partial_z {\rm V}_z +  
\right. \notag\\ \left. +
\tilde{\alpha}_3 {{1}\over{aH}} \left( \partial_x {\rm V}_x + \partial_y {\rm V}_y \right) + 
\right. \notag\\ \left. +
\tilde{\alpha}_4 {{1}\over{aH}} {{{\rm V}_z}\over{r_{{\rm H}}}} +
\tilde{\alpha}_5 r_{{\rm H}} \partial _ z \delta_m \right] ,
\end{align}
where $a$ is the scale factor, $\delta_m$ is the overdensity of matter in the universe, $r_{\rm H}$ is a length scale set the coefficients dimensionless and the parameters $\tilde{\alpha_i}$ are free parameters that quantify the coupling with the IGM properties.

From this expression, the monopole of the galaxy power spectrum can be expressed in real space as a function of the matter power spectrum $P_m$ :

\begin{align}
\label{eq:monopole_power_zz}
P_{0}(k) = \left[ \gamma_1 ^2 + 
{{2}\over{3}} \gamma_1 \gamma_2 + 
{{1}\over{5}}\gamma_2 ^2 + 
{{1}\over{3}} \gamma_3 ^2 
\right] 
b^{2} P_{m}(k),
\end{align}
where



\begin{align}
\gamma_1 = 
\left( 
1 + 
{{ \tilde{\alpha}_1 - \tilde{\alpha}_3 f}\over{b}}
\right) , 
\end{align}

\begin{align}
\gamma_2 = 
\left( 
\tilde{\alpha}_3 - \tilde{\alpha}_2
\right) 
\beta , 
\end{align}
and
\begin{align}
\gamma_3 = 
\left( 
\tilde{\alpha}_4 \beta {{1}\over{k r_{\rm H}}} +
\tilde{\alpha}_5 {{k r_{\rm H}}\over{b}}
\right),
\end{align}
where $f$ the growth factor and $\beta=f/b$. Meanwhile in redshift space, due to the redshift-space distortions \citep{kaiser87} $\gamma_2$ is rewritten as

\begin{align}
\label{eq:monopole_power_gamma2}
\gamma_2 = 
\left( 
1 + \tilde{\alpha}_3 - \tilde{\alpha}_2
\right) 
\beta .
\end{align}


\begin{figure*} 
\includegraphics[width=7.0in]{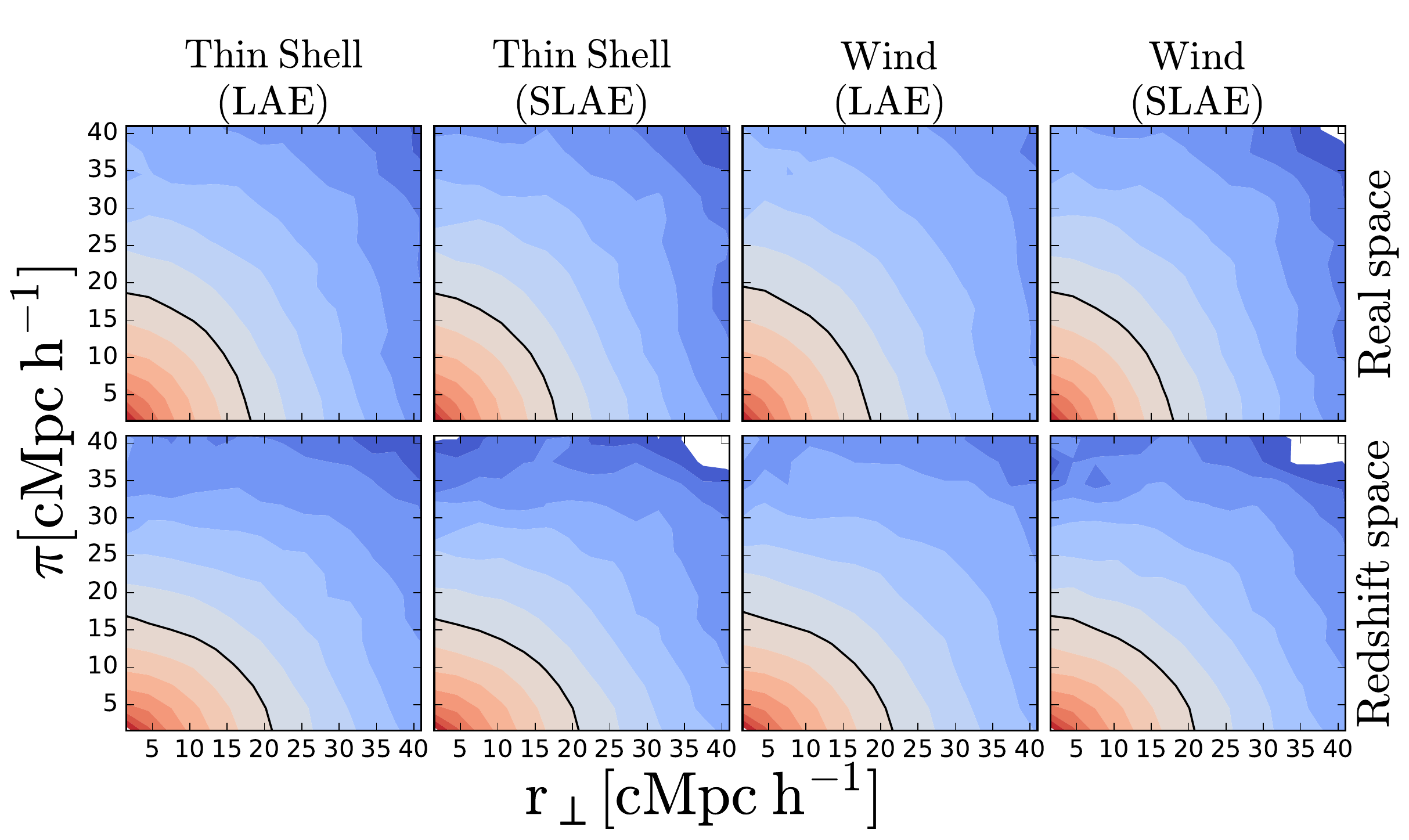} 
\caption{ LAE 2-point correlation function as a function of  parallel ($\pi$) and perpendicular ($\rm r_{\perp}$) distance to the line of sight in real space (top)  and redshift space (bottom) at redshifts 5.7. From left to right we show the \flae\ Thin Shell, the \slae Thin Shell, the \flae\ Galactic Wind and  the \slae\ Galactic Wind. The countors are curves of iso-clustering amplitude, divided in 15 bins from $\xi=10^{0.5}$ to $\xi=10^{-2.5}$ equispaced in logarithmic scale and are color-coded coherently. Additionally, the black curve indicate $\xi(\pi,r_{\perp})=0.01$. }\label{fig:RSD_real}
\end{figure*}


In this description the monopole ($P_0$) exhibits other additional  dependencies  on scale rather than $P_m$. There are two terms modifying the shape of the LAE power spectrum. First, a term proportional to $\tilde{\alpha}_4/k$ that enhances the clustering at large scales if the velocity field and LAE distribution are coupled. Second, a term proportional to $\tilde{\alpha}_5  k$ that amplifies the clustering in small scales if the LAEs are coupled to the gradient of density along the line of sight. Finally, the terms proportional to $\tilde{\alpha}_1$, $\tilde{\alpha}_2$ and $\tilde{\alpha}_3$ are scale independent and they only modify the clustering amplitude, leaving the monopole shape unaffected. 

We perform a MCMC fit to Eq.\ref{eq:monopole_power_zz} using our model predictions at z=5.7  in real space. We assume that the bias $b$ of the halo population hosting the \flae\ samples is the same as their derived \slae\ sample. In practice we compute $b$ for each geometry outflow comparing the \slae\ clustering to the matter clustering through Eq.\ref{eq:bias_from_xcor} and averaging between ${\rm 10 cMpc}h^{-1}$ and ${\rm 30 cMpc}h^{-1}$. Additionally, we  set $\tilde{\alpha}_1$, $\tilde{\alpha}_2$, $\tilde{\alpha}_3$ and $\tilde{\alpha}_5$ to zero, while $\rm \tilde{\alpha}_4/r_{H}$ is the only free parameter.  In fact, we checked that the quality of the fit does not improve when the other parameters are included in the analysis. Finally, we restrict our analysis to the linear regime, i.e., between 5  and  $60 { \rm cMpc } h^{-1}$.

The bias measurements and fitting results are listed in Table \ref{tab:ZZ}. In general, we find that the Thin Shell and Galactic Wind exhibit similar bias, as seen in Fig.\ref{fig:xcor}. Additionally, $\tilde{\alpha}_4$ is greater in the  Galactic Wind than in the Thin Shell, indicating that the LAE model with the first one is more coupled to the IGM.  

In Fig. \ref{fig:ZZ_fits}  we compare our \flae\ models (dots) with the MCMC output (solid lines). The \ZZ\ description matches our simulations remarkably well for both outflow geometries. The extension of our models to larger scales highlights the complicated shape of the clustering at cosmological scales. In general, the ratio between the $\xi_{\rm LAE}(r)$ and $\xi_{\rm DM}(r)$ shows a hill at $\sim 80 {\rm cMpc} h^{-1}$ and a valley at $\sim ~100 {\rm cMpc} h^{-1}$, while it increases  at even larger scales. 

Also, in Fig. \ref{fig:ZZ_fits}  we compare the matter 2-point correlation function (2PCF) with the analytic clustering description (calibrated with the MCMC analysis) of our Thin Shell and Galactic Wind \flae\ samples. For the sake of a better comparison, we have renormalized the clustering of the different samples so that the 2PCF maximum between 90 and 110${\rm cMpc }h^{-1}$ matches in all cases. In the matter 2PCF we find the baryon acustic oscilation (BAO) peak, which is produced by the balance between gravity and pressure in the early Universe \citep[e.g.][]{chavez_montero_2018}. However, in the analytical description calibrated using our \flae\ model, the shape of the BAO peak is distorted. In particular, it becomes broader in the Thin Shell geometry and even wider in the Galactic Wind geometry, as it is more affected by the IGM. Not only that, but also, the position of the maximum in our fits shifts by up to ${\rm ~1cMpc }h^{-1}$ and ${\rm ~3cMpc }h^{-1}$ respectively. 


\begin{table}
\centering
\caption{Parameters of the the \ZZ\ analytic expression (Eq.\ref{eq:over_density_laes}) to the clustering measurements of our model at redshift 5.7. }
\label{tab:ZZ}
\begin{tabular}{cccc}
Redshift & Geometry      &      $b$       &   $\log \; \tilde{\alpha_4}/r_{H}$ \\ \hline
$z=5.7$  & Thin Shell    & 4.82 & -0.44 \\
         & Galactic Wind & 4.87  & -0.2 
\end{tabular}
\end{table}



\begin{figure*} 
\includegraphics[width=7.0in]{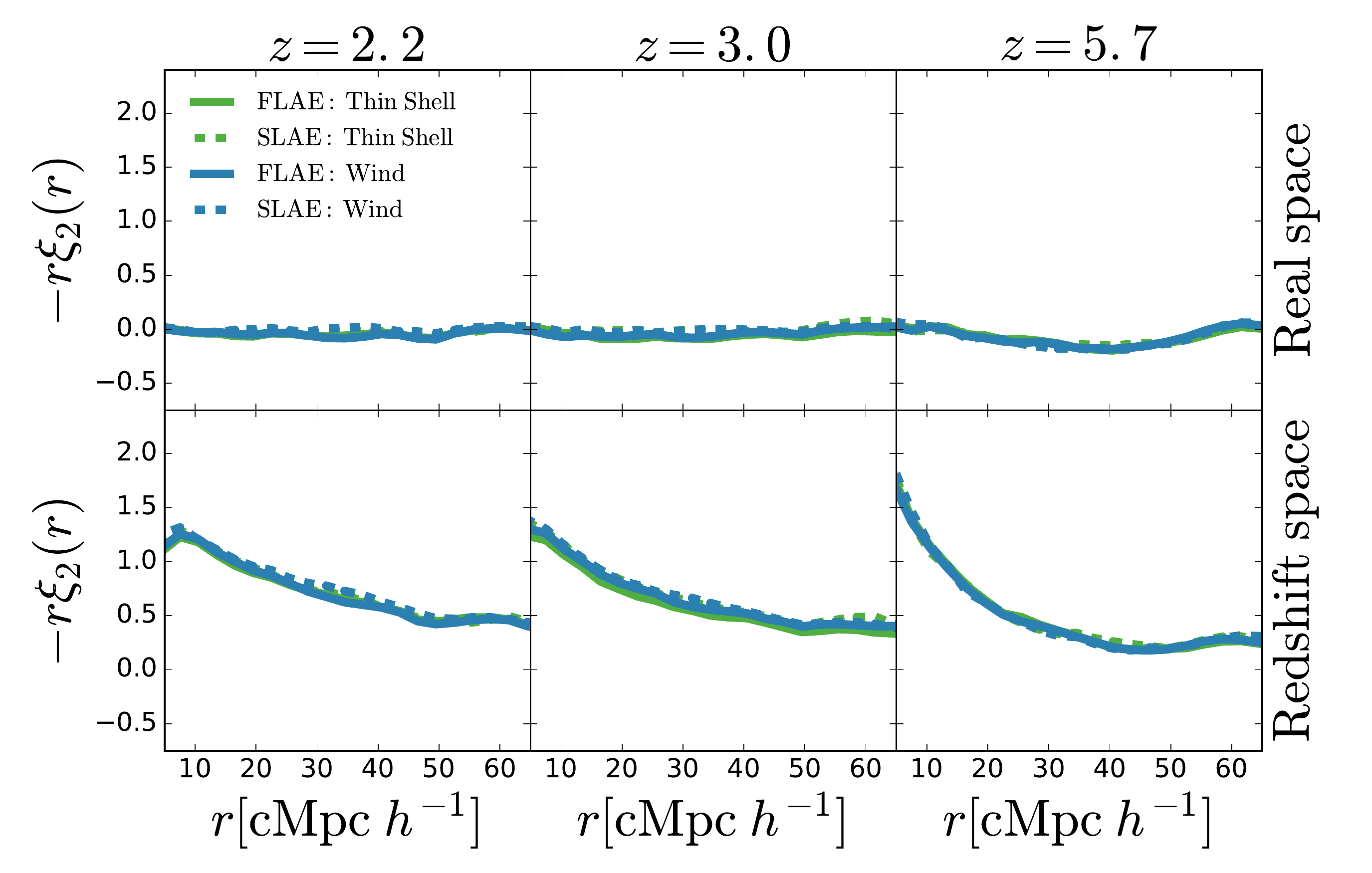} 
\caption{ Quadrupole of our different LAEs samples in real (top) and redshift space (bottom panels) at redshift 2.2, 3.0 and 5.7 from left to right. The Thin Shell and Galactic Wind are  displayed in green and blue respectively.  \flae\ population are shown in continuum lines while dashed lines represent the \slae\ samples. }\label{fig:quadrupole}
\end{figure*}


\subsection{ Clustering parallel and perpendicular to the line of sight. }

In this section we study the clustering of LAEs in the perpendicular and parallel directions to the line of sight. In order to understand the role of the IGM we compare directly our \flae\ and \slae\ samples. 

First, we make our analysis in real space to quantify the impact of the IGM on the LAE apparent spatial distribution. We perform our analysis at $z=5.7$, where the coupling with the IGM is strong.  Note that we find similar results at the other redshifts too. In Fig.\ref{fig:RSD_real} we show the real space clustering of our LAE populations split in parallel and perpendicular to the line of sight of the IGM. We find that the LAE parallel and perpendicular clustering are symmetric. No angular dependence is apparently found in any of our models. 

We also analyze the clustering of our LAE samples in redshift space. In order to convert from real space to redshift space we modify the galaxy positions with their peculiar motion along the LoS of the IGM ($\rm Z$ axis of our simulation) following

\begin{equation}\label{eq:redshift_space}
s = x_{\rm Z} +  \frac{ \rm v_{\rm Z}}{a(z)H(z)},
\end{equation} 
where $x_{\rm Z}$ is the galaxy position along the LoS and $\rm V_{\rm Z}$ is the galaxy peculiar velocity along the same direction.

In the bottom panel of Fig.\ref{fig:RSD_real} we show the 2D clustering of our LAE samples at $z=5.7$. In general, we find that in redshift space the clustering of all our models is suppressed along the LoS and conserved in the direction perpendicular to the LoS in every redshift bin. Additionally, the suppression along the LoS increases at lower redshift. In general, the peculiar velocity of galaxies points towards overdensities. Hence, the fluctuations along the line of sight are enhanced, which translates into the distortion of the clustering along the LoS.   

Furthermore, we also analyze the clustering quadrupole of our \flae\ and \slae\ samples. In particular, the quadrupole is computed as

\begin{equation}
\label{eq:quadrupole}
\xi_{\ell}(r) = {{2\ell+1}\over{2}} \int _{-1}^{1} d\mu \; \xi(r,\mu) \; \mathbcal{L} _{\ell} (\mu)
\end{equation} 
where $\ell=2$, $\mathbcal{L} _{\ell}$ is the Legendre's polynomial of degree $\ell$ and $\mu = \cos \theta$, where $\theta$ is the angle between the line of sight and the sources.

In Fig.\ref{fig:quadrupole} we show the quadrupole of the \flae\ (continuum lines) and \slae\ (dashed lines) samples at different redshifts in real space (top panels) and in redshift space (bottom panels). Overall, we find that at all redshifts the \flae\ and \slae\ quadrupoles are compatible in both, real and redshift space. Additionally, in real space (top panels) the quadrupole is null at all scales. This implies that the LAE-IGM coupling affects the parallel and perpendicular clustering in the same fashion, and no asymmetry is created. Then, in redshift space, we find a non-null quadrupole that evolves with redshift. The quadrupole becomes steeper on scales lower than $30 {\rm cMpc}h^{-1}$ at higher redshifts.

\section{ Discussion. }\label{sec:discussion}

In this section we compare our results to other theoretical works and current LAE observations.

\subsection{ Comparison with previous theoretical works. }

In \ZZ the authors studied the coupling of the LAE observability and the IGM large scale properties. Later, \cite{Behrens2017} (from now on \BC ) made the same analysis with similar techniques implemented in a simulation with higher spatial resolution. In this context, we compare our approach and results to these works. 

\subsubsection{Simulations}

\ZZ\ used a hybrid approach, where first a high resolution dark matter only N-body simulation was evolved while, on the fly, the hydrodynamic physics were run at lower resolution. The size of their simulation was $({\rm 100 cMpc}h^{-1})^3$ with a constant spatial resolution of the neutral hydrogen density field close to ${\rm 0.13 cMpc}h^{-1}$. For further detail about the simulation we refer the reader to \cite{zheng10}. 

Meanwhile, \BC\ implemented the IGM RT in the Illustris simulation \citep{Nelson2015}, a high resolution full hydrodynamic simulation using an adaptive mesh refinement approach. \BC\ rebinned Illustris to a uniform grid of resolution ${\rm 0.02 cMpc}h^{-1}$.  The size of this simulation was $({\rm 107 cMpc}h^{-1})^3$.  For further information see their original work \citep{Behrens2017}. 

Neither \ZZ\ or \BC\ were able to resolve the complex ISM structure due to its sub-$\rm kpc$ nature. Additionally, the volume of their simulation was not large enough to trace the large scale variation of the velocity field. In fact, their simulation could be enclosed within one of the many coherent motion regions populating our simulation (see Fig.\ref{fig:v_field}). 

In contrast, our work uses a dark matter N-body simulation and implements the baryons in a post-processing flavour. Our resolution for the neutral hydrogen density is ${\rm 0.2 cMpc}h^{-1}$. However, our simulation size is ${\rm 542.16 cMpc}h^{-1}$, which translates into more than 125 times more volume than previous studies. This allow us to make accurate clustering predictions up to ${\rm \sim 60 cMpc}h^{-1}$ and resolve the large scale structure of the velocity field of the Universe. 

\subsubsection{IGM radiative transfer methodology.}

Both, \ZZ\ and \BC\, implement the RT in the IGM using  post-processing Monte Carlo approaches similar to other works in the literature studying the \lya\ RT in the ISM \citep[e.g.][]{verhamme06,orsi12, Gronke_2016, Gurung_2018b}. In particular, in galaxy locations they generate photons in random directions and  frequencies assuming a Gaussian line profile centered on \lya . Then they track the photon's trajectory and changes in frequency. Finally, photons are collected and an $f_{\rm esc}^{\rm IGM}$ is computed for each galaxy comparing the number of emitted photons and the number of photons received within a given aperture centered on the galaxy position. Additionally, the \lya\ luminosity emitted to the IGM by each galaxy is assumed to be directly proportional to the SFR (although in a different way in each work). For further details, we refer the reader to \ZZ\ and \BC\ original works. 

Meanwhile, here we use a different approach also explored by \cite{Weinberger_2018}. We divide our simulation into thin cells along the line of sight and analytically compute the IGM transmission of each cells. Then we compute an absorption profile for each galaxy by summing all the small contribution of each cell. Finally, we convolve the IGM transmission with the \lya\ line profile that depends on galaxy ISM properties such as the cold gas mass or the metallicity, as well as the  \lya\ luminosity emitted to the IGM \citep{gurung18a}. 


\begin{table*}
\centering
\caption{Properties of our mock catalogs. In particular we list the redshift $z$, the redshift bin width $\delta z$, the size along and parallel to the line of sight ( $\rm L_{\parallel}$ and $\rm L_{\perp}$ respectively), the Number of mock catalogs within our simulation volume and the median number of LAE in the catalogs for different models as well as the $\pm1\sigma$ dispersions.}
\label{tab:survey_prop}
\begin{tabular}{cccccccccc}
Authors   &  $z$    & $\Delta z$ & $\rm L_{\parallel}$ & $\rm L_{\perp}$ & $\rm N_{mocks}$ & \multicolumn{3}{c}{$\rm \langle N_{LAE} \rangle$} \\ \cline{7-9}
          &         &            & $\rm (cMpc)$        & $\rm (cMpc)$    &                 & Survey       & Thin shell      & Galactic Wind             \\ \hline
\cite{Kusakabe2018} & 2.2 & 0.0773 & 104.9 & 93.6 & 448 & 1248 & $1197_{-86}^{+96}$ & $1193_{-76}^{+96}$ & \\
\cite{Bielby2016} & 3.0 & 0.0633 & 60.0 & 119.1 & 468 & 643 & $633_{-52}^{+49}$ & $633_{-47}^{+55}$ & \\
\cite{Ouchi2018a} & 5.7 & 0.0954 & 43.5 & 401.5 & 18 & 734 & $714_{-28}^{+53}$ & $723_{-51}^{+53}$ &
\end{tabular}
\end{table*}



\begin{table*}
\centering
\caption{Rest frame equivalent cut $\rm EW_{0}$ and \lya\ luminosity cut $\rm L_{Ly\alpha,cut}$ in the different surveys and \flae\ mocks.}
\label{tab:mocks_table}
\begin{tabular}{clccclccc}
Authors      &  & \multicolumn{3}{c}{$\rm EW_{0,cut} [$\AA{}$ ]$}      &  & \multicolumn{3}{c}{$\rm L_{Ly\alpha , cut } \left[ erg \ s^{-1} \right] $}       \\ \cline{3-5} \cline{7-9}
             &  &  Survey   &   Thin Shell &  Galactic Wind     &  &  Survey   & Thin Shell  &  Galactic Wind                           \\ \cline{1-9}
\cite{Kusakabe2018} &  & $20.0$ & $19.52$ & $20.3$ &  & $1.62\ 10 ^ {42}$ & $1.47\ 10 ^ {42}$ & $1.77\ 10 ^ {42}$ \\
\cite{Bielby2016} &  & $65.0$ & $38.45$ & $46.27$ &  & $1.62\ 10 ^ {42}$ & $1.32\ 10 ^ {42}$ & $1.52\ 10 ^ {42}$ \\
\cite{Ouchi2018a} &  & $20.0$ & $20.06$ & $20.06$ &  & $6.3\ 10 ^ {42}$ & $6.98\ 10 ^ {42}$ & $6.64\ 10 ^ {42}$ \\
\end{tabular}
\end{table*}


\subsubsection{IGM transmission}

In comparison, \ZZ\ and \BC\  predict an IGM escape fraction well below that in our models. \ZZ\ and \BC\ RT approaches greatly overestimate the IGM absorption in comparison to our work for two main reasons:

\begin{enumerate}
    \item The assumption of a Gaussian line profile centered on \lya\ in contrast to much observational evidence supporting that the \lya\ line profile is modified (and normally redshifted) by the ISM \citep[e.g.][]{verhamme08, Gronke2017, Sobral_2018a}. In this way, too much flux is put at bluer frequencies than \lya , where the IGM is more efficient at absorbing photons (see Fig. \ref{fig:transmission}).
    \item The assumption that the \llya\ emitted by a galaxy is directly proportional to the SFR. In fact, in our previous work \citep{gurung18a} we found that the \lya\ RT in the ISM breaks this relation. Their assumption places LAEs in more massive DM halos, hosted in denser environments, thus with lower median transmission (see  Fig.\ref{fig:density_trans}).
\end{enumerate}

However, our model also has limitations. For example, we assume that every photon that interacts with the IGM is lost. In this way, we underestimate $f_{\rm esc}^{\rm IGM}$ since we do not take into account photons that are scattered out of the LoS and thanks to consecutive scatters are  sent back in the LoS again. The contribution of these photons to the received \llya\ is predicted to be small \citep{zheng10}. 

We note that at $z=5.7$ the galaxy stellar mass distributions for our LAE samples truncates abruptly at $\rm M_{*}=10^{7}\Munits$. This is caused by the cut  in \galform\ at this stellar mass imposed to ensure a good resolution of galaxies. This suggests that a fraction of the LAE population at  $z=5.7$ would inhabit galaxies with $\rm M_{*}<10^{7}\Munits$. However, as our galaxy population lacks these low mass galaxies, other more massive galaxies are selected as LAEs. Computing the precise number of LAEs that should host galaxies with $\rm M_{*}<10^{7}\Munits$ is challenging. As a simple calculation, we rescale the stellar mass distribution at $z=3.0$  towards smaller $\rm M_{*}$ so that the peak of the distribution matches the one at $z=5.7$. Then, the fraction of galaxies below the resolution limit is $\sim 0.04$. This hints that only a small part of the LAE population would lie in galaxies not resolved in our model. However, this limitation of our model might cause an overestimation of the clustering bias at this epoch (see below).  

On one hand, \ZZ\ found a strong correlation between the IGM and the LAE population. Their model predicted a relatively small dependence on $\rm V_{z}$, $\rho$ and $\rm \partial_z \rho$ and a tight relation between $\rm \partial_z V_z$ and $f_{\rm esc}^{\rm IGM}$, causing differences in $f_{\rm esc}^{\rm IGM}$ greater than 1 order of magnitude  across the $\rm \partial_z V_z$ dynamical range. 

In their model, the strong coupling between LAEs and $\partial_z V_z$ has dramatic consequences in the clustering of LAEs at $z=5.7$. Their model predicts a scale independent enhancement in the clustering along the line of sight, which created an asymmetry between the clustering parallel and perpendicular to the LoS in real space. Moreover, in their model, the amplitude of this effect outpowers the Kaiser boost.  Hence, even in redshift space the clustering along the LoS is more powerful than in the perpendicular direction. 

On the other hand, \BC\ claimed to find only a marginal coupling between the LAE and the IGM large scale properties at $z=2.0$, 3.0, 4.0 and 5.85. Additionally, they studied the clustering of LAE samples and did not find any asymmetry between the directions parallel and perpendicular to the LoS clustering, nor any other strange clustering feature. Additionally, \BC\ claimed \ZZ\ results to be a consequence of poor spatial resolution. In particular, they recovered the strong selection on $\partial_z V_z$ after lowering their resolution to match \ZZ\ simulation. 

In this work we find a coupling between the \lya\ observability and the  large scale IGM properties. The amplitude of the coupling depends on redshift and on the outflow structure assumed in the ISM. Indeed, typical variations in $f_{\rm esc}^{\rm IGM}$ are $< 1\%$ , $\sim 2\%$ and $\sim 5\%$ at $z=2.2$, 3.0 and 5.7 respectively. In other words, we detect a LAE-IGM coupling but it is small, in particular at $z=2.2$ and 3.0 where it has negligible effect on the clustering.

We do not find the dramatic asymmetry of clustering parallel and perpendicular to the line of sight (see Fig. \ref{fig:RSD_real}) that \ZZ\ measured in their original work. Our models do not predict any other clustering modification on scales where \ZZ\ or \BC\ simulations allowed them to measure the clustering accurately (${\rm \sim 10 cMpc }h^{-1}$). Hence our clustering predictions are in agreement with \BC\ and differ from \ZZ .

Our model predicts a strong boost of the LAE clustering at scales ${\rm > 20 cMpc}h^{-1}$. This feature comes from the coupling with the IGM velocity field (see from eq.\ref{eq:over_density_laes} to eq.\ref{eq:monopole_power_gamma2}). We attribute the detection of this effect in our work and the non-detection in \BC\ to the difference in volume probed by the different simulations. Meanwhile, our simulation resolves the velocity large scale structure of the Universe, \ZZ\ and \BC\ simulations are small enough to be enclosed in one of the huge regions of coherent motion (see Fig. \ref{fig:v_field}). 

\subsection{Comparison with an analytical model}

In \cite{Wyithe_2011} (from now on WD11), authors studied the impact of the LAE-IGM coupling in the accuracy of LAE cosmological surveys. Here we make a brief comparison between this work and WD11 and we encourage the readers to visit their original article for full details. \\

The approach taken by WD11 is very different to the one developed through this work. On one side, we have created an LAE model over a cosmological dark matter simulation (\pmill) using a model of galaxy formation and evolution (\galform) and then we have implemented the RT in the ISM (through \flareon) and in the IGM by computing the IGM transmission a long the line of sight for each galaxy. On the other side, WD11 designed an analytical physically motivated model of the LAE power spectrum that took into account the coupling between the observability of \lya\ and the IGM conditions (similar to the one presented in  \ZZ). In particular, their model included the dependence with the density of the IGM, its velocity gradient along the line of sight and its ionization rate. In this work we have found that the coupling between the \lya\ observability and the IGM velocity along the line of sight  is the most relevant on large scales (see \S \ref{ssec:largE_scales}). Meanwhile we report that our model does not present a significant coupling with the ionization state of the IGM in the redshift range studied here.  

There are two main conclusions on WD11 regarding the accuracy of cosmological constrains using LAEs :

\begin{itemize}
    \item{ On one hand, the BAO wiggles are not heavily affected by the coupling of IGM and \lya\ observability. In fact, they reported that precision on cosmology given by BAO studies using LAEs should be similar to the one achieved using non-IGM-coupled galaxy populations.}
    \item{On the other hand, they argue that in such galaxy surveis most of the constrain power in the cosmology comes from the full shape of the power spectrum \citep[e.g.][]{Shoji_2009} . As a consequence, they found that the constrains in cosmology given by an LAE surveys  are weaker than in a normal galaxy survey with the same volume and number density. Moreover, the impact on the accuracy would depend on the level of coupling with the IGM and it could become highly significant if the IGM coupling was strong.}
\end{itemize}

Overall, we find a good agreement with the finding of WD11. First, we find that the BAO peak is shifted only $\sim$2\AA{} (depending on the outflow geometry) from its original position at redshift 5.7, while smaller displacement are found at lower redshifts. This would indicate that the BAO analysis would be affected, but not heavily by the LAE-IGM coupling predicted by our model. Additionally, our model includes different models for the escape of \lya\ photons from galaxies, which also might increase the uncertainty in BAO studies. 

Second, in contrast to our work, their model do not include the coupling with the velocity field, which, in our case is the one injecting more power to clustering of LAEs. Therefore, the impact of the LAE-IGM coupling in cosmological contains could be stronger in our models, since we find that our clustering is much more disrupted than the one explored in their work. However, this analysis needs a high level of detail and we will leave it to a future study. 

\begin{figure*} 
\includegraphics[width=6.9in]{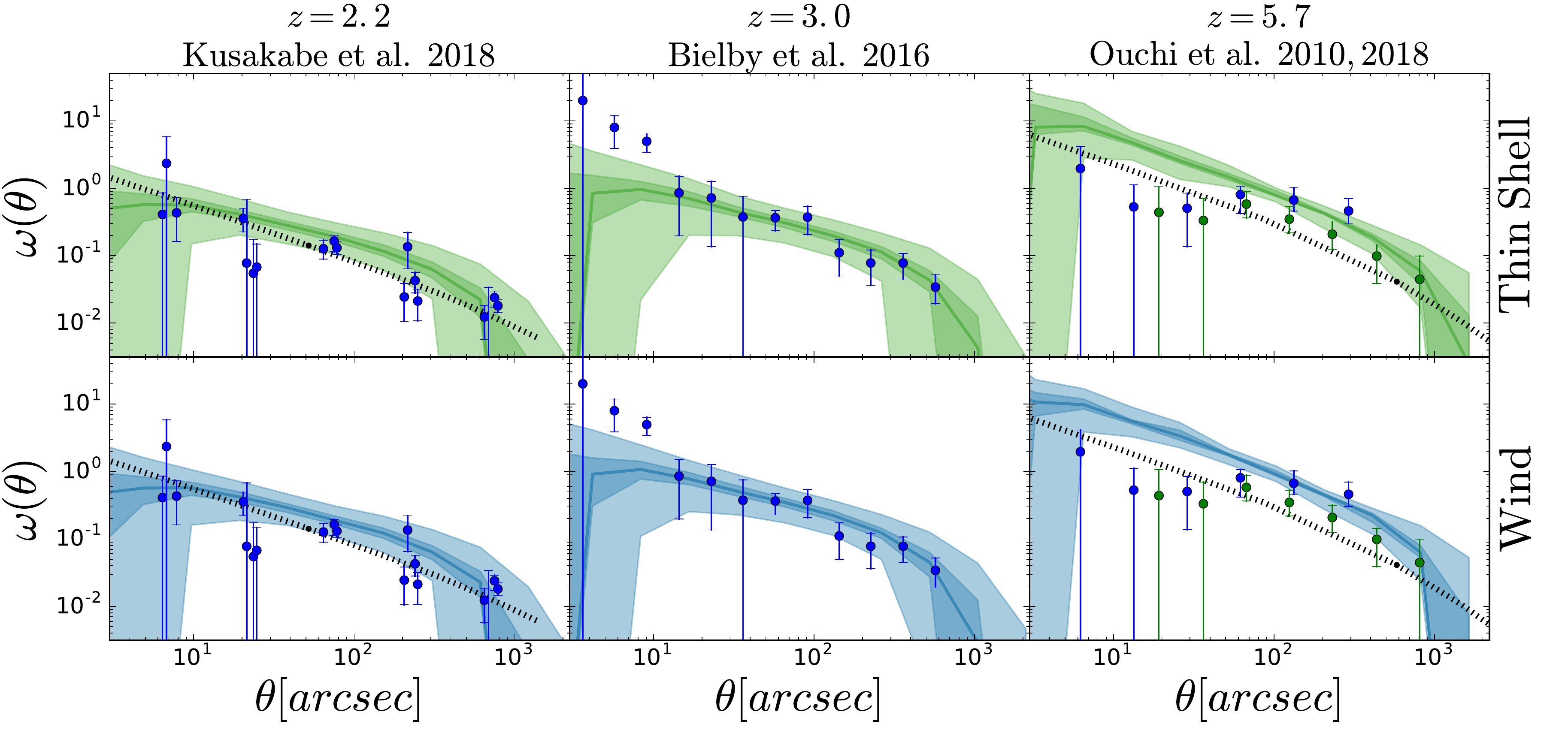} 
\caption{ Comparison of the clustering of our models (full RT Thin Shell, full RT Galactic Wind from top to bottom) and different observational data set at redshift 2.2, 3.0 and 5.7 from left to right. At redshift 5.7, SILVERRUSH \citep{Ouchi2018a} is shown in blue dots while its predecessor  \citep{ouchi10} is plotted in green dots. The solid colored line indicated the median angular 2PCF of our mocks while the $1\sigma$ and $2 \sigma$ are shown as shade regions. The different observations are shown as dots. Finally, the dashed black line indicate illustrate the best fitting clustering model presented within the original works. }\label{fig:mocks}
\end{figure*}


\subsection{ Comparison with current observations. }
    
In this section we compare our model's clustering with current LAE clustering observations across cosmic time. In order to make a fair comparison we mimic the observations performed by \cite{ouchi10,Ouchi2018a}, \cite{Bielby2016} and \cite{Kusakabe2018}. These surveys select LAE candidates by combining narrow and broad band photometric filters. The main differences between the different surveys are the flux depth, the sky coverage and the redshift of LAE candidates. The different properties of the observations used in this work are listed in Table \ref{tab:mocks_table} and Table \ref{tab:survey_prop}.

Here we summarize the mock construction and refer the readers to \cite{gurung18a} for a deeper explanation. In short:

\begin{enumerate}
    \item We convert the galaxy coordinates to redshift space along the same line of sight (LoS) used to compute the IGM absorption (Eq.\ref{eq:redshift_space}). 
    \item We divide our box into rectangular subvolumes. The faces perpendicular to the LoS  are squares with the same area as the survey sky coverage. The depth along the LoS is determined by the narrow band full width half maximum of each survey. 
    \item We make a rest frame equivalent width ($\rm EW_{0,cut}$) and luminosity cut ($\rm L_{Ly\alpha , cut}$). We choose the closest $\rm EW_{0,cut}$ and $\rm L_{Ly\alpha , cut}$ to the survey cuts that best match the observed LAE number density. These cuts are listed in Tab. \ref{tab:mocks_table}.
\end{enumerate}

The mock properties are listed in  Table \ref{tab:survey_prop}. At $z=2.2$ and $3.0$ the survey volumes are small in comparison with our simulation and there are about 450 mocks in each redshift bin. Hence, we get a good estimate of the cosmic variance of these observations. However, at $z=5.7$ the number of mock catalogs decreases to 18 and the constrain on cosmic variance is weaker. Additionally, the observed number density is well reproduced in our mocks. In particular, the median number of LAEs in each mock is within 1$\sigma$ of the observed value.

In Fig. \ref{fig:mocks} we compare our full RT model  mocks clustering with measurements at several redshifts. Overall, we find a good agreement with observations. In particular, at redshift $2.2$, our RT models match very well observation at all scales. Meanwhile, at $z=3.0$ the clustering is perfectly reproduced within the cosmic variance at angular separations higher than $\theta \gtrsim 10 \; arcsec $, while at smaller scales is underpredicted. Finally, at $z=5.7$ we find that the full RT LAE samples exhibit a higher bias at all scales. This small disagreement is not caused by LAE-IGM coupling predicted by our model. If this was the case, only the large scales would disagree. In fact, the clustering excess here is partially caused by the overpredicted stellar masses at $z=5.7$ in our \flae\ samples, as described above (see Fig.\ref{fig:galaxy_prop}).


The characteristic excess of power in the clustering of \flae s populations has been tentatively detected in \cite{Ouchi2018a}. For example, in Fig. \ref{fig:mocks} we show the LAE clustering measurements of SILVERRUSH \citep{Ouchi2018a} (in blue dots). The clustering at angular separations $>10^2 arcsec$ is boosted. However, this trend is not found in \cite{ouchi10} (green dots). We attribute this to the different volume traced by surveys. On the one hand, \cite{ouchi10} only covered $\rm \sim 1 \; deg^2$, an area  small enough to fall inside one region of coherent motion (see Fig.\ref{fig:v_field}). On the other hand \citep{Ouchi2018a} explored $\rm \sim 13 \; deg^2$ with contiguous patches of up to $\rm \sim 5 \; deg^2$, big enough to begin to resolve these regions.

\section{ Conclusions  }\label{sec:conclusion}
    
We have created a cosmological model of Lyman-$\alpha$ emitter galaxies that includes Lyman $\alpha$ radiative transfer physics in both the interstellar and intergalactic medium. The cosmological background is imprinted by the \pmill\ N-body simulation. Meanwhile, the semi-analytic model of galaxy of formation and evolution \galform\ populates the \pmill\ DM halos with galaxies. For the ISM transmission we used \flareon\ \citep{Gurung_2018b}, an open \texttt{Python} package based on a Monte Carlo RT code \citep{orsi12} that predicts the \lya\ line profiles and escape fractions of photons in outflows of different characteristics. Meanwhile, the RT in the IGM is implemented by computing the \lya\ transmission at each position of our simulation. Our main conclusions are:

\begin{enumerate}
    \item{ The RT in the ISM produces a strong selection effect over galaxy properties such as metallicity or SFR. Meanwhile, the RT in the IGM leave the galaxy property distributions nearly unchanged. In fact, LAEs tend to have a low-intermediate metallicity, moderate SFR and intermediate stellar mass. For further analysis on the galaxy properties of LAEs we refer the reader to \cite{gurung18a}. }
    
    \item{ Our models predict that the \lya\ IGM escape fraction depends on the large scale properties of the IGM such as the IGM density, motion, the density and velocity gradient along the line of sight, as first studied by \cite{zheng11}. While at low redshift ($z=2.2$ and 3.0) the correlations are weak and lead to variations in $f_{\rm esc}^{\rm IGM}$ of  $\sim 1\%$, it intensifies at higher redshifts ($z=5.7$), reaching variation on $f_{\rm esc}^{\rm IGM}$ around $\sim 5\%$.   }
    
    \item{The level of coupling between the LAE distribution and the  large scale IGM properties depends on the RT inside the ISM. Our model predicts that if the outflows driving the \lya\ photons escape from galaxies have a Thin Shell geometry the coupling is greater than if it is  driven by a Galactic Wind at low redshift. However, the opposite is found at $z=5.7$.}
    
    \item{The IGM-LAE coupling can have an impact on their  clustering. At redshift 5.7 the shape of the 2-point correlation function is modified, introducing extra power at scales larger than $\sim {\rm 20 cMpc}h^{-1}$ in a scale-dependent fashion. Meanwhile, at $z=2.2$ and z=3.0 we do not find any modification in the 2PCF, as the IGM-LAE coupling is too weak.}
    
    
    \item{ In order to study the LAE clustering at large scales we fit our model 2-point correlation function (2PCF) to an analytic clustering model including the IGM-LAE coupling (Eq.\ref{eq:monopole_power_zz}) introduced by \cite{zheng11}. The IGM-LAE coupling disrupts the LAE clustering and modify the shape of the 2PCF at the baryon acustic oscilation (BAO) peak scales, which becomes broader and modify the position of the maximum by $\sim 1 {\rm cMpc}h^{-1}$.   }
    
    \item{We have made a comparison between current LAE clustering measurements and our LAE models. Overall we find good agreement between our model and observations, even at $z=5.7$. This suggests that a greater sky coverage is necessary to detect the clustering excess presented in this work. However, hints of a scale dependent bias at $z=5.7$ can be found in the literature \citep{Ouchi2018a}.  }
    
    In future works we plan to implement the model presented in this work to a larger simulation to determine the IGM-LAE coupling impact on the clustering at low redshifts. Additionally, we will also implement the physics of reionization to understand the clustering of LAEs during this interesting epoch.
    
\end{enumerate}

\section*{Acknowledgements}

The authors acknowledge the useful discussions with Zheng Zheng, Mark Dijkstra and Anne Verhamme in addition to the whole CEFCA team. The authors also acknowledge the support of the Spanish Ministerio de Economia y Competividad project No. AYA2015-66211-C2-P-2. We acknowledge also STFC Consolidated Grants ST/L00075X/1 and ST/P000451/1 at Durham University. This work used the DiRAC@Durham facility managed by the Institute for
   Computational Cosmology on behalf of the STFC DiRAC HPC Facility
   (www.dirac.ac.uk). The equipment was funded by BEIS capital funding
   via STFC capital grants ST/P002293/1, ST/R002371/1 and ST/S002502/1,
   Durham University and STFC operations grant ST/R000832/1. DiRAC is
   part of the National e-Infrastructure.




\bibliographystyle{mnras}
\bibliography{ref} 

\begin{thebibliography}{}
\makeatletter
\relax
\def\mn@urlcharsother{\let\do\@makeother \do\$\do\&\do\#\do\^\do\_\do\%\do\~}
\def\mn@doi{\begingroup\mn@urlcharsother \@ifnextchar [ {\mn@doi@}
  {\mn@doi@[]}}
\def\mn@doi@[#1]#2{\def\@tempa{#1}\ifx\@tempa\@empty \href
  {http://dx.doi.org/#2} {doi:#2}\else \href {http://dx.doi.org/#2} {#1}\fi
  \endgroup}
\def\mn@eprint#1#2{\mn@eprint@#1:#2::\@nil}
\def\mn@eprint@arXiv#1{\href {http://arxiv.org/abs/#1} {{\tt arXiv:#1}}}
\def\mn@eprint@dblp#1{\href {http://dblp.uni-trier.de/rec/bibtex/#1.xml}
  {dblp:#1}}
\def\mn@eprint@#1:#2:#3:#4\@nil{\def\@tempa {#1}\def\@tempb {#2}\def\@tempc
  {#3}\ifx \@tempc \@empty \let \@tempc \@tempb \let \@tempb \@tempa \fi \ifx
  \@tempb \@empty \def\@tempb {arXiv}\fi \@ifundefined
  {mn@eprint@\@tempb}{\@tempb:\@tempc}{\expandafter \expandafter \csname
  mn@eprint@\@tempb\endcsname \expandafter{\@tempc}}}

\bibitem[\protect\citeauthoryear{{Ahn}}{{Ahn}}{2003}]{ahn03}
{Ahn} S.,  2003, Journal of Korean Astronomical Society, \href
  {http://adsabs.harvard.edu/abs/2003JKAS...36..145A} {36, 145}

\bibitem[\protect\citeauthoryear{{Baugh} et~al.,}{{Baugh}
  et~al.}{2019}]{Baugh_2019}
{Baugh} C.~M.,  et~al., 2019, \mn@doi [\mnras] {10.1093/mnras/sty3427}, \href
  {http://adsabs.harvard.edu/abs/2018MNRAS.tmp.3253B} {483, 4922}

\bibitem[\protect\citeauthoryear{{Behrens}, {Byrohl}, {Saito}  \&
  {Niemeyer}}{{Behrens} et~al.}{2017}]{Behrens2017}
{Behrens} C.,  {Byrohl} C.,  {Saito} S.,   {Niemeyer} J.~C.,  2017, preprint,
  \href {http://adsabs.harvard.edu/abs/2017arXiv171006171B} {} (\mn@eprint
  {arXiv} {1710.06171})

\bibitem[\protect\citeauthoryear{{Benitez} et~al.,}{{Benitez}
  et~al.}{2014}]{J-PAS}
{Benitez} N.,  et~al., 2014, preprint, \href
  {http://adsabs.harvard.edu/abs/2014arXiv1403.5237B} {} (\mn@eprint {arXiv}
  {1403.5237})

\bibitem[\protect\citeauthoryear{{Bielby} et~al.,}{{Bielby}
  et~al.}{2016}]{Bielby2016}
{Bielby} R.~M.,  et~al., 2016, \mn@doi [\mnras] {10.1093/mnras/stv2914}, \href
  {http://adsabs.harvard.edu/abs/2016MNRAS.456.4061B} {456, 4061}

\bibitem[\protect\citeauthoryear{{Caruana} et~al.,}{{Caruana}
  et~al.}{2018}]{Caruana_2018}
{Caruana} J.,  et~al., 2018, \mn@doi [\mnras] {10.1093/mnras/stx2307}, \href
  {http://adsabs.harvard.edu/abs/2018MNRAS.473...30C} {473, 30}

\bibitem[\protect\citeauthoryear{{Cassata} et~al.,}{{Cassata}
  et~al.}{2011}]{Cassata_2011}
{Cassata} P.,  et~al., 2011, \mn@doi [\aap] {10.1051/0004-6361/201014410},
  \href {http://adsabs.harvard.edu/abs/2011A%26A...525A.143C} {525, A143}

\bibitem[\protect\citeauthoryear{{Cazzoli}, {Arribas}, {Maiolino}  \&
  {Colina}}{{Cazzoli} et~al.}{2016}]{Cazzoli_2016}
{Cazzoli} S.,  {Arribas} S.,  {Maiolino} R.,   {Colina} L.,  2016, \mn@doi
  [\aap] {10.1051/0004-6361/201526788}, \href
  {http://adsabs.harvard.edu/abs/2016A%26A...590A.125C} {590, A125}

\bibitem[\protect\citeauthoryear{{Chaves-Montero}, {Angulo}  \&
  {Hern{\'a}ndez-Monteagudo}}{{Chaves-Montero}
  et~al.}{2018}]{chavez_montero_2018}
{Chaves-Montero} J.,  {Angulo} R.~E.,   {Hern{\'a}ndez-Monteagudo} C.,  2018,
  \mn@doi [\mnras] {10.1093/mnras/sty924}, \href
  {http://adsabs.harvard.edu/abs/2018MNRAS.477.3892C} {477, 3892}

\bibitem[\protect\citeauthoryear{{Cole}, {Lacey}, {Baugh}  \& {Frenk}}{{Cole}
  et~al.}{2000}]{cole00}
{Cole} S.,  {Lacey} C.~G.,  {Baugh} C.~M.,   {Frenk} C.~S.,  2000, \mn@doi
  [\mnras] {10.1046/j.1365-8711.2000.03879.x}, \href
  {http://adsabs.harvard.edu/abs/2000MNRAS.319..168C} {319, 168}

\bibitem[\protect\citeauthoryear{{Contreras}, {Zehavi}, {Padilla}, {Baugh},
  {Jim{\'e}nez}  \& {Lacerna}}{{Contreras} et~al.}{2019}]{Contreras_2019}
{Contreras} S.,  {Zehavi} I.,  {Padilla} N.,  {Baugh} C.~M.,  {Jim{\'e}nez} E.,
    {Lacerna} I.,  2019, \mn@doi [\mnras] {10.1093/mnras/stz018}, \href
  {http://adsabs.harvard.edu/abs/2019MNRAS.484.1133C} {484, 1133}

\bibitem[\protect\citeauthoryear{{Cooray} et~al.,}{{Cooray}
  et~al.}{2016}]{Cooray_2016}
{Cooray} A.,  et~al., 2016, arXiv e-prints, \href
  {http://adsabs.harvard.edu/abs/2016arXiv160205178C} {}

\bibitem[\protect\citeauthoryear{{Dijkstra} \& {Loeb}}{{Dijkstra} \&
  {Loeb}}{2009}]{dijkstra09}
{Dijkstra} M.,  {Loeb} A.,  2009, \mn@doi [\mnras]
  {10.1111/j.1365-2966.2009.15533.x}, \href
  {http://adsabs.harvard.edu/abs/2009MNRAS.400.1109D} {400, 1109}

\bibitem[\protect\citeauthoryear{{Dijkstra}, {Lidz}  \& {Wyithe}}{{Dijkstra}
  et~al.}{2007}]{dijkstra07}
{Dijkstra} M.,  {Lidz} A.,   {Wyithe} J.~S.~B.,  2007, \mn@doi [\mnras]
  {10.1111/j.1365-2966.2007.11666.x}, \href
  {http://adsabs.harvard.edu/abs/2007MNRAS.377.1175D} {377, 1175}

\bibitem[\protect\citeauthoryear{{Foreman-Mackey}, {Hogg}, {Lang}  \&
  {Goodman}}{{Foreman-Mackey} et~al.}{2013}]{emcee}
{Foreman-Mackey} D.,  {Hogg} D.~W.,  {Lang} D.,   {Goodman} J.,  2013, \mn@doi
  [\pasp] {10.1086/670067}, \href
  {http://adsabs.harvard.edu/abs/2013PASP..125..306F} {125, 306}

\bibitem[\protect\citeauthoryear{{Garel}, {Blaizot}, {Guiderdoni}, {Schaerer},
  {Verhamme}  \& {Hayes}}{{Garel} et~al.}{2012}]{garel12}
{Garel} T.,  {Blaizot} J.,  {Guiderdoni} B.,  {Schaerer} D.,  {Verhamme} A.,
  {Hayes} M.,  2012, \mn@doi [\mnras] {10.1111/j.1365-2966.2012.20607.x}, \href
  {http://adsabs.harvard.edu/abs/2012MNRAS.422..310G} {422, 310}

\bibitem[\protect\citeauthoryear{{Giallongo} et~al.,}{{Giallongo}
  et~al.}{2019}]{Giallongo_2019}
{Giallongo} E.,  et~al., 2019, arXiv e-prints, \href
  {https://ui.adsabs.harvard.edu/abs/2019arXiv190900702G} {p. arXiv:1909.00702}

\bibitem[\protect\citeauthoryear{{Granato}, {Lacey}, {Silva}, {Bressan},
  {Baugh}, {Cole}  \& {Frenk}}{{Granato} et~al.}{2000}]{granato00}
{Granato} G.~L.,  {Lacey} C.~G.,  {Silva} L.,  {Bressan} A.,  {Baugh} C.~M.,
  {Cole} S.,   {Frenk} C.~S.,  2000, \mn@doi [\apj] {10.1086/317032}, \href
  {http://adsabs.harvard.edu/abs/2000ApJ...542..710G} {542, 710}

\bibitem[\protect\citeauthoryear{{Gronke}}{{Gronke}}{2017}]{Gronke2017}
{Gronke} M.,  2017, \mn@doi [\aap] {10.1051/0004-6361/201731791}, \href
  {http://adsabs.harvard.edu/abs/2017A%26A...608A.139G} {608, A139}

\bibitem[\protect\citeauthoryear{{Gronke}, {Dijkstra}, {McCourt}  \&
  {Oh}}{{Gronke} et~al.}{2016}]{Gronke_2016}
{Gronke} M.,  {Dijkstra} M.,  {McCourt} M.,   {Oh} S.~P.,  2016, \mn@doi
  [\apjl] {10.3847/2041-8213/833/2/L26}, \href
  {http://adsabs.harvard.edu/abs/2016ApJ...833L..26G} {833, L26}

\bibitem[\protect\citeauthoryear{{Gurung-Lopez}, {Orsi}  \&
  {Bonoli}}{{Gurung-Lopez} et~al.}{2018a}]{Gurung_2018b}
{Gurung-Lopez} S.,  {Orsi} A.~A.,   {Bonoli} S.,  2018a, preprint, \href
  {http://adsabs.harvard.edu/abs/2018arXiv181109630G} {} (\mn@eprint {arXiv}
  {1811.09630})

\bibitem[\protect\citeauthoryear{{Gurung L{\'o}pez}, {Orsi}, {Bonoli}, {Baugh}
  \& {Lacey}}{{Gurung L{\'o}pez} et~al.}{2018b}]{gurung18a}
{Gurung L{\'o}pez} S.,  {Orsi} {\'A}.~A.,  {Bonoli} S.,  {Baugh} C.~M.,
  {Lacey} C.~G.,  2018b, preprint, \href
  {http://adsabs.harvard.edu/abs/2018arXiv180700006G} {} (\mn@eprint {arXiv}
  {1807.00006})

\bibitem[\protect\citeauthoryear{{Haardt} \& {Madau}}{{Haardt} \&
  {Madau}}{2012}]{Haardt_2012}
{Haardt} F.,  {Madau} P.,  2012, \mn@doi [\apj] {10.1088/0004-637X/746/2/125},
  \href {http://adsabs.harvard.edu/abs/2012ApJ...746..125H} {746, 125}

\bibitem[\protect\citeauthoryear{{Harrington}}{{Harrington}}{1973}]{harrington73}
{Harrington} J.~P.,  1973, \mnras, \href
  {http://adsabs.harvard.edu/abs/1973MNRAS.162...43H} {162, 43}

\bibitem[\protect\citeauthoryear{{Henry}, {Berg}, {Scarlata}, {Verhamme}  \&
  {Erb}}{{Henry} et~al.}{2018}]{Henry_2018}
{Henry} A.,  {Berg} D.~A.,  {Scarlata} C.,  {Verhamme} A.,   {Erb} D.,  2018,
  \mn@doi [\apj] {10.3847/1538-4357/aab099}, \href
  {http://adsabs.harvard.edu/abs/2018ApJ...855...96H} {855, 96}

\bibitem[\protect\citeauthoryear{{Hill} et~al.,}{{Hill}
  et~al.}{2008a}]{Hill2008}
{Hill} G.~J.,  et~al., 2008a, in {Kodama} T.,  {Yamada} T.,   {Aoki} K.,  eds,
  Astronomical Society of the Pacific Conference Series Vol. 399, Panoramic
  Views of Galaxy Formation and Evolution. p.~115 (\mn@eprint {arXiv}
  {0806.0183})

\bibitem[\protect\citeauthoryear{{Hill} et~al.,}{{Hill} et~al.}{2008b}]{hill08}
{Hill} G.~J.,  et~al., 2008b, in {T.~Kodama, T.~Yamada, \& K.~Aoki} ed.,
  Astronomical Society of the Pacific Conference Series Vol. 399, Astronomical
  Society of the Pacific Conference Series. pp 115--+ (\mn@eprint {arXiv}
  {0806.0183})

\bibitem[\protect\citeauthoryear{{Hu}, {Cowie}  \& {McMahon}}{{Hu}
  et~al.}{1998}]{hu98}
{Hu} E.~M.,  {Cowie} L.~L.,   {McMahon} R.~G.,  1998, \mn@doi [\apjl]
  {10.1086/311506}, \href {http://adsabs.harvard.edu/abs/1998ApJ...502L..99H}
  {502, L99+}

\bibitem[\protect\citeauthoryear{{Kaiser}}{{Kaiser}}{1987}]{kaiser87}
{Kaiser} N.,  1987, \mnras, \href
  {http://adsabs.harvard.edu/abs/1987MNRAS.227....1K} {227, 1}

\bibitem[\protect\citeauthoryear{{Kimm} \& {Cen}}{{Kimm} \&
  {Cen}}{2014}]{kimm_2014}
{Kimm} T.,  {Cen} R.,  2014, \mn@doi [\apj] {10.1088/0004-637X/788/2/121},
  \href {http://adsabs.harvard.edu/abs/2014ApJ...788..121K} {788, 121}

\bibitem[\protect\citeauthoryear{{Konno}, {Ouchi}, {Nakajima}, {Duval},
  {Kusakabe}, {Ono}  \& {Shimasaku}}{{Konno} et~al.}{2016}]{Konno2016}
{Konno} A.,  {Ouchi} M.,  {Nakajima} K.,  {Duval} F.,  {Kusakabe} H.,  {Ono}
  Y.,   {Shimasaku} K.,  2016, \mn@doi [\apj] {10.3847/0004-637X/823/1/20},
  \href {http://adsabs.harvard.edu/abs/2016ApJ...823...20K} {823, 20}

\bibitem[\protect\citeauthoryear{{Konno} et~al.,}{{Konno}
  et~al.}{2018}]{Konno_2018}
{Konno} A.,  et~al., 2018, \mn@doi [\pasj] {10.1093/pasj/psx131}, \href
  {http://adsabs.harvard.edu/abs/2018PASJ...70S..16K} {70, S16}

\bibitem[\protect\citeauthoryear{{Kormendy}}{{Kormendy}}{2013}]{Kormendy_2013}
{Kormendy} J.,  2013, {Secular Evolution in Disk Galaxies}.
p.~1

\bibitem[\protect\citeauthoryear{{Kusakabe} et~al.,}{{Kusakabe}
  et~al.}{2018}]{Kusakabe2018}
{Kusakabe} H.,  et~al., 2018, \mn@doi [\pasj] {10.1093/pasj/psx148}, \href
  {http://adsabs.harvard.edu/abs/2018PASJ..tmp...11K} {}

\bibitem[\protect\citeauthoryear{{Lacey} et~al.,}{{Lacey}
  et~al.}{2016}]{lacey16}
{Lacey} C.~G.,  et~al., 2016, \mn@doi [\mnras] {10.1093/mnras/stw1888}, \href
  {http://adsabs.harvard.edu/abs/2016MNRAS.462.3854L} {462, 3854}

\bibitem[\protect\citeauthoryear{{Laursen}, {Sommer-Larsen}  \&
  {Razoumov}}{{Laursen} et~al.}{2011}]{laursen11}
{Laursen} P.,  {Sommer-Larsen} J.,   {Razoumov} A.~O.,  2011, \mn@doi [\apj]
  {10.1088/0004-637X/728/1/52}, \href
  {http://adsabs.harvard.edu/abs/2011ApJ...728...52L} {728, 52}

\bibitem[\protect\citeauthoryear{{Leclercq} et~al.,}{{Leclercq}
  et~al.}{2017}]{Leclercq_2017}
{Leclercq} F.,  et~al., 2017, \mn@doi [\aap] {10.1051/0004-6361/201731480},
  \href {http://adsabs.harvard.edu/abs/2017A%26A...608A...8L} {608, A8}

\bibitem[\protect\citeauthoryear{{Malhotra} \& {Rhoads}}{{Malhotra} \&
  {Rhoads}}{2002}]{malhotra02}
{Malhotra} S.,  {Rhoads} J.~E.,  2002, \mn@doi [\apjl] {10.1086/338980}, \href
  {http://adsabs.harvard.edu/abs/2002ApJ...565L..71M} {565, L71}

\bibitem[\protect\citeauthoryear{{Matthee}, {Sobral}, {Best}, {Smail}, {Bian},
  {Darvish}, {R{\"o}ttgering}  \& {Fan}}{{Matthee}
  et~al.}{2017}]{Matthee_2017_boot}
{Matthee} J.,  {Sobral} D.,  {Best} P.,  {Smail} I.,  {Bian} F.,  {Darvish} B.,
   {R{\"o}ttgering} H.,   {Fan} X.,  2017, \mn@doi [\mnras]
  {10.1093/mnras/stx1569}, \href
  {http://adsabs.harvard.edu/abs/2017MNRAS.471..629M} {471, 629}

\bibitem[\protect\citeauthoryear{{Nelson} et~al.,}{{Nelson}
  et~al.}{2015}]{Nelson2015}
{Nelson} D.,  et~al., 2015, \mn@doi [Astronomy and Computing]
  {10.1016/j.ascom.2015.09.003}, \href
  {http://adsabs.harvard.edu/abs/2015A%26C....13...12N} {13, 12}

\bibitem[\protect\citeauthoryear{{Neufeld}}{{Neufeld}}{1991}]{neufeld91}
{Neufeld} D.~A.,  1991, \mn@doi [\apjl] {10.1086/185983}, \href
  {http://adsabs.harvard.edu/abs/1991ApJ...370L..85N} {370, L85}

\bibitem[\protect\citeauthoryear{{Orlitov{\'a}}, {Verhamme}, {Henry},
  {Scarlata}, {Jaskot}, {Oey}  \& {Schaerer}}{{Orlitov{\'a}}
  et~al.}{2018}]{Orlitova_2018}
{Orlitov{\'a}} I.,  {Verhamme} A.,  {Henry} A.,  {Scarlata} C.,  {Jaskot} A.,
  {Oey} M.~S.,   {Schaerer} D.,  2018, \mn@doi [\aap]
  {10.1051/0004-6361/201732478}, \href
  {http://adsabs.harvard.edu/abs/2018A%26A...616A..60O} {616, A60}

\bibitem[\protect\citeauthoryear{{Orsi}, {Lacey}  \& {Baugh}}{{Orsi}
  et~al.}{2012}]{orsi12}
{Orsi} A.,  {Lacey} C.~G.,   {Baugh} C.~M.,  2012, \mn@doi [\mnras]
  {10.1111/j.1365-2966.2012.21396.x}, \href
  {http://adsabs.harvard.edu/abs/2012MNRAS.425...87O} {425, 87}

\bibitem[\protect\citeauthoryear{{Ouchi} et~al.,}{{Ouchi}
  et~al.}{2008}]{ouchi08}
{Ouchi} M.,  et~al., 2008, \mn@doi [\apjs] {10.1086/527673}, \href
  {http://adsabs.harvard.edu/abs/2008ApJS..176..301O} {176, 301}

\bibitem[\protect\citeauthoryear{{Ouchi} et~al.,}{{Ouchi}
  et~al.}{2010}]{ouchi10}
{Ouchi} M.,  et~al., 2010, \mn@doi [\apj] {10.1088/0004-637X/723/1/869}, \href
  {http://adsabs.harvard.edu/abs/2010ApJ...723..869O} {723, 869}

\bibitem[\protect\citeauthoryear{{Ouchi} et~al.,}{{Ouchi}
  et~al.}{2018}]{Ouchi2018a}
{Ouchi} M.,  et~al., 2018, \mn@doi [\pasj] {10.1093/pasj/psx074}, \href
  {http://adsabs.harvard.edu/abs/2018PASJ...70S..13O} {70, S13}

\bibitem[\protect\citeauthoryear{{Oyarz{\'u}n}, {Blanc}, {Gonz{\'a}lez},
  {Mateo}  \& {Bailey}}{{Oyarz{\'u}n} et~al.}{2017}]{oyarzun17}
{Oyarz{\'u}n} G.~A.,  {Blanc} G.~A.,  {Gonz{\'a}lez} V.,  {Mateo} M.,
  {Bailey} III J.~I.,  2017, \mn@doi [\apj] {10.3847/1538-4357/aa7552}, \href
  {http://adsabs.harvard.edu/abs/2017ApJ...843..133O} {843, 133}

\bibitem[\protect\citeauthoryear{{Parsa}, {Dunlop}  \& {McLure}}{{Parsa}
  et~al.}{2018}]{Parsa_2018}
{Parsa} S.,  {Dunlop} J.~S.,   {McLure} R.~J.,  2018, \mn@doi [\mnras]
  {10.1093/mnras/stx2887}, \href
  {http://adsabs.harvard.edu/abs/2018MNRAS.474.2904P} {474, 2904}

\bibitem[\protect\citeauthoryear{{Planck Collaboration} et~al.,}{{Planck
  Collaboration} et~al.}{2016}]{Planck_2016}
{Planck Collaboration} et~al., 2016, \mn@doi [\aap]
  {10.1051/0004-6361/201525830}, \href
  {http://adsabs.harvard.edu/abs/2016A%26A...594A..13P} {594, A13}

\bibitem[\protect\citeauthoryear{{Rhoads}, {Malhotra}, {Dey}, {Stern},
  {Spinrad}  \& {Jannuzi}}{{Rhoads} et~al.}{2000}]{rhoads00}
{Rhoads} J.~E.,  {Malhotra} S.,  {Dey} A.,  {Stern} D.,  {Spinrad} H.,
  {Jannuzi} B.~T.,  2000, \mn@doi [\apjl] {10.1086/317874}, \href
  {http://adsabs.harvard.edu/abs/2000ApJ...545L..85R} {545, L85}

\bibitem[\protect\citeauthoryear{{Rudie}, {Steidel}  \& {Pettini}}{{Rudie}
  et~al.}{2012}]{Rudie_2012}
{Rudie} G.~C.,  {Steidel} C.~C.,   {Pettini} M.,  2012, \mn@doi [\apj]
  {10.1088/2041-8205/757/2/L30}, \href
  {https://ui.adsabs.harvard.edu/abs/2012ApJ...757L..30R} {757, L30}

\bibitem[\protect\citeauthoryear{{Shibuya} et~al.,}{{Shibuya}
  et~al.}{2018}]{Shibuya_2018}
{Shibuya} T.,  et~al., 2018, \mn@doi [\pasj] {10.1093/pasj/psx107}, \href
  {http://adsabs.harvard.edu/abs/2018PASJ...70S..15S} {70, S15}

\bibitem[\protect\citeauthoryear{{Shimakawa} et~al.,}{{Shimakawa}
  et~al.}{2017}]{Shimakawa_2017}
{Shimakawa} R.,  et~al., 2017, \mn@doi [\mnras] {10.1093/mnras/stx091}, \href
  {http://adsabs.harvard.edu/abs/2017MNRAS.468.1123S} {468, 1123}

\bibitem[\protect\citeauthoryear{{Shoji}, {Jeong}  \& {Komatsu}}{{Shoji}
  et~al.}{2009}]{Shoji_2009}
{Shoji} M.,  {Jeong} D.,   {Komatsu} E.,  2009, \mn@doi [\apj]
  {10.1088/0004-637X/693/2/1404}, \href
  {https://ui.adsabs.harvard.edu/abs/2009ApJ...693.1404S} {693, 1404}

\bibitem[\protect\citeauthoryear{{Sobral}, {Matthee}, {Darvish}, {Schaerer},
  {Mobasher}, {R{\"o}ttgering}, {Santos}  \& {Hemmati}}{{Sobral}
  et~al.}{2015}]{Sobral_2015_CR7}
{Sobral} D.,  {Matthee} J.,  {Darvish} B.,  {Schaerer} D.,  {Mobasher} B.,
  {R{\"o}ttgering} H.~J.~A.,  {Santos} S.,   {Hemmati} S.,  2015, \mn@doi
  [\apj] {10.1088/0004-637X/808/2/139}, \href
  {http://adsabs.harvard.edu/abs/2015ApJ...808..139S} {808, 139}

\bibitem[\protect\citeauthoryear{{Sobral} et~al.,}{{Sobral}
  et~al.}{2017}]{Sobral2017}
{Sobral} D.,  et~al., 2017, \mn@doi [\mnras] {10.1093/mnras/stw3090}, \href
  {http://adsabs.harvard.edu/abs/2017MNRAS.466.1242S} {466, 1242}

\bibitem[\protect\citeauthoryear{{Sobral} et~al.,}{{Sobral}
  et~al.}{2018a}]{Sobral_2018_dust}
{Sobral} D.,  et~al., 2018a, \mn@doi [\mnras] {10.1093/mnras/sty782}, \href
  {http://adsabs.harvard.edu/abs/2018MNRAS.477.2817S} {477, 2817}

\bibitem[\protect\citeauthoryear{{Sobral} et~al.,}{{Sobral}
  et~al.}{2018b}]{Sobral_2018a}
{Sobral} D.,  et~al., 2018b, \mn@doi [\mnras] {10.1093/mnras/sty782}, \href
  {http://adsabs.harvard.edu/abs/2018MNRAS.477.2817S} {477, 2817}

\bibitem[\protect\citeauthoryear{{Spinoso}, {Bonoli}, {Dotti}, {Mayer}, {Madau}
   \& {Bellovary}}{{Spinoso} et~al.}{2017}]{Spinoso_2017}
{Spinoso} D.,  {Bonoli} S.,  {Dotti} M.,  {Mayer} L.,  {Madau} P.,
  {Bellovary} J.,  2017, \mn@doi [\mnras] {10.1093/mnras/stw2934}, \href
  {http://adsabs.harvard.edu/abs/2017MNRAS.465.3729S} {465, 3729}

\bibitem[\protect\citeauthoryear{{Spitzer}}{{Spitzer}}{1978}]{Spitzer_1978}
{Spitzer} Jr. L.,  1978, \jrasc, \href
  {http://adsabs.harvard.edu/abs/1978JRASC..72..349S} {72, 349}

\bibitem[\protect\citeauthoryear{{Steidel}, {Giavalisco}, {Pettini},
  {Dickinson}  \& {Adelberger}}{{Steidel} et~al.}{1996}]{steidel96}
{Steidel} C.~C.,  {Giavalisco} M.,  {Pettini} M.,  {Dickinson} M.,
  {Adelberger} K.~L.,  1996, \mn@doi [\apjl] {10.1086/310029}, \href
  {http://adsabs.harvard.edu/abs/1996ApJ...462L..17S} {462, L17+}

\bibitem[\protect\citeauthoryear{{Verhamme}, {Schaerer}  \&
  {Maselli}}{{Verhamme} et~al.}{2006}]{verhamme06}
{Verhamme} A.,  {Schaerer} D.,   {Maselli} A.,  2006, \mn@doi [\aap]
  {10.1051/0004-6361:20065554}, \href
  {http://adsabs.harvard.edu/abs/2006A%26A...460..397V} {460, 397}

\bibitem[\protect\citeauthoryear{{Verhamme}, {Schaerer}, {Atek}  \&
  {Tapken}}{{Verhamme} et~al.}{2008}]{verhamme08}
{Verhamme} A.,  {Schaerer} D.,  {Atek} H.,   {Tapken} C.,  2008, \mn@doi [\aap]
  {10.1051/0004-6361:200809648}, \href
  {http://adsabs.harvard.edu/abs/2008A%26A...491...89V} {491, 89}

\bibitem[\protect\citeauthoryear{{Wang} et~al.,}{{Wang}
  et~al.}{2018}]{Atlas_2018}
{Wang} Y.,  et~al., 2018, arXiv e-prints, \href
  {http://adsabs.harvard.edu/abs/2018arXiv180201539W} {}

\bibitem[\protect\citeauthoryear{{Weinberger}, {Kulkarni}, {Haehnelt},
  {Choudhury}  \& {Puchwein}}{{Weinberger} et~al.}{2018}]{Weinberger_2018}
{Weinberger} L.~H.,  {Kulkarni} G.,  {Haehnelt} M.~G.,  {Choudhury} T.~R.,
  {Puchwein} E.,  2018, \mn@doi [\mnras] {10.1093/mnras/sty1563}, \href
  {http://adsabs.harvard.edu/abs/2018MNRAS.479.2564W} {479, 2564}

\bibitem[\protect\citeauthoryear{{Wyithe} \& {Dijkstra}}{{Wyithe} \&
  {Dijkstra}}{2011}]{Wyithe_2011}
{Wyithe} J. S.~B.,  {Dijkstra} M.,  2011, \mn@doi [\mnras]
  {10.1111/j.1365-2966.2011.19007.x}, \href
  {https://ui.adsabs.harvard.edu/abs/2011MNRAS.415.3929W} {415, 3929}

\bibitem[\protect\citeauthoryear{{Zheng} \& {Miralda-Escud{\'e}}}{{Zheng} \&
  {Miralda-Escud{\'e}}}{2002}]{zheng02}
{Zheng} Z.,  {Miralda-Escud{\'e}} J.,  2002, \mn@doi [\apj] {10.1086/342400},
  \href {http://adsabs.harvard.edu/abs/2002ApJ...578...33Z} {578, 33}

\bibitem[\protect\citeauthoryear{{Zheng}, {Cen}, {Trac}  \&
  {Miralda-Escud{\'e}}}{{Zheng} et~al.}{2010}]{zheng10}
{Zheng} Z.,  {Cen} R.,  {Trac} H.,   {Miralda-Escud{\'e}} J.,  2010, \mn@doi
  [\apj] {10.1088/0004-637X/716/1/574}, \href
  {http://adsabs.harvard.edu/abs/2010ApJ...716..574Z} {716, 574}

\bibitem[\protect\citeauthoryear{{Zheng}, {Cen}, {Trac}  \&
  {Miralda-Escud{\'e}}}{{Zheng} et~al.}{2011}]{zheng11}
{Zheng} Z.,  {Cen} R.,  {Trac} H.,   {Miralda-Escud{\'e}} J.,  2011, \mn@doi
  [\apj] {10.1088/0004-637X/726/1/38}, \href
  {http://adsabs.harvard.edu/abs/2011ApJ...726...38Z} {726, 38}

\makeatother
\end{thebibliography}




\appendix
\section{The $\rm N_{H}$, $\rm V_{exp}$ and $\rm \tau_a$ distributions}\label{Ap:A}

In our model, the \lya\ radiative transfer physics inside galaxies are integrated through \flareon. \flareon\ is open source code that predicts escape fraction and the emerging \lya\ line profile from different outflow configurations among several gas geometries. In particular,   \flareon\ is based on pre-computed \{$\rm N_{H}$, $\rm V_{exp}$, $\rm \tau_a$ \} grids of the full radiative transfer Monte Carlo code \texttt{LyaRT} \citep{orsi12}. Then, different algorithms such as multidimensional interpolation are used to obtain the line profile and \fesc . Hence,  the high performance of \flareon\ is limited to the space covered by the grid. In the following we study the fraction of LAEs in our model that fall within the \flareon\ range. 

\begin{figure*}
    \includegraphics[width=7.0in]{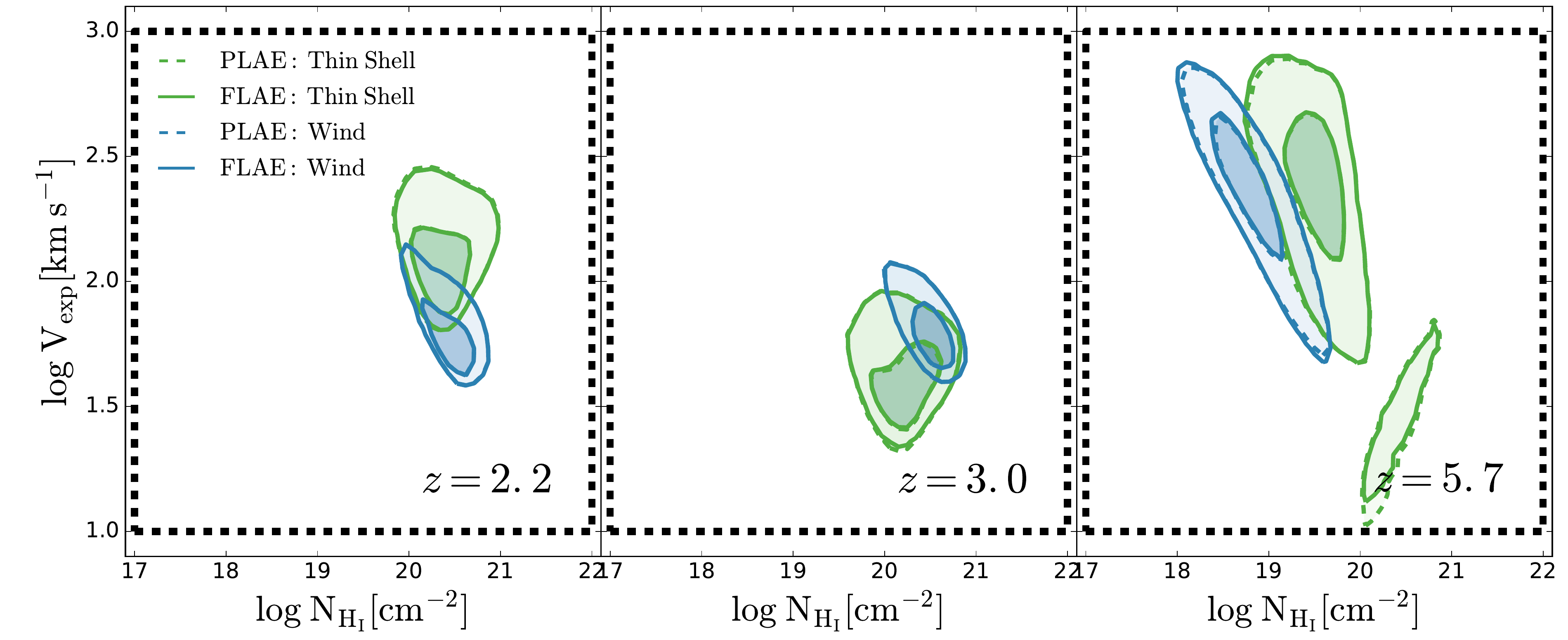}
    \caption{ Distribution of LAE in the $\rm V_{exp}$-$\rm N_{H_I}$ space at redshift 2.2, 3.0 and 5.7 from left to right. The Thin Shell and Galactic Wind model are shown in green and blue. The dark countors represent the $1\sigma$  while the light the $2\sigma$. The full RT models are shown in solid lines while the models with RT in the ISM but not in the IGM are plotted in dashed lines. In dashed black line we show \flareon\  borders.}\label{fig:NH_Vexp}
\end{figure*}

In Fig. \ref{fig:NH_Vexp} we show the $\rm N_{H}-V_{exp}$ distributions at different redshifts (solid colored lines) and the \flareon\ accuracy boundaries (black lines). Over all, the Thin Shell and the Galactic Wind models behave likely and there is always some overlap between them. Additionally, as described above, there is little variation between the properties of the samples including the full RT (\flae) and galaxies with only RT in in the ISM (\nlae). At low redshifts the $\rm N_{H}-V_{exp}$ distributions are quite compact. In particular, at redshift 2.2 both geometries have a similar $\rm N_H$ distribution that peaks at $\sim 10^{20.5} cm^{-2}$, while the velocities of the Thin Shell ($\sim 10^{2} km \;s^{-1}$) are slightly above the Galactic Wind distribution ($\sim 10^{1.8} km \;s^{-1}$). Meanwhile, $z=3.0$ the velocity rounds $\sim 10^{1.5} km \;s^{-1}$ and $\sim 10^{1.8} km \;s^{-1}$ in the Thin Shell and Galactic Wind respectively, while $\rm N_H$ are close to $\sim 10^{20.2} cm^{-2}$ and $\sim 10^{20.8} cm^{-2}$. 

Finally, at $z=5.7$ the distributions become broader. The velocities are higher and similar in both geometries, rounding $\sim 10^{2.4} km \;s^{-1}$, while the Thin Shell have higher $\rm N_H$ ($\sim 10^{19.8} cm^{-2}$) than the Galactic Wind ($\sim 10^{18.8} cm^{-2}$). We find that at that the fraction of LAEs outside the \flareon\ range is less than a 3\% in any of our samples.


\begin{figure*}
    \includegraphics[width=7.0in]{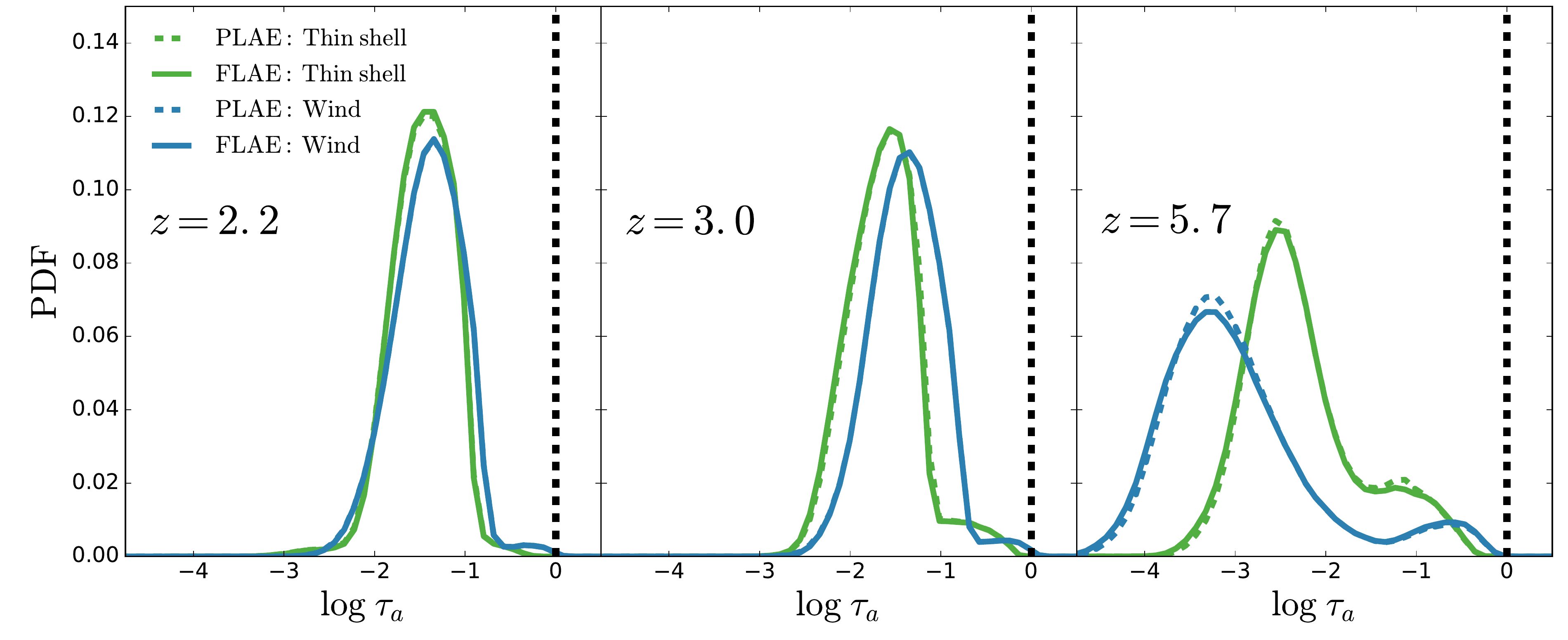}
    \caption{ Distribution of the dust optical depth our models including RT in the ISM at redshift 2.2, 3.0 and 5.7 from left to right. The color code of the lines is the same as in Fig.\ref{fig:NH_Vexp}. }\label{fig:tau_dust}
\end{figure*}


In Fig. \flareon\ we show the resulting $\tau_a$ distribution after calibration for our LAE samples. The $\tau_a$ distribution both geometries at $z=2.2$ are narrow and centered at $\sim 10^{-1.5}$. At redshift 3.0 the distribution becomes slightly broader. Additionally, the Thin Shell exhibits lower dust optical depth ($\sim 10^{-1.8}$) than the Galactic Wind ($\sim 10^{-1.5}$). Meanwhile, at $z=5.7$ the distribution become wider and the Galactic Wind have lower $\tau_a$ ($\sim 10^{-3.5}$) than the Thin Shell ($\sim 10^{-2.5}$). We also show the accuracy border of \flareon\ at $\log \tau_a = 0.0$. We find that less than 1\% of the LAE in our sample lie outside \flareon\ $\tau_a$ dynamical range. All together we find that the fraction of LAEs outside \flareon\ accuracy range is negligible.


\bsp	
\label{lastpage}
\end{document}